\documentclass[aip,jap,amsmath,amssymb,preprint]{revtex4-1}
\usepackage{graphicx}
\usepackage{multirow}
\usepackage{subfigure}
\usepackage{dcolumn}
\usepackage{bm}
\usepackage[utf8]{inputenc}
\usepackage[T1]{fontenc}
\usepackage{mathptmx}
\usepackage{textcomp}
\usepackage{etoolbox}
\usepackage{ulem}
\usepackage{float}
\makeatletter
\def\@email#1#2{%
	\endgroup
	\patchcmd{\titleblock@produce}
	{\frontmatter@RRAPformat}
	{\frontmatter@RRAPformat{\produce@RRAP{*#1\href{mailto:#2}{#2}}}\frontmatter@RRAPformat}
	{}{}
}%
\makeatother
\begin{document}
	\title{Phase-field study of surface diffusion enhanced break-ups of nanowire junctions}
	\author{Abhinav Roy}
	\affiliation{Department of Metallurgical and Materials Engineering, National Institute of Technology Rourkela, Odisha, 769008 INDIA }
	\author{Arjun Varma R.}
	\author{M. P. Gururajan}
	\email{gururajan.mp@gmail.com; guru.mp@iitb.ac.in}
	\affiliation{Department of Metallurgical Engineering and Materials Science, Indian Institute of Technology Bombay, Powai, Mumbai, Maharashtra, 400076 INDIA}
	
	\begin{abstract}
		Using a phase-field model which incorporates enhanced diffusion at the nanowire surfaces, we study the effect of different parameters on the stability of intersecting nanowires. Our study shows that at the intersection of nanowires, sintering (curvature driven material flow) leads to the formation of junctions. These junctions act the initiators of nanowire break-up. The subsequent break-ups take place due to Rayleigh instability at the arms away from these junctions. Finally, at long time scales, the fragments coarsen due to the differences in sizes. The radii of the nanowires that form the junction, the difference in size of the intersecting nanowires and the angle of intersection play a dominant role in determining the kinetics of break-up while the density of intersections has little or no effect on the kinetics. We rationalise
our results using maps of (i) mean curvatures (and, hence, chemical potentials), and, (ii) Interfacial Shape Distributions (ISDs) (which are
based on probability densities associated with different combinations of the two principal curvatures). Finally, we use the moment of 
inertia tensor to characterise the (non-spherical) shapes and morphologies of (central) nanowire fragments at the junctions.
	\end{abstract}
	\maketitle
	\section{Introduction}
	\label{section1}
	
	Metallic nanowires are used in a wide range of applications: for example, their ductility and bendability allows for their use in solar cells, flexible and transparent electronic devices, light emitting diodes, and so on -- see, for example,~\cite{Lee2012, Sannicolo2016, Lee2020, Gonzalez-Garcia2016}. Regular ordered array of metallic nanodots have been used in biosensing applications~\cite{Vazquez-Mena2011}. In particular, network of Ag nanowires are considered as the next generation of transparent conducting electrodes due to their enhanced electrical and optical properties~\cite{Song2015, Langley2014}. Given such widespread use, their stability at elevated temperatures under standard operating conditions is an important area of study. Thus, an understanding of the driving forces for morphological changes and control of the parameters which affect those driving forces are of significant academic and technological importance. 
	
	The effect of nanowire size, network density and temperature on the stability of metallic nanowires have been investigated experimentally~\cite{Shin2007, Rauber2012, Lagrange2015}. The structural stability of the metallic nanowires at the nanoscale is governed by numerous factors \textemdash~the most dominant factors being Rayleigh instability~\cite{Rayleigh1878, Nichols1965, Ma1998} and anisotropy in interfacial energy~\cite{Cahn1979, Stolken1992, Gurski2003, Kim2015, Gorshkov2020}. Thermally accelerated surface diffusion plays a central role in the resulting change in morphology of the metallic nanowires due to Rayleigh instability, and their subsequent break-up into nanodots~\cite{Molares2004, Karim2006, Karim2007, Huang2010, Beavers2010, Hsiung2010, Oh2018}. Rayleigh instability mediated nanowire fragmentation has also been observed in non-metallic nanowires~\cite{Kolb2005, Bechelany2012}. Quantum effects at the nanoscale also play a role in morphological evolution in conjunction with Rayleigh instability and have been investigated both experimentally and numerically~\cite{Burki2005, Kassubek2001}. 
	
	It has been  observed experimentally that nanowire junctions have a drastic effect on the break-up kinetics of Au and Ag nanowires when annealed at elevated temperatures much below their melting point~\cite{Vigonski2017}. Presence of such junctions between the nanowires lead to initial break-up preferentially at the junction, followed by break-up at remainder of the nanowires. Such fragmentation behaviour of nanowires can be exploited to produce ordered array of nanodots for different technological applications~\cite{Xue2016,Zhu2019}.
	
	In the past, various modelling approaches (mostly atomistic) have been adopted to simulate the nanoparticle morphologies and various effects at the nanoscale \textemdash~a review of such approaches can be found in~\cite{Barnard2010} and references therein. Linear stability analysis of Rayleigh instability is also well understood; for example, a linear stability analysis of the effect of general surface energy anisotropy on Rayleigh instability has been carried out in~\cite{Gurski2003} (extending the work of Cahn~\cite{Cahn1979}).
	
	Phase-field modelling is an efficient numerical technique for simulating microstructural evolution at the mesoscopic length scales and diffusive time scales~\cite{Chen2002, Steinbach2011}. There exist several phase-field models to study variable mobility; for example, variable mobilities have been introduced in the past  to study phenomena such as domain growth in binary mixtures~\cite{Lacasta1992}, late-stage coarsening in phase separating systems~\cite{Bray1995, Puri1997}, and for determining the coarsening kinetics of bulk-diffusion-controlled and interface-diffusion-controlled growth in systems with interconnected phases~\cite{Zhu1999}. The use of variable mobility phase-field models has also been made for effectively incorporating the effect surface diffusion, assuming isotropic interfacial energy~\cite{Geslin2019, Salvalaglio2020, Andrews2020}. Such variable mobility phase-field models have been used in simulating the Rayleigh instability in the solid-state~\cite{Joshi2016}, destabilisation of nanoporus membranes by grain boundary grooving~\cite{Joshi2017}, thermal stability of nanoporus aggregates~\cite{Mukherjee2011},  and, instability in multi-layer nanocrystalline thin-films due to Rayleigh instability driven by grain boundary~\cite{Chakrabarti2017}. 
	
	Phase-field models have been developed to incorporate anisotropy in interfacial energy using either trigonometric functions for the interfacial energy coefficient~\cite{Debierre2003, Ji2018}, or using higher order tensor terms which has benefits over the former approach~\cite{Abinandanan2002}. Phase-field model for stability of nanowire fragmentation with regularized trigonometric function of interfacial energy along with finite amplitude axisymmetric perturbations has been developed previously~\cite{Wang2011}. Higher order tensor terms in the free energy functional of phase-field models have been used to study the faceting of precipitates due to interfacial energy anisotropy~\cite{Roy2017}. 
	
	Our aim, in this paper, is to implement
	a continuum model (based on the Cahn-Hilliard equation~\cite{Cahn1958}) for long time evolution of the morphologies of the nanowires with enhanced surface diffusion. Further, the formulation and its implementation are capable of accommodating surface 
	energy anisotropy. Specifically,
	we use the extended Cahn-Hilliard equation which consists of a fourth order tensor term in order to incorporate cubic surface energy anisotropy, and, following 
	the approach adopted in some of the previous phase-field studies mentioned above, we define an order-parameter dependent mobility function. This helps us incorporate enhanced surface diffusion in our model.
Using this model, we carry out a systematic study of various factors such as wire diameters, the angles
	of intersection, and density of intersections in the simulation cell on 
	fragmentation of intersecting nanowires; our preliminary, 2-D results on the effect of anisotropy are given
	as an appendix.	
	
	At this point, we want to note that there exist phase-field studies pertaining to the phenomena of solid-state dewetting in nanowires~\cite{Dziwnik2017,Jiang2012, Kim2015}. Different phenomena like the formation of nanoparticles through solid-state dewetting of a thin-film on a substrate~\cite{Verma2020}, stress effects on solid-state dewetting of thin-films~\cite{Cheynis2012}, solid-state dewetting of Au aggregates on titanium oxide nanorods~\cite{Liu2017}, and effect of surface energy anisotropy on Rayleigh-like solid-state dewetting~\cite{Kim2015} have been studied in the past. Note that there is also the possibility of Asaro-Tiller-Grinfeld instabilities in nanowires
	on substrates -- see, for example~\cite{LiuEtAl2019}. However, unlike these models which include the substrate on which dewetting takes place, in our model, the wires are free-standing. 
	
	The rest of this paper is organized as follows: we describe (albeit briefly) the  formulation and numerical implementation of the phase-field model in section~\ref{section2}, which also contains the simulation details. Results and discussion follow in section~\ref{section3} where we discuss the effect of various parameters on the kinetics of junction break-up systematically, followed by a discussion on the morphology
	of the fragments~\ref{section4}; we conclude the paper with a summary of our salient conclusions in section~\ref{section5}. 
	
	\section{Phase-field model}
	\label{section2}
	
	As indicated in the introduction, our phase-field model is a combination of extended Cahn-Hilliard
	model for cubic anisotropy in interfacial energy~\cite{Abinandanan2002} coupled with enhanced 
	surface diffusion implemented using a variable mobility~\cite{Zhu1999}. Since these models are well 
	known in the literature, we briefly describe the models and other details in this section, for the sake 
	of completion. 
	
	\subsection{Formulation}	
	
	We consider a conserved, non-dimensionalised order parameter, denoted by $c(x,t)$ to describe our system. This order parameter takes a value of zero in vacuum and unity in  the (nanowire) material albeit across a flat interface;  in the circular cross-section nanowire-vacuum geometry, the order parameter value shifts from unity and zero to account for Gibbs-Thomson effect.
	
	The free energy functional of the system is given by the expression: 
	\begin{equation}
		F = N_{V} \int_{\Omega}  \left\{f_0(c) + \frac{\kappa_{c}}{2} (|\nabla c|)^2 + \frac{\gamma_{\langle hkl \rangle}}{2} (\nabla^2 c)^2 \right\} d\Omega,
		\label{eq2}
	\end{equation}
	where, $N_V$ denotes the number of atoms per unit volume (assumed to be constant), $f_0(c)$ is
	the bulk free energy density per atom, $\kappa_{c}$ is the gradient energy coefficient (assumed
	to be a scalar and hence gives rise to an isotropic
	interfacial free energy), and, the third term accounts for cubic interfacial free energy anisotropy as explained below. 
	
	The cubic anisotropy in the interfacial energy of the nanowires is incorporated into the model using the extended Cahn-Hilliard formulation~\cite{Abinandanan2002}, which is the third term in the equation. In Eq.~\ref{eq2}, $\gamma_{\langle hkl \rangle}$ is the coefficient of a fourth order term which is defined for a particular crystallographic orientation as:
	\begin{equation}
		\gamma_{\langle hkl \rangle} = \gamma_{I} + \gamma_{A}\left(h^{4} + k^{4} + l^{4}\right), 
		\label{eq3}
	\end{equation}
where, $\gamma_{A}$ and $\gamma_I$ are the anisotropic and isotropic contributions from the fourth rank term, and $h,\;k,\;l$ denote the Miller Indices of the normal to the interface. 
Abinandanan and Haider~\cite{Abinandanan2002} have also shown that the interfacial 
energy scales as $\gamma_{\langle hkl \rangle}^{\frac{1}{4}}$. The values of $\gamma_{A}$ and $\gamma_{I}$ can be chosen using the scaling curve 
given by Abinandanan and Haider~\cite{Abinandanan2002}, and knowing the ratio of interfacial energies along different directions.
	
	In Eq.~\ref{eq2}, the function $f_0(c)$ represents the bulk free energy density per atom, which is given by the polynomial, 
	\begin{equation}
		f_0(c)=A_{c}\left[c^2(1-c)^2\right].
		\label{eq4}
	\end{equation}
	This polynomial produces a double well potential with the energy minima at $c=0$ (vacuum) and $c=1$ (nanowire material). 
	$A_c$ is a coefficient that determines the height of the potential energy barrier. Together with the gradient energy coefficient 
	$\kappa_c$, $A_c$ determines the interfacial energy and width in the system  (in the absence of the $\gamma_{<hkl>}$ term; 
	when present, $\gamma_{<hkl>}$ along with $A_c$
	and $\kappa_c$ determine the interfacial properties, namely, energy and width).
	
	The chemical potential is derived from the variational derivative of the free energy functional Eq.\ref{eq2}:
	\begin{equation}
		\mu = \frac{1}{N_V}\frac{\delta F}{\delta c} = \left[ \frac{\partial f_0(c)}{\partial c} - \kappa_{c} \nabla^2 c + \gamma_{<hkl>}\nabla^4 c \right].
		\label{eq5}
	\end{equation}
	
	The evolution of the system is described by the modified Cahn-Hilliard equation, which is given by the expression:
	\begin{equation}
		\displaystyle{\frac{\partial c}{\partial t} = \nabla \cdot \left[ M(c) \cdot \nabla\left(\\frac{\partial f_0(c)}{\partial c} - \kappa_{c}\nabla^2c + \gamma_{<hkl>}\nabla^4 c \right)\right]},
		\label{eq6}
	\end{equation}
	where, $M(c)$ denotes the mobility as a function of the order parameter, and is used to capture the effect of surface diffusion. 
	
	Various polynomials have been suggested in the literature for use as variable mobility functions in the Cahn-Hilliard equation. Most commonly used mobility functions in previous studies have been quadratic and quartic function~\cite{Cahn1996, Gugenberger2008, Dziwnik2019, Shin2019, Hoffrogge2021} of order parameter, to incorporate the dominating role of surface diffusion in the models. The mobility function used
	in our model serves the same purpose, and has an added advantage of reducing the stiffness of the equations in conjunction with the semi-implicit Fourier spectral numerical method. The function is defined as follows:
	\begin{equation}
		M(c) = \left[c(1-c)\right]^\frac{1}{2}.
		\label{eq7}
	\end{equation}
	 
	Substituting Eq. \ref{eq7} into Eq. \ref{eq6}, the evolution equation can be written as: 
	\begin{equation}
		\displaystyle{\frac{\partial c}{\partial t} = \nabla \cdot \left[ \left\{c(1-c)\right\}^\frac{1}{2} \cdot \nabla\left(\frac{\partial f_0(c)}{\partial c} - \kappa_{c}\nabla^2c + \gamma_{<hkl>}\nabla^4 c \right)\right]}.
		\label{eq8}
	\end{equation}

	Note that the polynomial~\eqref{eq7} gives the maximum mobility along the surface of the wire ($c=0.5$) and is 
zero within the nanowire ($c=1$) and outside it ($c=0$). Thus, in these simulations, surface diffusion plays a dominant role in material 
transport. In order to make sure that zero mobility in the bulk and vacuum does not produce any numerical instabilities, we have carried out 
simulations taking the bulk and vacuum mobilities to be two orders of magnitude less than that of surface mobility; we have found that the time 
for the first pinch-off at the junction remains the same in both these cases: see the Supplementary Information.

	\subsection{Numerical implementation}
	
	The evolution equation~(Eq.~\ref{eq8}) is solved numerically in order to track the morphological evolution of both finite and infinite cylindrical nanowires (in both 2-D and 3-D). We use the 
	semi-implicit Fourier spectral technique for solving the evolution equation. This method is known to be efficient for solving non-linear partial differential equations and also eliminates the severe time-step constraint~\cite{Zhu1999,Chen1998}. We follow the same method of discretisation used in \cite{Zhu1999}. We transform equation Eq. \ref{eq8} to Fourier space and perform first order forward finite-difference discretisation in time. 
	
	\begin{eqnarray}
		\frac{{\tilde{c}({\bf k })}^{t+\Delta t} - {\tilde{c}({\bf k })}^t}{\Delta t} = i {\bf k } \{\{c(1-c)\}^\frac{1}{2} \cdot [i {\bf k }^{\prime} \left(\tilde{g}(c^t) 
		+ \tilde{c}({\bf k^{\prime}})^t\chi \right) ]_{r}\}_{k},
		\label{eq9}
	\end{eqnarray}
	where, $i$ is the pure imaginary number,
	$$
	\chi = \left(\kappa_{c} k^{\prime2}  + \gamma_{I}k^{\prime 4} + \gamma_{A}\left(k_{x}^{\prime 4} + k_{y}^{\prime 4} + k_{z}^{\prime 4}\right)\right),
	$$
	$\tilde{c}({\bf k }, t)$ is the non-dimensionalised order parameter field in the Fourier space and, ${\bf k}$ and ${\bf k^{\prime}}$ are the Fourier space vectors. ${\bf |k|} = k = \sqrt{k_{x}^{2} + k_{y}^{2} + k_{z}^{2}}$, where $k_{x}$, $k_{y}$ and $k_{z}$ are the three components of the Fourier space vector. 
	The operation $[\cdot]_{r}$ represents the inverse spatial Fourier transform of the quantity in square brackets to the real space, while $\{\cdot\}_{k}$ represents the forward spatial Fourier transform of the quantity in curly brackets. The function $g(c)$ represents the derivative of the bulk free energy density function with respect to $c$ and is defined as:
	\begin{equation}
		g(c) = \frac{\partial f_0(c)}{\partial c} = 2A_{c}\left\{c(1-c)(1-2c)\right\}.
		\label{eq10}
	\end{equation}
	The severe time step constraint associated with the explicit solution of the equation is circumvented by introducing a suitable stabilizing constant, which separates the mobility function into two parts: $\xi$ and $\{c(1-c)\}^\frac{1}{2} - \xi$, after Zhu et al. \cite{Zhu1999}. Therefore, the modified evolution equation becomes: 
	\begin{eqnarray}
		\frac{{\tilde{c}({\bf k })}^{t+\Delta t} - {\tilde{c}({\bf k })}^t}{\Delta t} = i {\bf k } \left\{\left[\xi + \left \{\{c(1-c)\}^\frac{1}{2} - \xi \right \} \right] \cdot [i {\bf k }^{\prime}(\tilde{g}(c^t) + \tilde{c}({\bf k^{\prime} })^t \chi) ]_r\right\}_{k}.
		\label{eq11}
	\end{eqnarray}
It is known that the choice 
of $\xi$ that lies in the middle of the maximum and 
minimum values of the mobility gives the best performance 
for the numerical scheme~\cite{Zhu1999}. Hence,
in these simulations, we have taken $\xi$ as $\frac{1}{2} (\mathrm{max} [M(c)] + \mathrm{min} [M(c)]) = \frac{1}{2} (0.5+0) = 0.25$.
	
	After some algebraic manipulation, the evolution equation can be written as:
	\begin{eqnarray}
		\beta \tilde{c}({\bf k})^{t+\Delta t} = \beta \tilde{c}({\bf k })^{t} + i {\bf k } \Delta t\{{c(1-c)}^\frac{1}{2}\cdot [i {\bf k }^{\prime} \left(\tilde{g}(c^t) +  \tilde{c}({\bf k^{\prime}}\right)^t \chi ]_{r}\}_{k}
		\label{eq12}
	\end{eqnarray}
	where, $\beta = \left(1+\xi\Delta t\chi \right)$.
	
	Thus, knowing the $c$ at a given time $t$, the $c$ at time $t+\Delta t$ can be obtained
	in the Fourier space using Eq.~\ref{eq12}.
	
	\subsection{Simulation details}
	
	We have carried out simulations of finite and infinite nanowires in both 2- and 3-D. In this paper, we present 3-D results for wires with isotropic interfacial energy (and 2-D results for wires with (cubic) anisotropic interfacial energy, in the Appendix A.). The deployment of Fourier spectral technique for the numerical solution implies imposition of periodic boundary conditions.  We introduce a small noise in the scaled order parameter, which adequately simulates the thermal noise in the system. We used the software package Fastest Fourier Transform in the West (FFTW3)~\cite{Frigo2005} for computing the discrete Fourier transforms. We use the same non-dimensionalisation as described in Abinandanan and Haider~\cite{Abinandanan2002}. The non-dimensional simulation parameters used in our simulations
	are tabulated in Table \ref{Table1}. Note that the lengths are reported in units of grid lengths for the rest of 
	this paper; that is, an $R$ value of 16 corresponds to a non-dimensional length of 8 in the 3-D isotropic case, and so on.
	\begin{table}
		\caption{Non-dimensional simulation parameters}
		\begin{ruledtabular}
			\begin{tabular}{lll}
				\textbf{Simulation Case}  & \textbf{Simulation Parameters} & \textbf{Values}\\ \hline
				\multirow{3}{*}{General parameters}    & $\kappa_{c}$ &   1.0\\ 
				& $A_{c}$  &   1.0\\
				& $\xi$   &   0.5\\	\hline 
				\multirow{4}{*}{Isotropic interfacial energy case}
				& $Nx \times Ny \times Nz$ (3D)   &   $768 \times 768 \times 96$\\ 
				& $\Delta x = \Delta y = \Delta z$    &   0.5\\
				& $\Delta t$ &   1.0\\ 
				& $\gamma_{I} = \gamma_{A}$ &   0\\ \hline
				\multirow{3}{*}{Anisotropic interfacial energy case}
				& $Nx \times Ny$ (2D) &   $512 \times 512$\\ 
				& $\Delta x = \Delta y$   &   1.0\\ 
				& $\Delta t$ &   0.01\\ 
				& $\gamma_{I}$ &   -50 \cite{Roy2015}\\ 
				& $\gamma_{A}$    &   107.69 \cite{Roy2015}\\
			\end{tabular}
		\end{ruledtabular}
		\label{Table1}
	\end{table}
	\section{Results and discussion}
	\label{section3}

	The angle of intersection of the nanowires,
	the radii of the nanowires and the size difference in the radii between the nanowires
	that form the intersections, and
	the density of intersections are the parameters
	that are of interest to us. In this section, we
	present our results for all these scenarios. Before we do so,
	in order to benchmark our implementation, we first present results
	from single nanowire simulations and show that the results are in good agreement
	with known analytical results for Rayleigh instability in infinite rods.

	\subsection{Single nanowire simulations}

	We have carried out simulations on free standing single finite and infinite nanowires of radii of 12, 13, 14, 15, and 16
	in 3D. The interfacial energy in all these cases in isotropic.	The morphological snapshots for radius 12 nanowire for both finite and 
	infinite cases are shown in Fig.~\ref{single_nanowires}. The morphological evolution for the other radii are given in the supplementary
	information.

	From the morphological snapshots at different timesteps it can be seen that the wires fragment from the edges in the case of a 
	finite wire. The free ends retract, and bulge to form constrictions along the body of the nanowires, which lead to further break-up. 
	In the case of infinite nanowires, the perturbations set in and lead to break-up of nanowire into fragments; the first break-up, however,
	in contrast to finite nanowire, occurs at a random location along the body of the nanowire.

    It is well known in the literature that in case of infinite solid cylindrical rods with isotropic surface energy, the fragment separation is dominated by the maximally growing wavelength ($\lambda_{max}$). This maximally growing wavelength is related to initial radius of the cylinder as $\lambda_{max} = 8.89 R$ (See Ref. \cite{Karim2006, Mullins1957}) in case of purely surface diffusion mediated instability. From this expression, the average separation between nanoparticles formed after fragmentation should be 106.68 and 124.46 units for radius 12 and 14, respectively. It is observed from simulations that when the fragmentation starts, the average separation between particles in case of radius 12 is about 108 $\pm$ 19 and in case of radius 14 is about 129 $\pm$ 54 -- indicating that the separation between the fragmented nanoparticles in our simulations is 
    broadly consistent with the maximally growing wavelength obtained from the analytical expressions for Rayleigh instability. The large error 
    bars are because of subsequent break-ups and coarsening of the initial fragments.
     
	\begin{figure}[!h]
        \centering
        \includegraphics[scale=0.8]{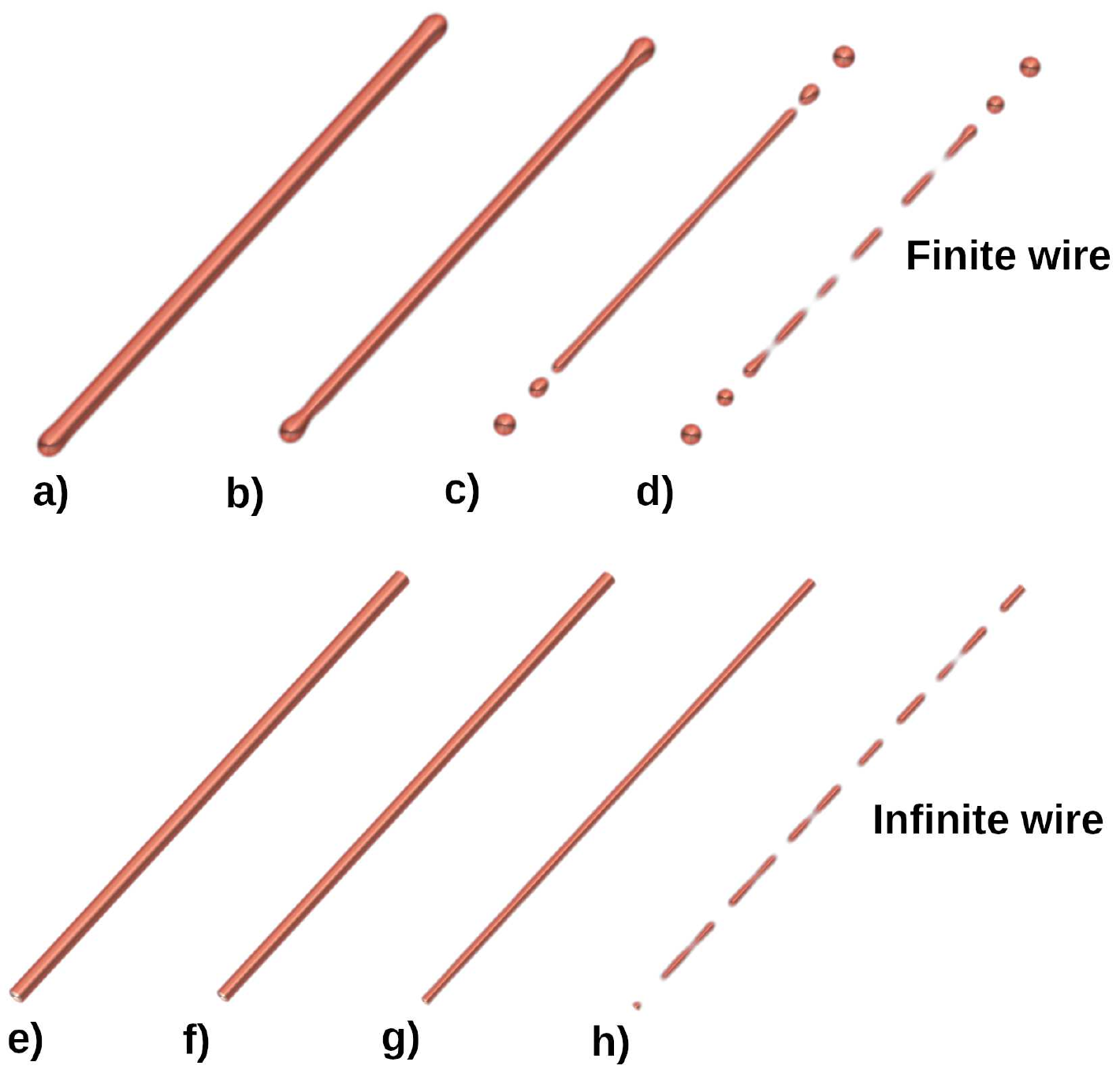}
        \caption{Single nanowire simulations for finite nanowires of radius 12 at (a) t=500 (b) t=1500 (c) t=2700 (d) t=2800 and infinite nanowires at (e) t=500 (f) t=1500 (g) t=2700 (h) t=2800 respectively.}
        \label{single_nanowires}
    \end{figure}
    
\begin{figure}
       \centering
       \includegraphics{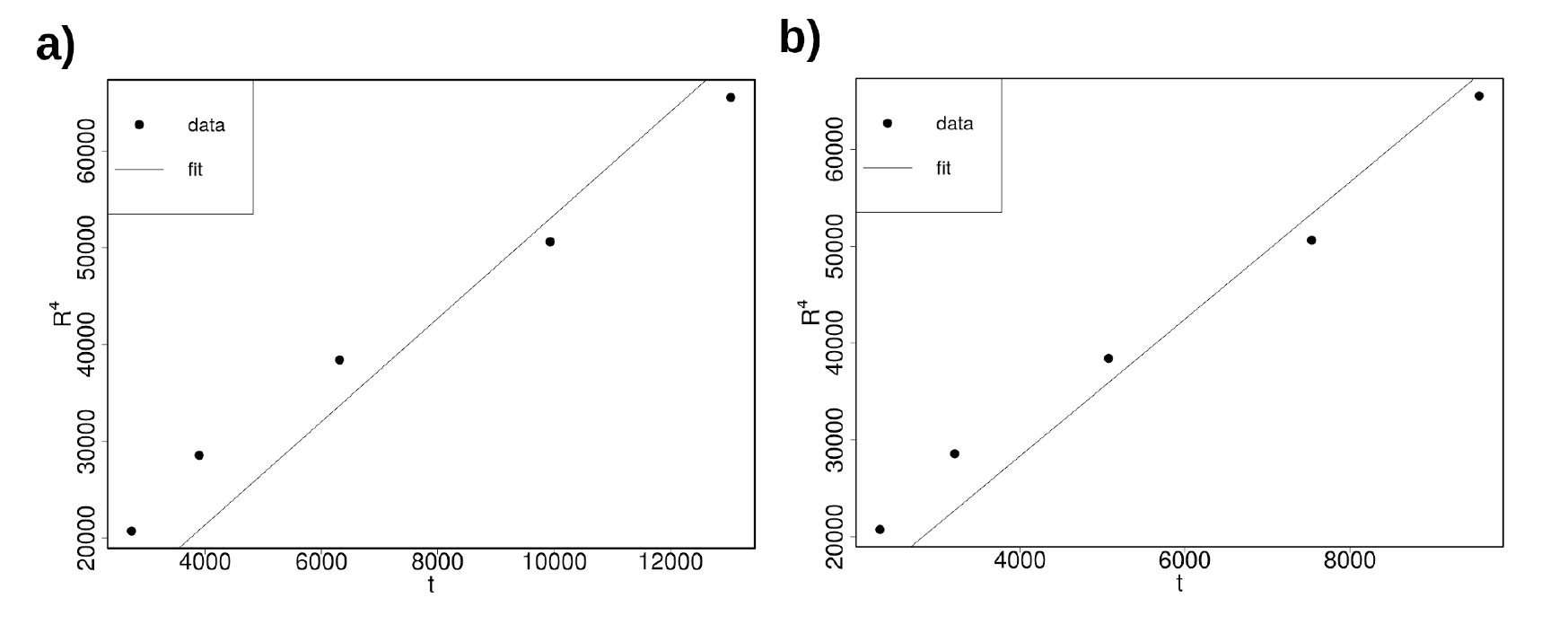}
       \caption{The plot of first pinch-off time ($t_f$) against R$^4$. We fit a line ($R^4 = m t_f$) using non-linear least squares method 
       to the data points. The value of slopes are: $m = 5.34 \pm 0.33$ ($R^2=0.98$) for the infinite wire. In the case of the finite wire, $m = 7.07 \pm 0.30$ ($R^2=0.99$).} 
       \label{pinchoff}
   \end{figure}

In order to confirm that it is the Rayleigh instability that causes for nanowire fragmentation in our simulations, 
we have plotted the time for first pinch-off $t_f$ against $R^4$ in Fig.~\ref{pinchoff}. We fit a straight line of the form 
$R^4 = m t_f$ using non-linear least squares method 
       to the data points. The value of slopes are: $m = 5.34 \pm 0.33$ ($R^2=0.98$) for the infinite wire. In the case of the finite wire, $m = 7.07 \pm 0.30$ ($R^2=0.99$).  
       As seen in the figure, the fit is good and in line with the expectation from 
theoretical studies of surface diffusion mediated Rayleigh instability~\cite{Nichols1965}. 

    Finally, using these single wire simulations, we also calculate the change composition due to the Gibbs-Thomson 
    effect. For our case, where the
    equilibrium compositions of the nanowire and the vacuum across a planar interface is kept at unity and zero respectively, this
    change is given by (see, for example, Appendix E of Ref.~\cite{gururajan2006})
    \begin{equation}
    \Delta c = \frac{\chi \gamma}{\frac{\partial^2 f_0}{\partial c^2}}
    \end{equation}
     In this equation, $\Delta c$ is the difference between the equilibrium composition for a flat interface and the curved interface with mean
     curvature $\chi$; the mean curvature of the interface ($\chi = 1/R$), where R is the instantaneous radius of the nanowire; $\gamma$ is
     the interfacial free energy and the denominator is the second derivative of the bulk free energy density evaluated at the equilibrium 
     composition. The (scaled) interfacial energy ($\gamma$) in our model takes a value of  0.33. The second derivative of free energy in our model takes a value of 2. Using these values, for a single nanowire of radius 16, the change in composition ($\Delta c$) is analytically
     evaluated to be $\Delta c = 0.0175$ and we obtain a $\Delta c$ of $0.017$ from our simulations. Similarly, for a nanowire
     of radius 12, the analytically calculated value is $\Delta c = 0.018$ and in our simulations of $90^{\circ}$ configuration 
     with radii 12 for both the nanowires, we see $\Delta c = 0.019$. Thus, our phase field model accurately captures
     the change in composition in the system due to Gibbs-Thomson effect.

	\subsection{Effect of relative orientation}
	
\begin{figure}
		\centering
		\includegraphics[width=\textwidth]{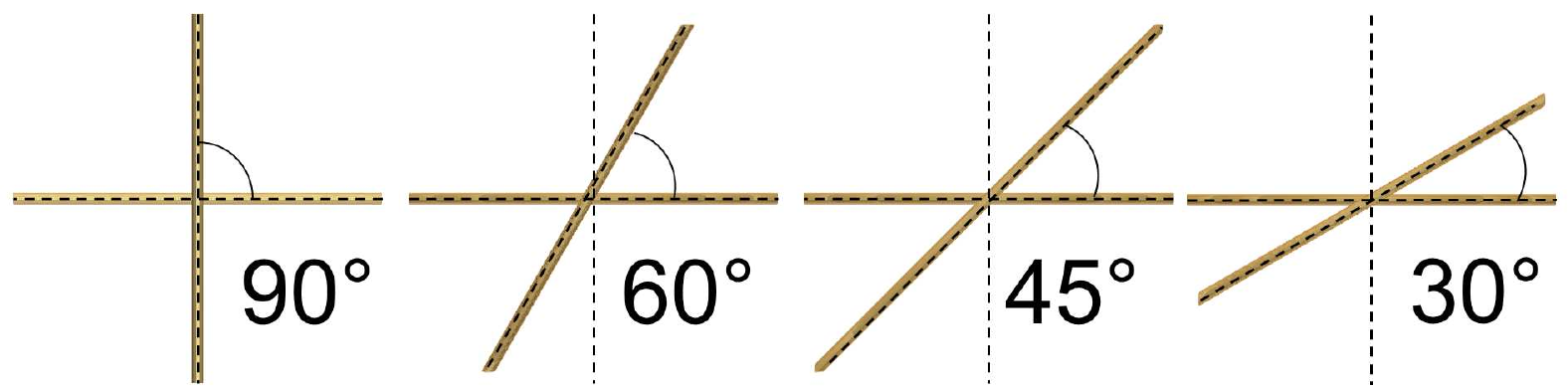}
		\caption{Schematic (top view) of the relative orientation between the wires used in the current study}
		\label{Fig1}
	\end{figure}

	We consider a system with isotropic interfacial
	energy (in 3D) with a single intersection of 
	the nanowires (of circular cross-section) in 
	the simulation cell. We assume
	periodic boundary conditions. As seen from the
	schematic top view of the relative 
	orientations in Fig.\ref{Fig1}, in the case of an 
	intersection
	angle of $90^{\circ}$, both the wires are 
	infinite. However, for all the other three 
	angles, the nanowire parallel to the $x$-axis
	of the simulation cell is infinite while the 
	other one is finite. Having said that, we have 
	chosen
	big enough system sizes so that the break-up at
	the edges of the finite nanowire does not 
	affect the break-up at the intersection.
	
\begin{figure}[!h]
	    \includegraphics[width=\textwidth]{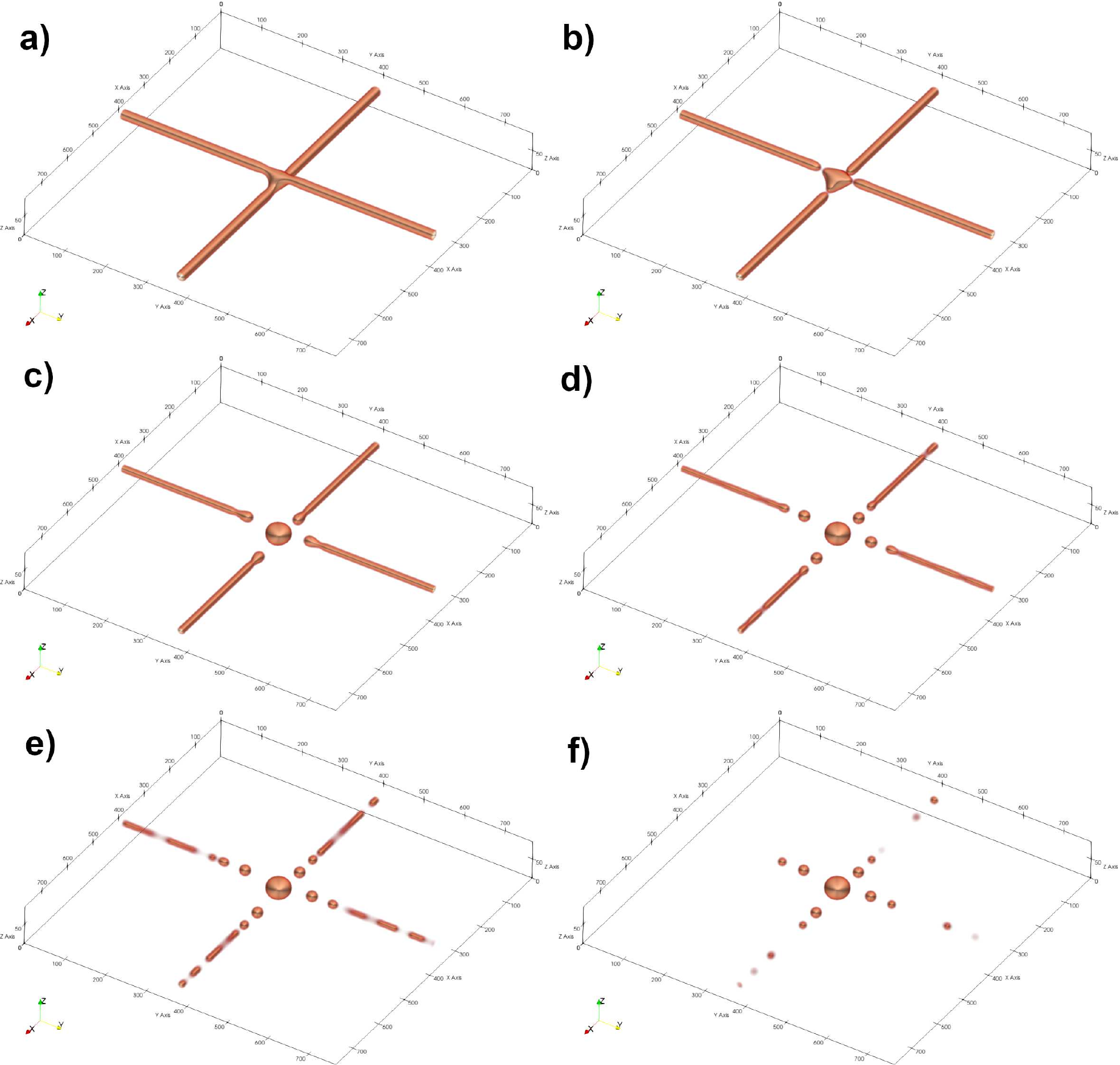}	
		\caption{Morphological evolution of two nanowires intersecting at $90^{\circ}$ at different times:
		(a) 1000, (b) 1750, (c) 2500, (d) 2750, (e) 2800 and (f) 2900 time units. The video of the break-up is available as supplementary data.}
		\label{Fig2}
	\end{figure}

	In Fig.\ref{Fig2}, we show the time evolution of 
	the nanowires of equal initial radii 
	($\mathrm{R_1} = \mathrm{R_2} = 12$ (non-dimensional) length units) with
	an angle of $90^{\circ}$ at the intersection to
	begin with. Note that in 3D, one nanowire (in
	this case, the one along the x-axis)
	is at the bottom and the other on top (in this 
	case, the one along the y-axis); in this and the subsequent cases, $\mathrm{R_1}$ represents the radius of the nanowire at the bottom and
	$\mathrm{R_2}$ represents the radius of the nanowire on top. 
	As can be seen from the figure, due to the high curvatures at the point of contact at the intersection, initially, a junction forms at the intersection (Fig.~\ref{Fig2}(a), corresponding to a (non-dimensional) time of 1000 units). The material accumulation at the junction leads to the break-up of the nanowire
	(Fig.~\ref{Fig2}(b), corresponding to a (non-dimensional) time of 1750 units). This, in turn, leads to subsequent break-up of the nanowire as seen in Fig.~\ref{Fig2}(c)-(e), corresponding to a (non-dimensional) times of 2500, 2750 and 2800 units. As can be seen from Fig.~\ref{Fig2}(f), corresponding to a (non-dimensional) time of 2900 units, the
	broken pieces of the nanowire coarsen due,
	primarily to, the difference in sizes between the central particle and the others along the wire length. 
	
	After fragmentation of the junction, the free ends of the nanowire retract and subsequently break-up. After the break-up
	at the junction, the arms of the wire are similar to a single finite nanowire in their geometry. Hence, one can ask if their break-up is also 
	similar to finite nanowires. As we can see, although the sequence of fragmentation is the same, the kinetics is not the same
	in these systems. In case of a single finite nanowire of radius 12, the first pinch-off event (ovulation) occurs at 2303 time units, 
	the second pinch-off occurs at 2315, the third pinch-off occurs at 2666, and the fourth pinch-off occurs at 2680. Therefore, the first set of 
	two pinch-off events are closely spaced because these events occur at the two free ends, and similarly for the next set of two pinch-off 
	events which again occur at the free ends. In case of the $90^{\circ}$ configuration, first fragmentation occurs at the junction at 1719 time
	 units, and subsequently the second pinch-off event along the arm of nanowire occurs are 2549 units. Thus, in the 
	 presence of the junction, the time to first failure is faster as compared to a single, isolated finite nanowire.
	
\begin{figure}
		\centering
        \includegraphics[width=\textwidth]{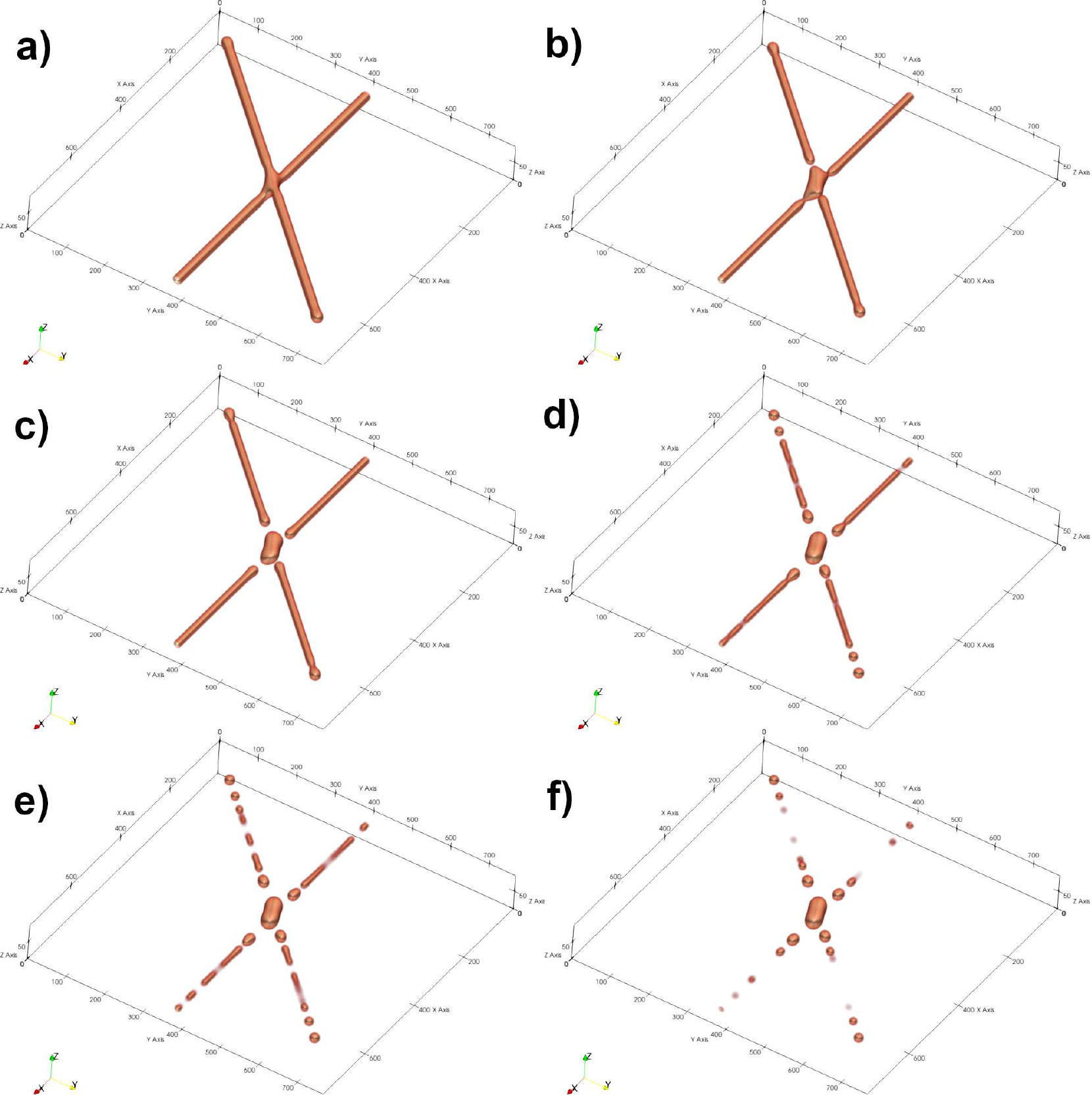}
		\caption{Morphological evolution of two nanowires intersecting at $45^{\circ}$ at different times:
		(a) 1000, (b) 2000, (c) 2250, (d) 2750, (e) 2800 and (f) 2900 time units. The video of the break-up is available as supplementary data.}
		\label{Fig3}
	\end{figure}

	In order to better understand the effect of
	orientation, in Fig.~\ref{Fig3}, we show the morphological evolution in the case of
	an angle of intersection of 45$^{\circ}$
	between the two nanowires (of the same radii, namely, 12 units). 
	In this case also, there is junction formation
	at the intersection; however, unlike the previous case where the contact between the nanowires is  at a point, in this case, the contact
	between the two wires is along a line. Hence, the differences in curvatures at different points at the intersection lead to more material filling in at the sites which make smaller angle with the wire along the $x$-axis. Thus, when the
	central break-up takes place, the morphology
	of the central particle is not spherical; it is elongated and is aligned closer to (the infinite wire along) the $x$-axis. The subsequent break-ups and the coarsening are
	similar to the earlier case.
	
\begin{figure}
		\centering
		\includegraphics[scale=1.0]{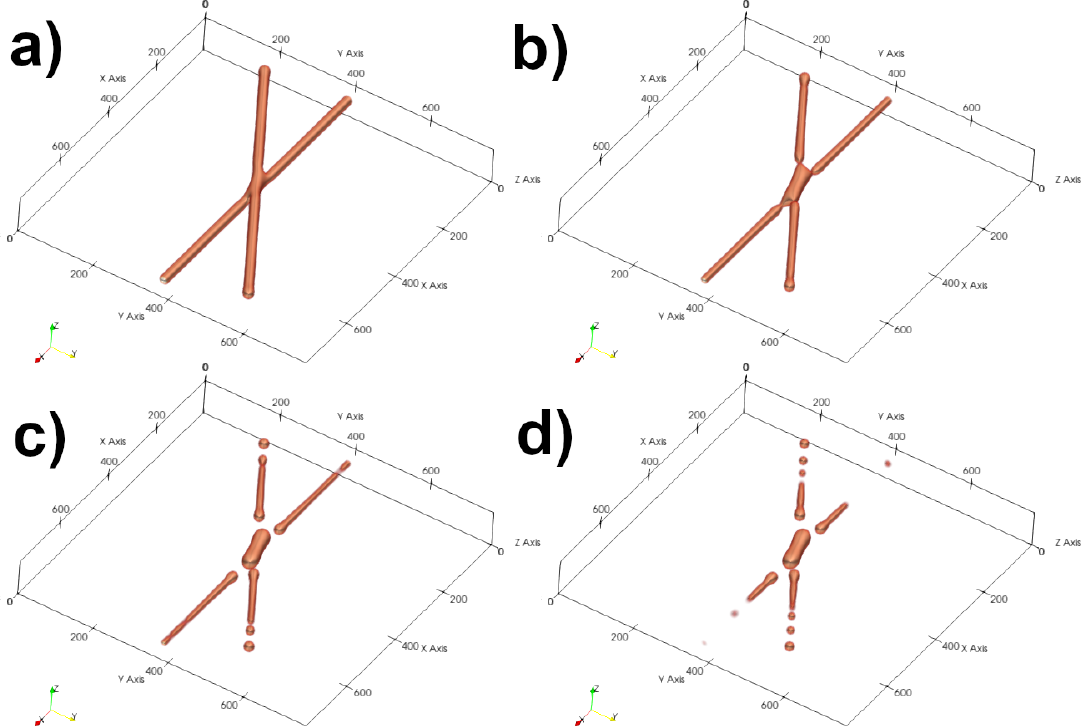}
		\caption{Morphological evolution of two nanowires intersecting at $30^{\circ}$ at different times:
		(a) 750, (b) 2125, (c) 2750, and (d) 2950 time units. The video of the break-up is available as supplementary data.}
		\label{Fig4}
	\end{figure}

\begin{figure}
		\centering
        \includegraphics[scale=1.0]{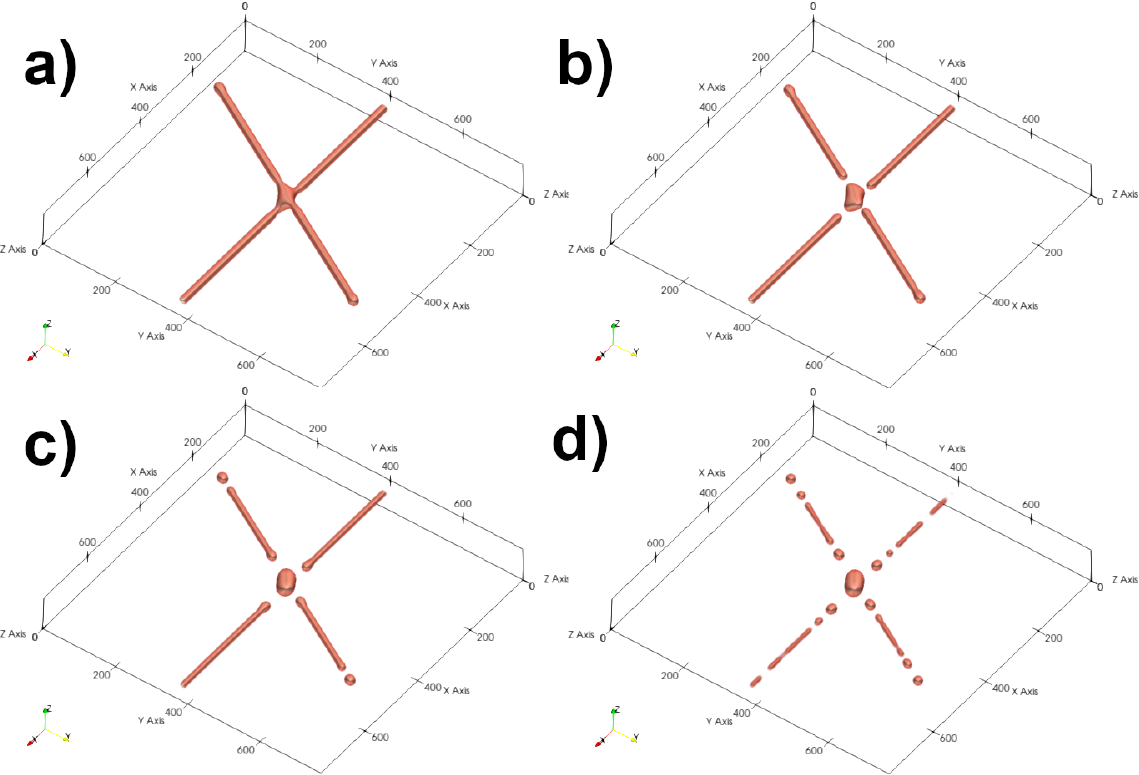}
		\caption{Morphological evolution of two nanowires intersecting at $60^{\circ}$ at different times:
		(a) 1500, (b) 2000, (c) 2500, and (d) 2800 time units. The video of the break-up is available as supplementary data.}
		\label{Fig5}
	\end{figure}

	In Fig.~\ref{Fig4} and Fig.~\ref{Fig5}, we show the morphological evolution for the cases of nanowires (of radii 12 units) which make angles of intersection of $30^{\circ}$ and  $60^{\circ}$, respectively. These morphologies are
	qualitatively similar to that in Fig.~\ref{Fig3}
	in terms of the morphology of the central
	particle.
	
	These microstructural features are in good agreement with experimental observations. For example, all the features noticed in the simulations above, namely, the formation of junction, the first break-up at 
	the junction, and elongated central particle when the angle of intersection of wires is not 90$^{\circ}$
	are seen in experiments -- specifically, see the Figures 3 and 4 of Ref. \cite{Vigonski2017}. Interestingly, our simulation results also resemble nano-welds generated experimentally using different techniques like furnace annealing (Fig.3(c) of Ref.\cite{Bellew2015}), and laser nano-welding of long Ag nanowires (Fig.~4 of Ref.\cite{Lee2012}). However, it is not clear to us if these welding are a result of local melting at the junction. In our simulations, however, there is no local melting and all the morphological changes are through
	mass transport by enhanced surface diffusion. 
	
	Phase-field modelling has been used previously to simulate the sintering of Ag nanoparticles \cite{Chockalingam2016}. The effect of local sintering at the junction of nanowires have also been studied using atomistic simulation methods, where it was observed that the nanowires undergo self-limiting rotation during neck growth, which is a result of complex interaction of surface diffusion and dislocation growth \cite{Jahangir2020}. We also observe the nanowire junction break-up in all cases described above as a result of local sintering at the junctions of nanowires. In isolated nanowires, the primary cause of fragmentation is the Rayleigh instability. But, the presence of junctions lead to modification of curvature and chemical potential in the intersection region, leading to junction formation followed by preferential break-up at the junctions.
	
	Our simulation results are also in good agreement not only with the experimental results but also with the kinetic Monte Carlo simulation studies of Vigonski et al~\cite{Vigonski2017}. They reported a similar observation of preferential fragmentation of nanowires initially at the junction. Vigonski et al, proposed that intersecting surfaces of nanowires act as sites of defects promoting atomic diffusion and attributed break-up at the junction primarily to the mechanism of diffusion of surface atoms. In this study, we show that the geometric factors in terms of the curvature leads to sintering at the junctions, of course, assisted by faster surface diffusion. Further, unlike the results from the Monte Carlo simulations, the phase-field model allow us to explore long time dynamics and morphological evolution such as the subsequent nanodot formation and the coarsening of the dots. 

\begin{figure}
		\centering
        \includegraphics[scale=1.0]{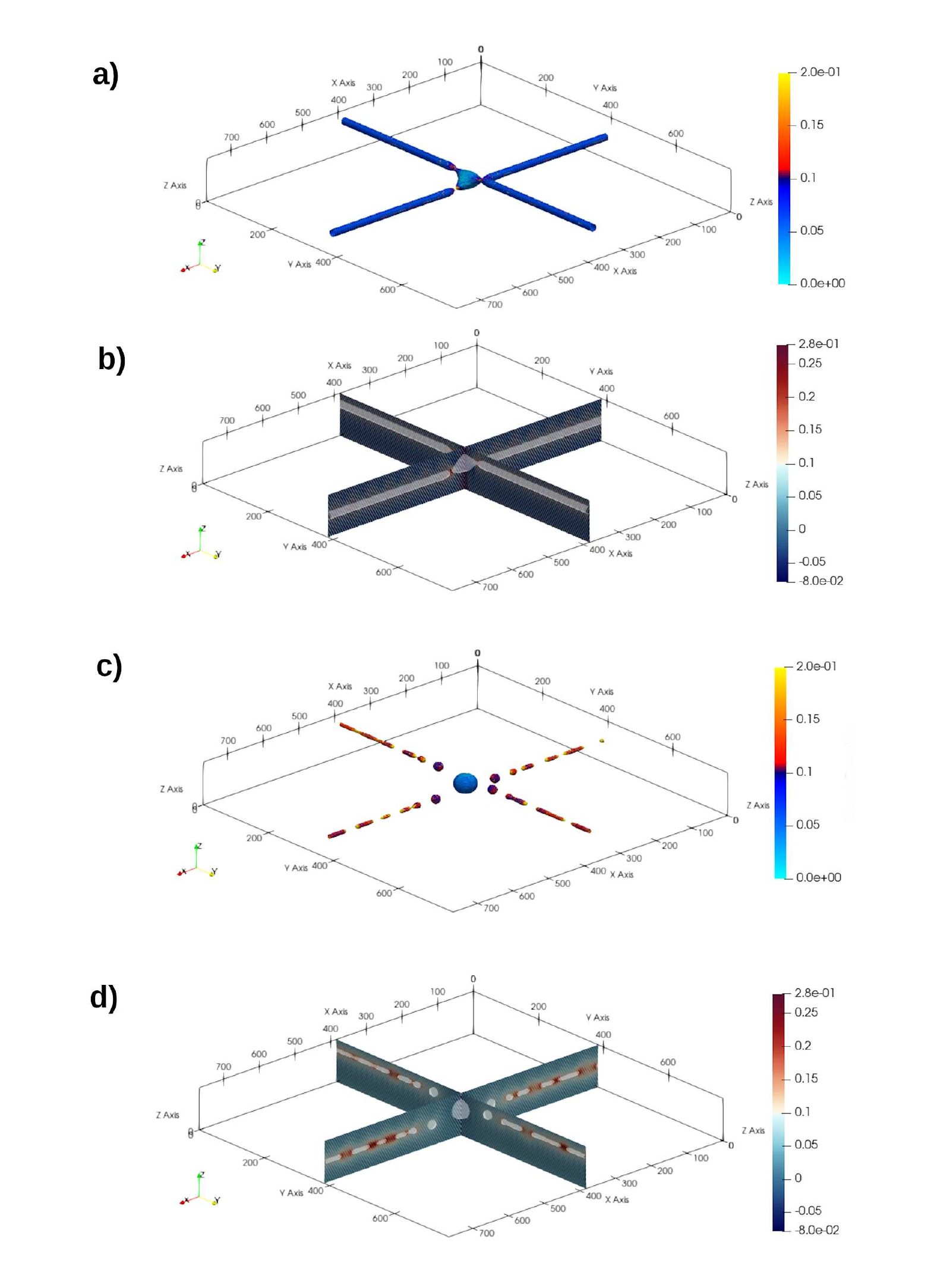}
		\caption{Mean curvature maps at two different times (1700 and 2800 time unites) for the $90^{\circ}$ configuration, along with the corresponding chemical potential maps.}
		\label{Fig6}
	\end{figure}

	In order to better understand the junction	formation and the subsequent break-up at the junctions, we have mapped the mean curvature and chemical potential in these systems. As shown by Nichols and Mullins~\cite{Nichols1965}, the surface movement in these systems 
	is driven by the the mean curvature of the interface. In Fig. \ref{Fig6}, we show the mean curvature ((a) and (b)) and chemical potential ((c) and (d)) maps for the 90\textdegree wire configuration, at the instant of break up and after the formation of nanodots. In this plot, the mean curvature is visualized as a colour map (at the order parameter isosurface of $c=0.5$) overlaid on the nanowire assembly. The chemical potential is visualized as a colour map in a plane which cuts through the cross section of nanowires (the nanowires are superimposed for visual clarity). The colour maps for the chemical potential correspond to excess chemical potential above its mean value. Not surprisingly, the chemical potential maps have one-to-one correspondence with the mean curvature maps indicating that the driving forces for atomic diffusion at different regions of the nanowire assembly are due to the surface energy considerations.

\begin{figure}
	    \centering
	    \includegraphics[scale=1.05]{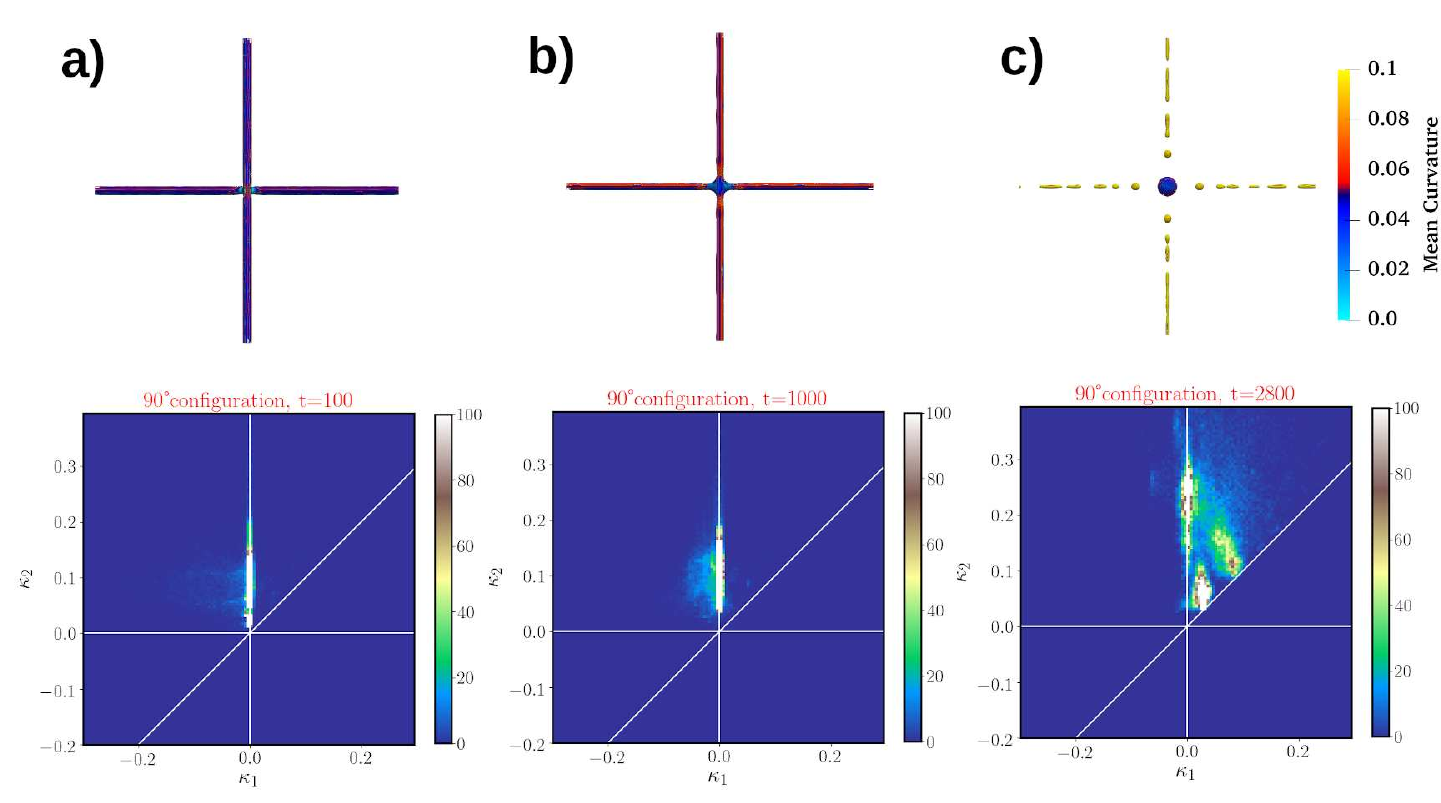}
	    \caption{The maps of mean curvature (top row) and the ISD maps (bottom row) for the 90\textdegree wire configuration at timesteps (a) 100, (b) 1000 and (c) 2800 respectively. The mean curvature maps show the morphology of the wires corresponding to the ISD map. 
	    The upper limit of probability density for the colour map has been fixed at 100 to resolve the smaller values.}
	    \label{90degISD}
    	\end{figure}

\begin{figure}
	    \centering
	    \includegraphics[scale=1.0]{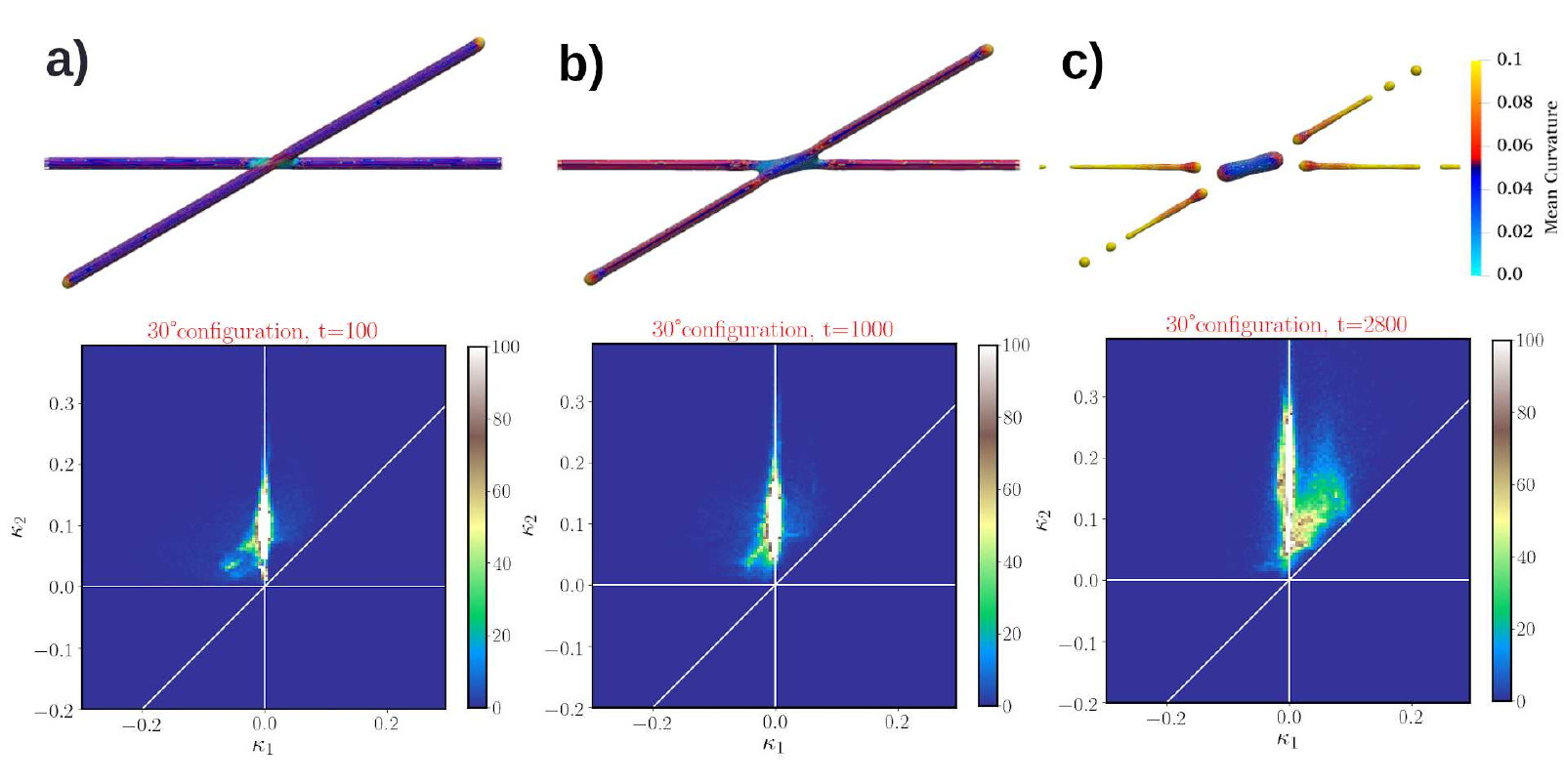}
	    \caption{The maps of mean curvature (top row) and the ISD maps (bottom row) for the 30\textdegree wire configuration at timesteps (a) 100, (b) 1000 and (c) 2800 respectively.}
	    \label{30degISD}
	    \end{figure}

\begin{figure}
	    \centering
	    \includegraphics[scale=1.0]{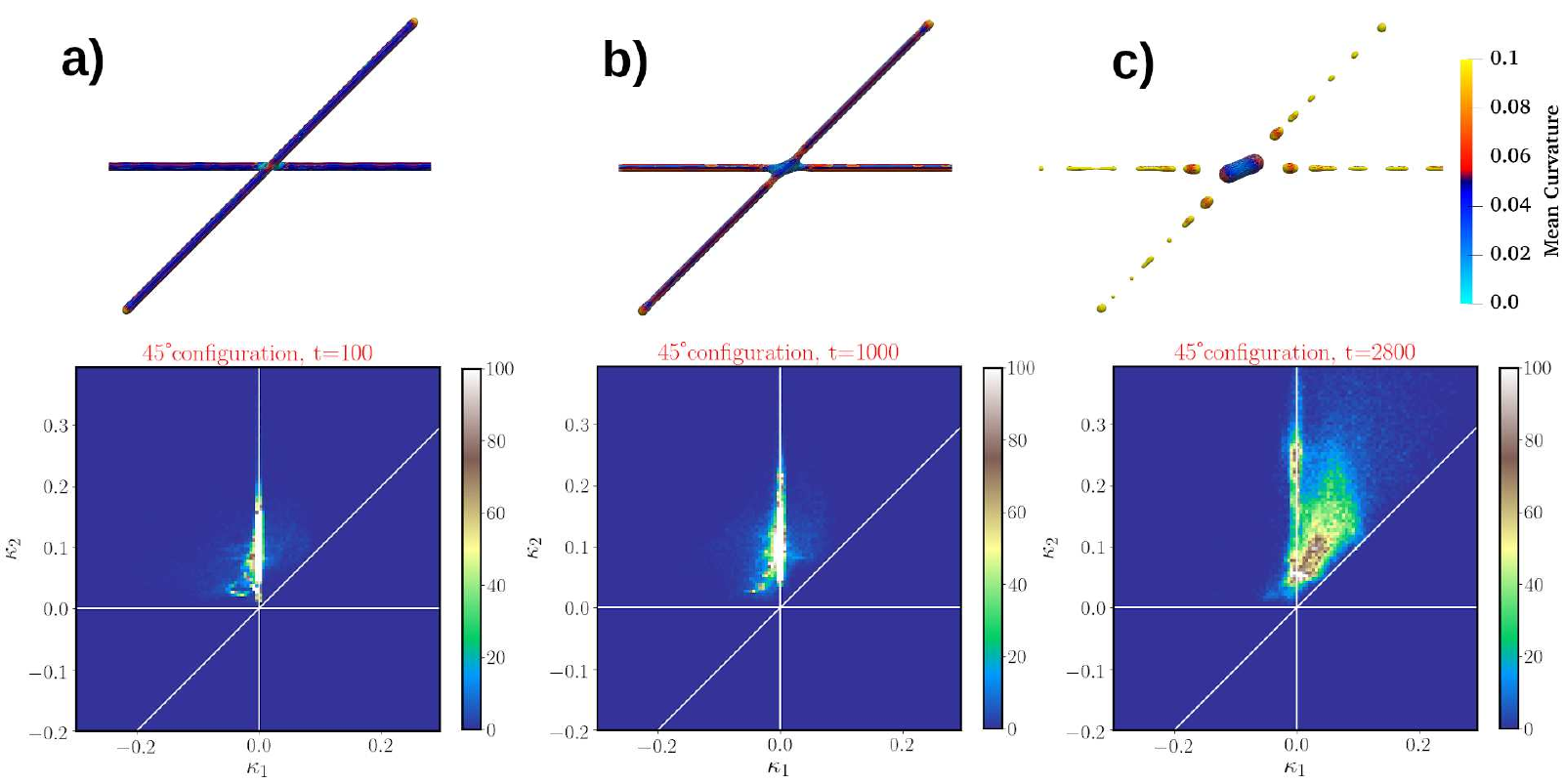}
	    \caption{The maps of mean curvature (top row) and the ISD maps (bottom row) for the 45\textdegree wire configuration at timesteps (a) 100, (b) 1000 and (c) 2800 respectively.}
	    \label{45degISD}
    	\end{figure}

\begin{figure}
	    \centering
	    \includegraphics[scale=1.05]{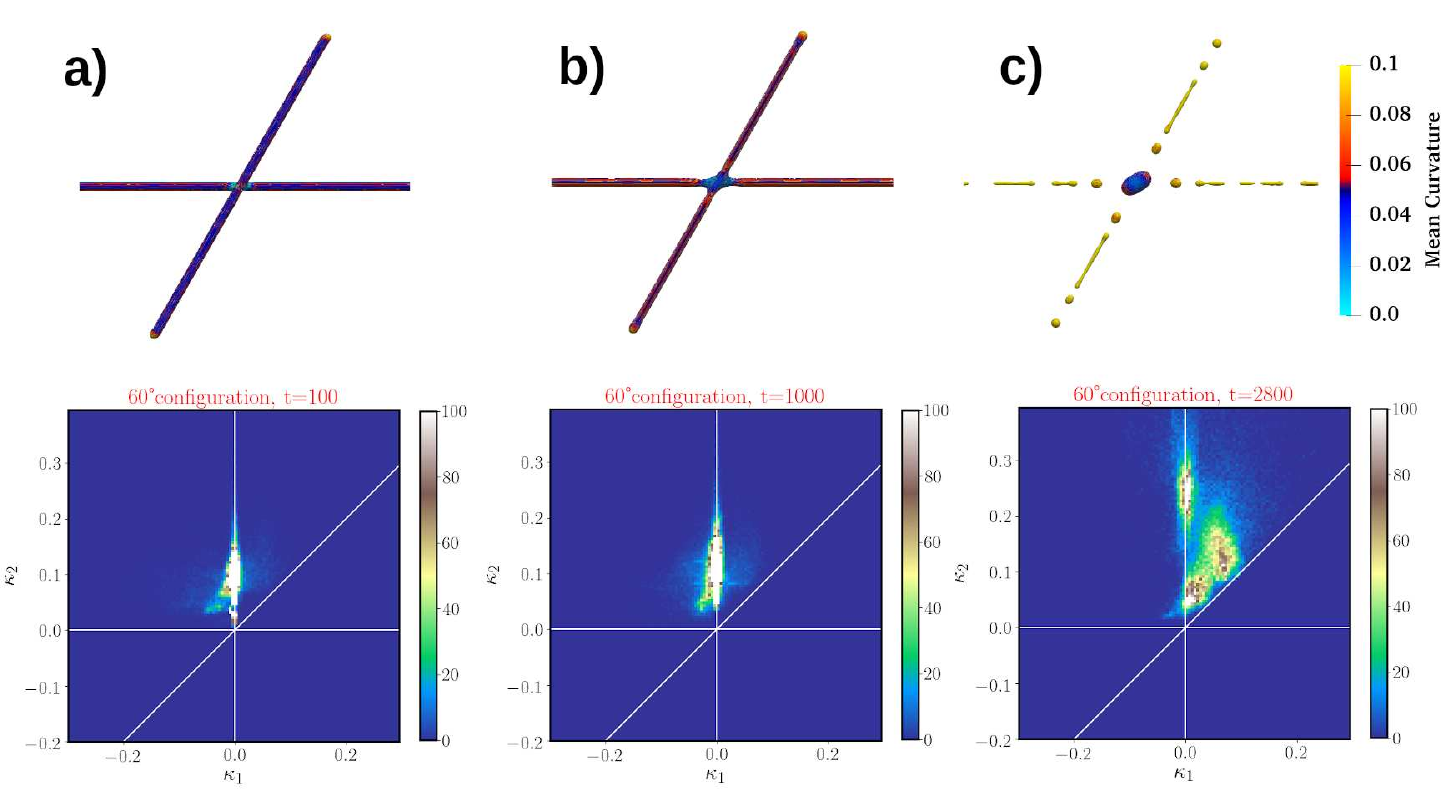}
	    \caption{The maps of mean curvature (top row) and the ISD maps (bottom row) for the 60\textdegree wire configuration at timesteps (a) 100, (b) 1000 and (c) 2800 respectively.}
	    \label{60degISD}
    	\end{figure}

In Fig.~\ref{90degISD}-~\ref{60degISD}, we show the mean curvature maps (in the first row). In all cases, as can be seen, initially, 
	the high curvatures at the constrictions formed near the junction increase the chemical potential leading to higher atomic transport. 
	Hence, atoms diffuse from these narrow regions towards the central nanoparticle and arms of the nanowires. Therefore, these constrictions 
	get narrower finally causing detachment of nanowires from the junction. At this point, we wish to note that the saddle point, where the 
	two principal curvatures assume values of opposite sign during the sintering of the nanowires  at the junction is better seen in the 
	Gaussian curvature maps of these morphologies as shown in the Supplementary Information.
	
	Thereafter, the free ends of the nanowires retract due to high curvature at the tip of the broken junction. As these nanowire free ends retract, more matter accumulates and these tips get blunted. Subsequently these tips acquire a spherical shape and get detached; this mechanism
	is the same as observed in Rayleigh instability driven break-up. This process continues leading to formation of further constrictions 
	along the arms of the nanowires, which are regions of higher chemical potential. This leads to subsequent formation of nanodots. 
	These nanodots show a mean curvature of about 0.05 which occurs due to the spheriodisation of the nanoparticles -- in order to reduce the (isotropic) interfacial energy.

	The mean curvature is the mean of the two principal curvatures ($\kappa_1$ and $\kappa_2$) of the interface, one of which tries to stabilize the perturbation while the other tries to de-stabilize it.
	The interplay between the two principal curvatures which determines the evolution of the system is captured well using interfacial shape distribution (ISD) maps~\cite{Mendoza2003,Kwon2010,Park2014,Park2017}. The probability density of the principal curvatures are calculated, following Kwon et al~\cite{Kwon2010} as
	\begin{equation}
	P(\kappa_1,\kappa_2) = \frac{A(\kappa_1,\kappa_2)}{A_{total}\Delta(\kappa)^2}
	\end{equation}
	where $\Delta \kappa$ was taken as 0.005, $A(\kappa_1,\kappa_2)$ is the sum of area of all patches having principal curvatures between $\kappa_1$, $(\kappa_1 + \Delta \kappa)$ and $\kappa_2$, $(\kappa_2 + \Delta \kappa)$ and $A_{total}$ is the total area of the interface. In Fig.~\ref{90degISD}-~\ref{60degISD}, we show the ISD maps (in the second row) corresponding to the mean curvature maps shown above
	them.
		
	In Fig.\ref{90degISD}, for example, we show the ISD maps for 
	90\textdegree case, at time steps of 100, 1000 and 2800 
	respectively, below the corresponding wire morphologies. The 
	maximum value of probability density in all ISD maps
	reported is fixed at 100, to get a better resolution 
	of the smaller values. Initially, there is a higher 
	density along the $\kappa_1=0$ line, which corresponds
	to the larger fraction of cylindrical regions in the morphology, where 
	one of the 	principal curvatures will be zero. The points in the $\kappa_1<0,\;\kappa_2>0$ regions correspond to the saddle shapes 
	formed at the intersection of the two nanowires. These points are more pronounced at $t=1000$ where the constrictions at the 
	intersection become more pronounced. It is clear that the sign of the two curvatures in this region are opposite, which leads 
	to a negative sign of the Gaussian curvature, making it a metric to identify the saddle shaped regions; as indicated above, the 
	Gaussian curvature maps for all	configurations have been provided in the Supplementary Information. 
	The loss of material from the saddle shaped regions result in the break-up of the wire and the formation of nanodots, which evolve 
	to be spherical in shape. At $t=2800$ there is a large density of points in the $\kappa_1>0,\;\kappa_2>0$ domain, corresponding to the convex 
	surfaces. Two patches near the $\kappa_1=\kappa_2$ line correspond to the near-spherical nanodots of different sizes, formed after 
	the wire break-up. As the wire breaks up, the density of points at the $\kappa_1=0$ line reduces due to the smaller fraction of 
	cylindrical region in the microstructure.
	
	The ISD maps for the $30$\textdegree, $45$\textdegree 
	and $60$\textdegree configurations have also been made
    in Fig.\ref{30degISD}, Fig.\ref{45degISD} and
    Fig.\ref{60degISD} respectively. The underlying
    trend in these cases are also the same as that of 
    90\textdegree configuration. However, the 
    increased overlap of the wires at the junctions for
    the angular configurations causes a higher density
    in the $\kappa_1 < 0,\;\kappa_2 > 0$ region of the map. This is due to the zipping of the two wires at the junction, 
    resulting in a larger fraction of saddle shaped regions as compared to the 90\textdegree case.
    This is most pronounced in the 30\textdegree case, and becomes less and less pronounced for the 45\textdegree and 60\textdegree cases. 
    As the wire breaks-up and starts to disintegrate, the density of points in the convex region ($\kappa_1>0,\;\kappa_2>0$) 
    increases for all cases. 
    	
\begin{figure}
		\centering
		\includegraphics[scale=0.2]{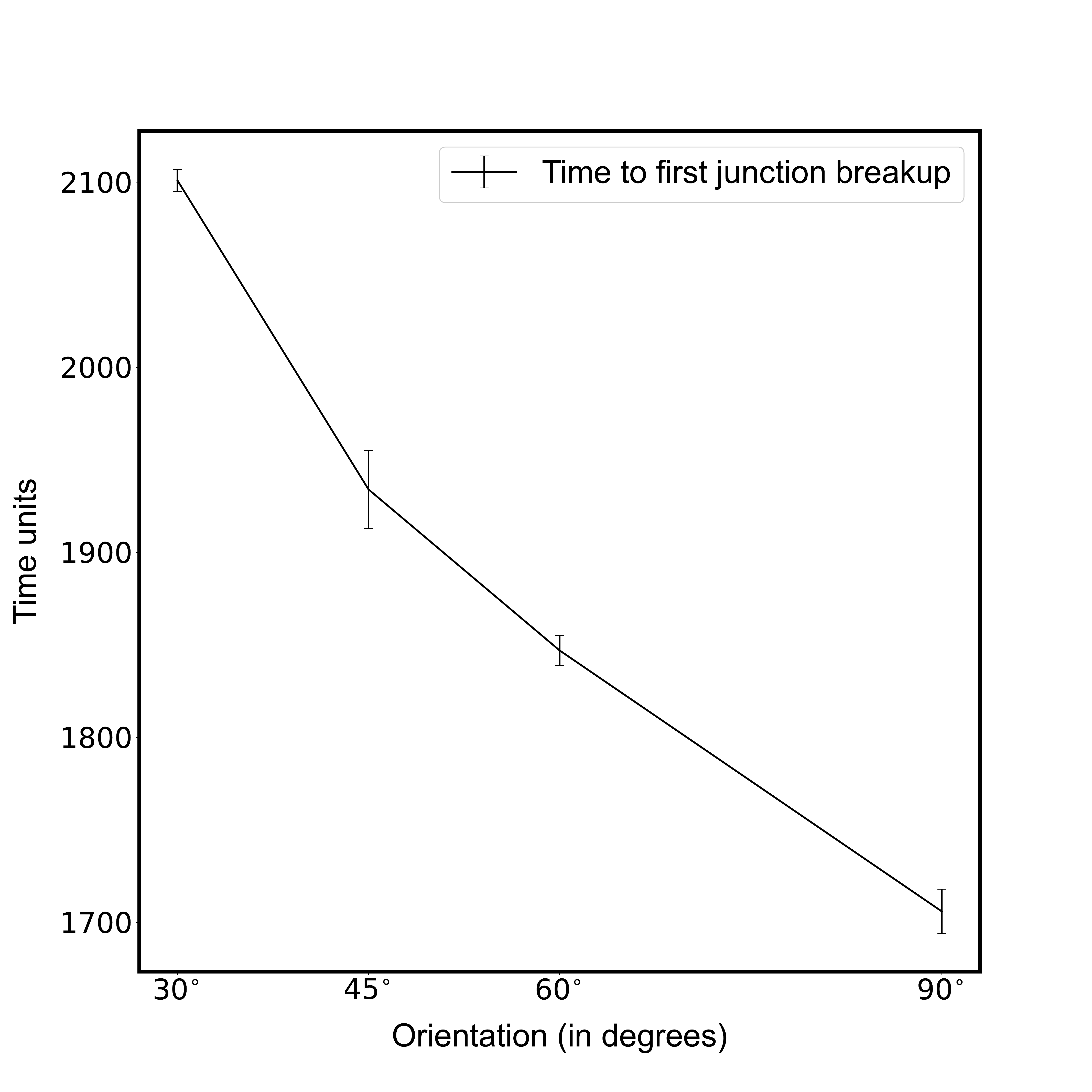}
		\caption{Plot showing the dependence of time to first break-up at the junction on the 
			relative wire orientation. The line connecting the data points is drawn only as a guide to the eye.}
		\label{Fig7}
	\end{figure}

	The angle between the nanowires does not affect the sequence of the nanowire break-up; the first fragmentation always occurs at the junction; but, it does affect the break-up kinetics. The time to first break-up (at the junction) is calculated for different orientations, and plotted in Fig.~\ref{Fig7}. Here, we have assumed that there is a break-up at the junction if two or more of the arms at a junction (consisting of four arms) detach. The data points are obtained from a set of three different simulations for each configuration (that is, by using three different seed values of pseudo-random number generator). It is observed that the system is relatively more stable as the acute angle between the nanowires decrease. The time to first junction break-up is highest for the $30^{\circ}$ configuration system, and the least for $90^{\circ}$ configuration system, when only the effect of relative wire orientation is considered in isolation. This observation can be correlated with the average distance between nearest nanodots from the central nanoparticle along the two nanowires, formed subsequently after the junction break-up (values given in Table \ref{Table2}). From the values presented in Table \ref{Table2}, it can be inferred that the average separation between the nanodots from the central nanoparticle along the two wires increases as the angle between the nanowires decrease. These values were calculated by averaging the results from three different set of simulations. The correlation between time to first junction break-up and average separation between the nanodots from the junction center can be explained with the mean curvature maps; as the angle between the nanowire decrease, the curvature increases at the junction which leads to enhanced atomic transport in between the two wires. Therefore, more material accumulates at the central agglomerate before constriction occurs at the junction, leading to nanowire junction break-up. For this reason, the central nanoparticle acquires an ellipsoidal shape as the the angle between nanowire is reduced. The end result is the junction break-up substantially delayed along with the particles being placed afar from the center of the junction.     
	
	\begin{table}
		\begin{ruledtabular}
			\begin{tabular}{ccc}
				\textbf{Wire configuration}  & \textbf{Distance along wire 1} & \textbf{Distance along wire 2} \\ \hline
				$30^{\circ}$ configuration & $110.9 \pm 0.4$ & $111.5 \pm 0.8$ \\
				$45^{\circ}$ configuration & $94.1 \pm 0.9$ & $93.9 \pm 0.6$ \\
				$60^{\circ}$ configuration & $87 \pm 0$ & $85.3 \pm 1.4$  \\
				$90^{\circ}$ configuration & $83 \pm 0$ & $83 \pm 0$  
			\end{tabular}
		\end{ruledtabular}
		\caption{Average distance of first nanodots from the central nanoparticle (wire 1 corresponds to the infinite length nanowire at the bottom, and wire 2 corresponds to the finite length inclined nanowire on top of the infinite length nanowire).}
		\label{Table2}
	\end{table} 
	
	\subsection{Effect of relative wire diameter}
	
	It has been observed experimentally that there is on average about 25\% variation in the diameter of the fabricated metallic nanowires \cite{Karim2006}. Therefore, in order to study the effect of relative variation in diameters of the nanowires, we use three different assemblies of nanowires in $90^{\circ}$ configuration, with initial radii of $\mathrm{R_1},\mathrm{R_2} = 12,14,16$. We study the three different combinations of radii \textemdash the first combination with $\mathrm{R_1}=14, \mathrm{R_2}=16$, the second combination with $\mathrm{R_1}=12, \mathrm{R_2}=14$, and the final combination with $\mathrm{R_1}=12, \mathrm{R_2}=16$; with the relative radius variations of 14\%, 17\%, and 33\%, respectively. We have carried out a set of three different simulations for each case, and the average from these simulations are presented here.   
	
	It is observed that for all three combinations, the nanowires with smaller radius breaks up at the junction and the central agglomerate becomes part of the larger radius nanowires. The time to first junction break-up (values given in Table \ref{Table3}) within a given margin of error is dependent primarily on the radius of the smaller diameter nanowire. Having said that, the difference in the radii of the nanowires has a small but definite effect on kinetics. The kinetics when the radii are different is slightly faster than when the radii are the same; for example, for
	the cases of $\mathrm{R_1}=12$, $\mathrm{R_2}=14$ and $\mathrm{R_1}=12$, $\mathrm{R_2}=16$ are slightly faster than $R_1=R_2=12$ case.
	
	\begin{table}
		\begin{ruledtabular}
			\begin{tabular}{cc}
				\textbf{Wire configuration}  & \textbf{Time to first junction break-up} \\ \hline
				$\mathrm{R_1} = 12, \mathrm{R_2} = 12$	&	$1706 \pm 12$	\\
				$\mathrm{R_1} = 12, \mathrm{R_2} = 14$	&	$1623 \pm 12$	\\
				$\mathrm{R_1} = 12, \mathrm{R_2} = 16$	&	$1622 \pm 10$	\\
				$\mathrm{R_1} = 14, \mathrm{R_2} = 14$	&   $4000 \pm 5$  \\
				$\mathrm{R_1} = 14, \mathrm{R_2} = 16$	&	$3907 \pm 15$	\\
				$\mathrm{R_1} = 16, \mathrm{R_2} = 16$	&   $8142 \pm 14$  \\
				4-junction ($\mathrm{R_1} = 12, \mathrm{R_2} = 12$)	&	$1694 \pm 9$	\\
				9-junction ($\mathrm{R_1} = 12, \mathrm{R_2} = 12$)	&	$1691 \pm 6$	
			\end{tabular}
		\end{ruledtabular}
		\caption{Time units to first break-up at the junctions for different configurations.}
		\label{Table3}
	\end{table} 
	
	\begin{figure}
		\centering
		\includegraphics[scale=0.2]{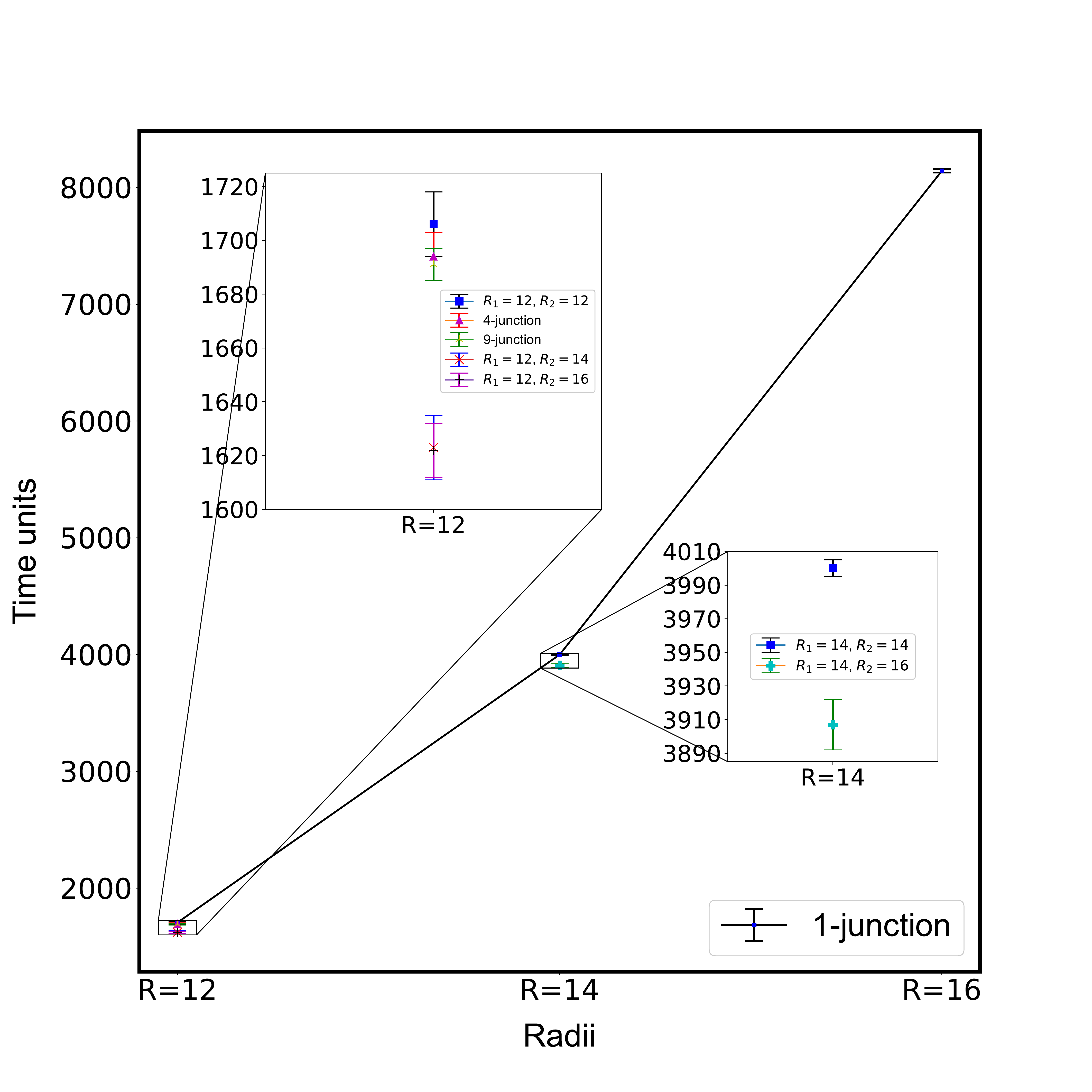}
		\caption{Plot showing the dependence of time to first break-up at the junctions on the relative wire radii. The values for time to first junction break-up for 4-junction and 9-junction grid, as well different radii combinations are given for comparison (the magnified inset is provided for better visibility). The line connecting the data points is drawn only as a guide to the eye.}
		\label{Fig8}
	\end{figure}
	We have summarised in Fig.\ref{Fig8} the data given in Table \ref{Table3}.  We have used the $90^{\circ}$ configuration for $\mathrm{R_1},\mathrm{R_2} = 14$ and $\mathrm{R_1},\mathrm{R_2} = 16$ and compare it with the $90^{\circ}$ case with $\mathrm{R_1},\mathrm{R_2} = 12$. From Fig.\ref{Fig8}, it can be observed that the time to first junction break-up increases drastically as the radius of the nanowires is increased. This difference, in our opinion,
	is related to the reduction in driving force due to increasing size of the nanowires which leads to smaller driving forces and the fact that in our simulations while the radii are increased, the inter-junction distance is maintained a constant.  
	
	\subsection{Effect of the density of junctions}
	
\begin{figure}
		\centering
        \includegraphics[scale=1.0]{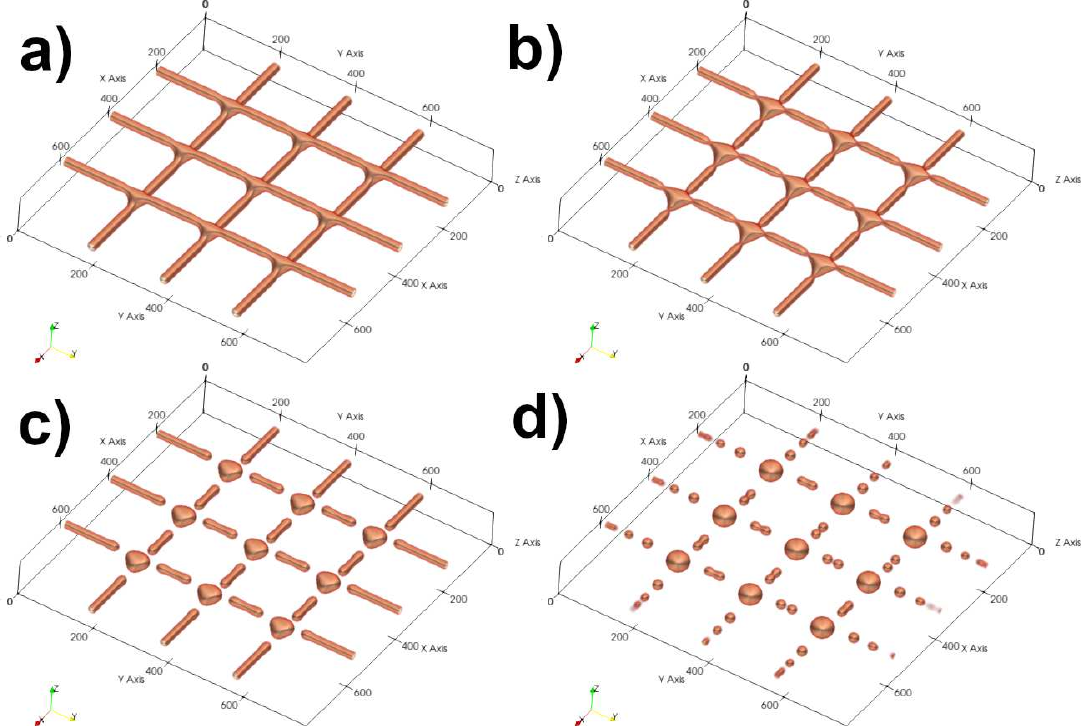}
		\caption{Morphological evolution of nanowires with 9 junctions in the simulation cell at different times. The video of the break-up is available as supplementary data.}
		\label{Fig9}
	\end{figure}

	During fabrication of nanowires, it is seldom possible to fabricate a nanowire in isolation. In most practical cases, they are fabricated in large numbers with the nanowire network resembling structures best approximated as a grid pattern \cite{Li2018}. It has been observed that local sintering at the junction of nanowires has drastic effects on the optimization of electrical and optical properties of the nanowire networks~\cite{Langley2014, Bellew2015, Lagrange2015, Hu2019}. The network density (no of junctions in close proximity) also plays a role in optical transmittance of the assembly. Therefore, we have have carried out systematic study of nanowire assemblies consisting of grids with 1, 4, and 9 junctions in the simulation cell. Since the 4 and 9-junctions behave qualitatively the same, we only show the results for the 9-junction configuration in Fig.\ref{Fig9}. In all the three cases, we observe the first fragmentation of the nanowires to occur at the junctions; however, the specific junction at which the first detachment occurs is stochastic in nature. We track the time to first junction break-up for all the three cases, and correlate the junction break-up kinetics with the number of junctions in close proximity, the values of which are given in Table \ref{Table3}. It is observed that within the given margin of error, there is no substantial difference between the time to first junction break-up. Therefore, close proximity of the junctions does not effect the the break-up kinetics of individual junctions in the network. The kinetics remains unaffected, but the morphological evolution plays a role in the network electrical resistance which falls during annealing to reach a stable minima followed by an abrupt rise, as reported in~\cite{Langley2014, Lagrange2015}. This experimental observation can be directly correlated with the morphological evolution of the nanowire assembly of the 9-junction grid as seen in Fig.\ref{Fig9}(a) to (d). As the sintering occurs, the network resistance is expected to decrease continuously due to junction formation, followed by an abrupt increase due to fragmentation at the junctions. 

	\section{Characterization of the central nanoparticle morphology} \label{section4}

	In this section, we analyse the morphology of the central nanoparticles formed as a result of junction break-up. The nanoparticles which are formed as a result of detachment of nanowires from the junction assume different morphologies. These morphologies are primarily dependent on the angle of intersection between the nanowires. In order to quantify the morphologies of the non-spherical nanoparticles at the centres (formed 
	as a result of the break-up of the junctions), we first use a 3D implementation of the Hoshen-Kopelman algorithm~\cite{hoshen1976} and label the different fragments. The central nanoparticles is isolated using the algorithm by doing an analysis of the microstructures at a time which is later than
	the junction break-up time for that configuration. As an example, we show the labelled clusters for the 90\textdegree configuration 
	in Fig.~\ref{clusters}. Similar labelled clusters for other configurations are shown in the Supplementary Information.
	
	\begin{figure}
	    \centering
	    \includegraphics[width=0.7\textwidth]{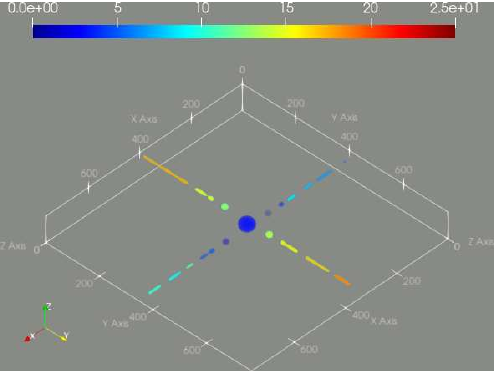}
	    \caption{Labelled clusters for the $90^{\circ}$ configuration. Each cluster is denoted by a specific colour from the 
	    corresponding colormap shown in the image.}
	    \label{clusters}
	\end{figure}

\begin{figure}
	    \centering
	    \includegraphics[scale=0.5]{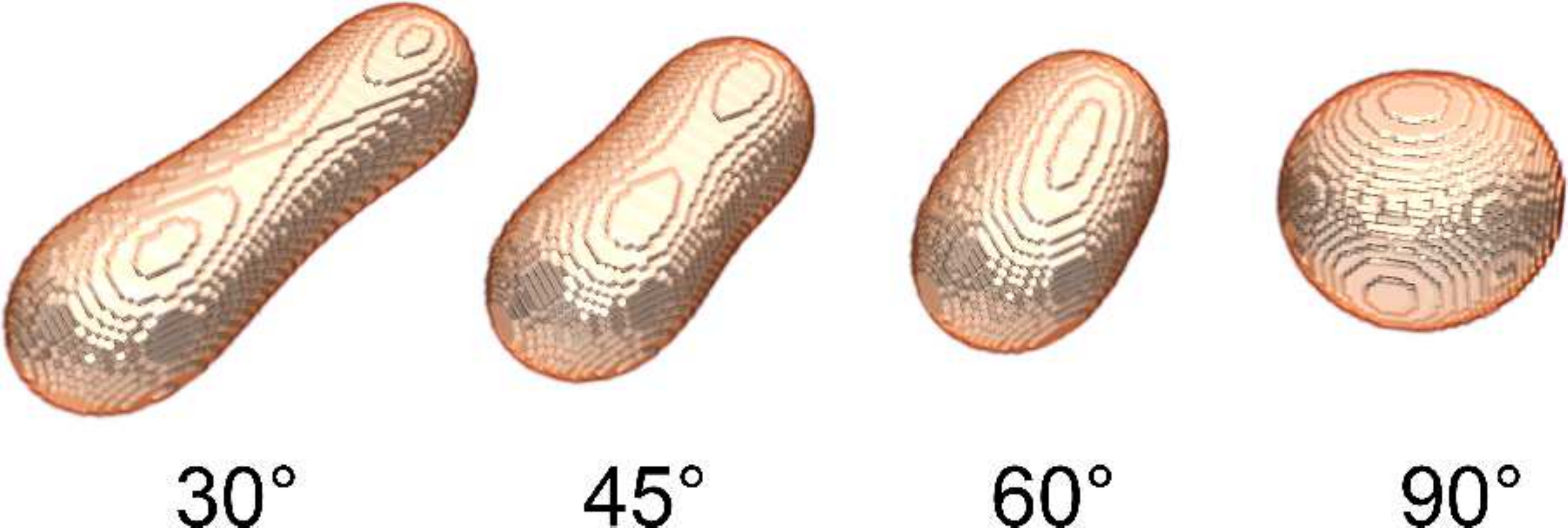}
	    \caption{Nanoparticle morphologies of the isolated central nanoparticles for four different configurations.}
	    \label{nanoparticles}
	\end{figure}

\begin{figure}
	    \centering
	    \includegraphics[scale=0.2]{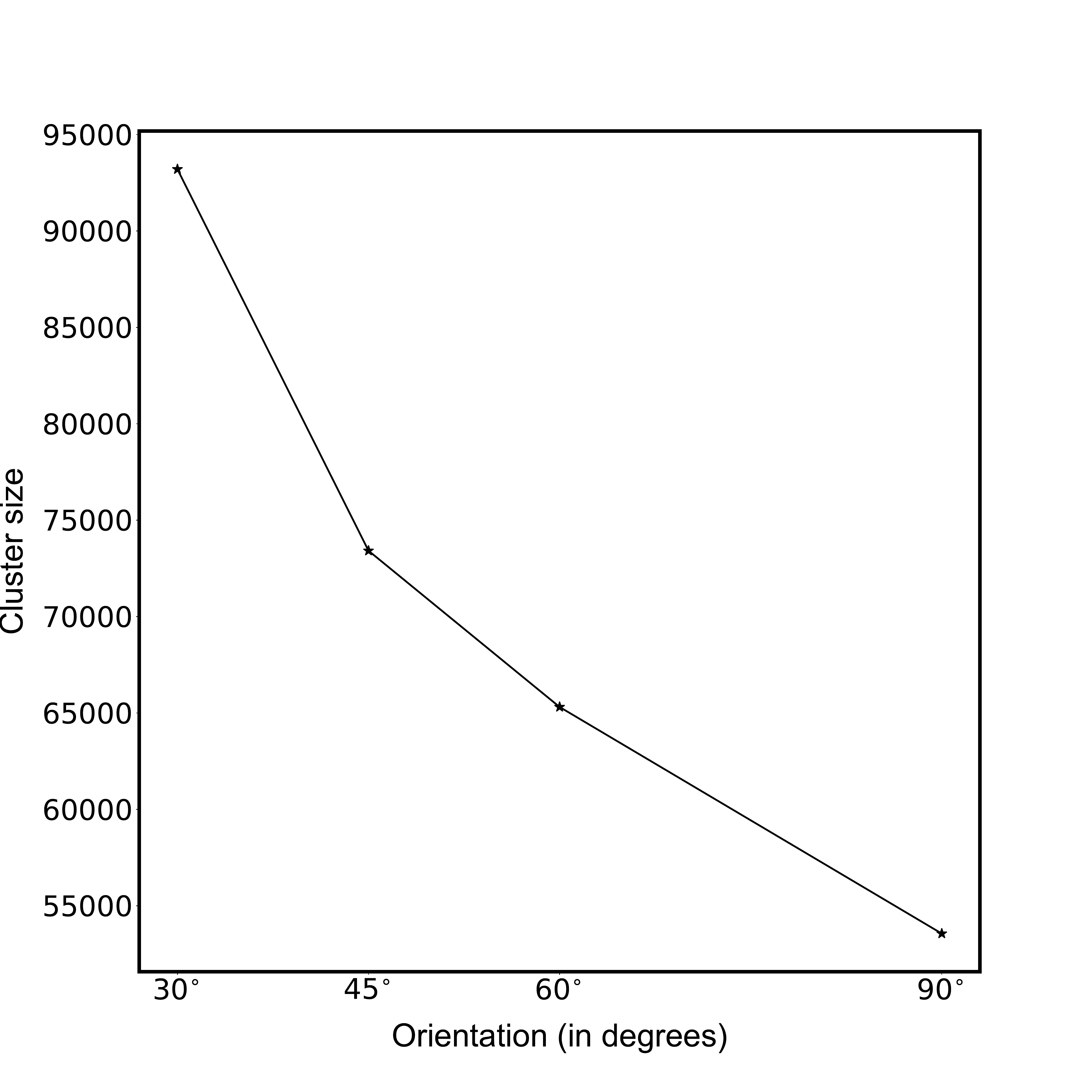}
	    \caption{Cluster size for the four different configurations.}
	    \label{cluster_size}
	\end{figure}

After labelling the clusters, we extract the data points corresponding to the label of the central nanoparticle and subsequently characterize its morphology. The isolated nanoparticles for four different configurations are shown in Fig. \ref{nanoparticles}. It is clearly seen that as the initial angle between the nanowires decreases, the central nanoparticle diverges significantly from spherical shape. The central nanoparticle assumes an oblate spheroidal shape for the $90^{\circ}$ configuration, and it diverges from sphericity as the angle is decreased. For the $30^{\circ}$ configuration, it assumes a nearly ellipsoidal shape. We also calculate the cluster size (nanoparticle volume) for all four configurations as shown in Fig \ref{cluster_size}.

We can make the characterization of central nanoparticle quantitative using the moment of inertia tensor (See Appendix B) of the particles. 
	The moment of inertia tensor (calculated about the center of mass of the nanoparticle) is a real, symmetric tensor. We obtain the three eigenvalues and eigenvectors of this tensor. The three eigenvectors give the directions of the three principal axes of the nanoparticle (corresponding to the three axes of rotation), and, the three eigenvalues describe the distribution of mass around the three 
	corresponding principal axes. 	
	\begin{figure}
	    \centering
	    \includegraphics[scale=0.25]{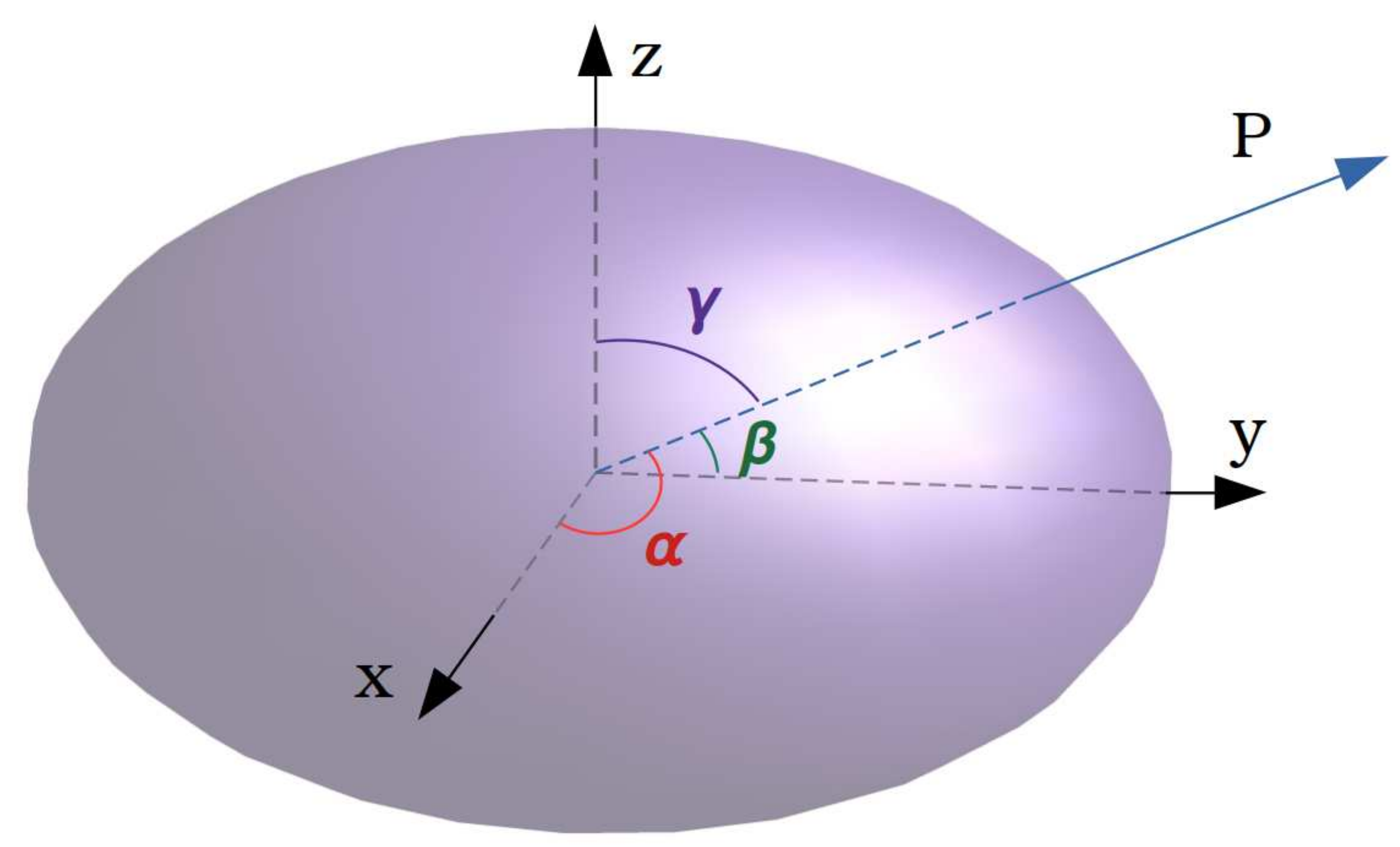}
	    \caption{Schematic of an ellipsoidal body with its three principal axes is shown. One of the principal axis (P-axis) makes an angle of $\alpha$ with the x-axis, $\beta$ with the y-axis, and $\gamma$ with the z-axis. }
	    \label{ellipsoid}
	\end{figure}

In the schematic of an ellipsoidal body shown in Fig. \ref{ellipsoid}, the principal axis (P) makes an angle of $\alpha$ with the x-axis, $\beta$ with the y-axis, and $\gamma$ with the z-axis, respectively. The three principal axes are denoted by P1, P2 and P3, respectively. We use this notation convention for the angles and report the results of principal axis calculation in Table \ref{table_paxis}.
	\begin{table}[H]
	    \caption{Angles the three principal axes make with the coordinate axes for all four configurations.}
	    \begin{ruledtabular}
	        \begin{tabular}{ccccc}
	        Configuration   &   principal Axes  &   $\alpha$    &   $\beta$ &   $\gamma$ \\ \hline
	         \multirow{3}{*}{$30^{\circ}$ configuration} & P1   &   14.74   &   75.26   &    89.92   \\
	         &  P2  &   75.26   &   14.26   &   89.29   \\ 
	         &  P3  &   89.89   &   89.28   &   0.72    \\ 
	         \hline
	         \multirow{3}{*}{$45^{\circ}$ configuration} & P1   &   22.02   &   67.98   &    89.95   \\
	         &  P2  &   67.99   &   22.01   &   89.64   \\ 
	         &  P3  &   89.82   &   89.65   &   0.35    \\ 
	         \hline
	         \multirow{3}{*}{$60^{\circ}$ configuration} & P1   &   30.70   &   59.30   &    89.86   \\
	         &  P2  &   59.30   &   30.70   &   89.77   \\ 
	         &  P3  &   90.00   &   89.73   &   0.27    \\ 
	         \hline
	         \multirow{3}{*}{$90^{\circ}$ configuration} & P1   &   83.34   &   6.67   &    89.95   \\
	         &  P2  &   6.66   &   83.33   &   89.57   \\ 
	         &  P3  &   89.55   &   89.80   &   0.45    \\ 
	    \end{tabular}
	    \end{ruledtabular}
	    \label{table_paxis}
	\end{table}
	
\begin{figure}
	    \centering
	    \includegraphics[scale=0.2]{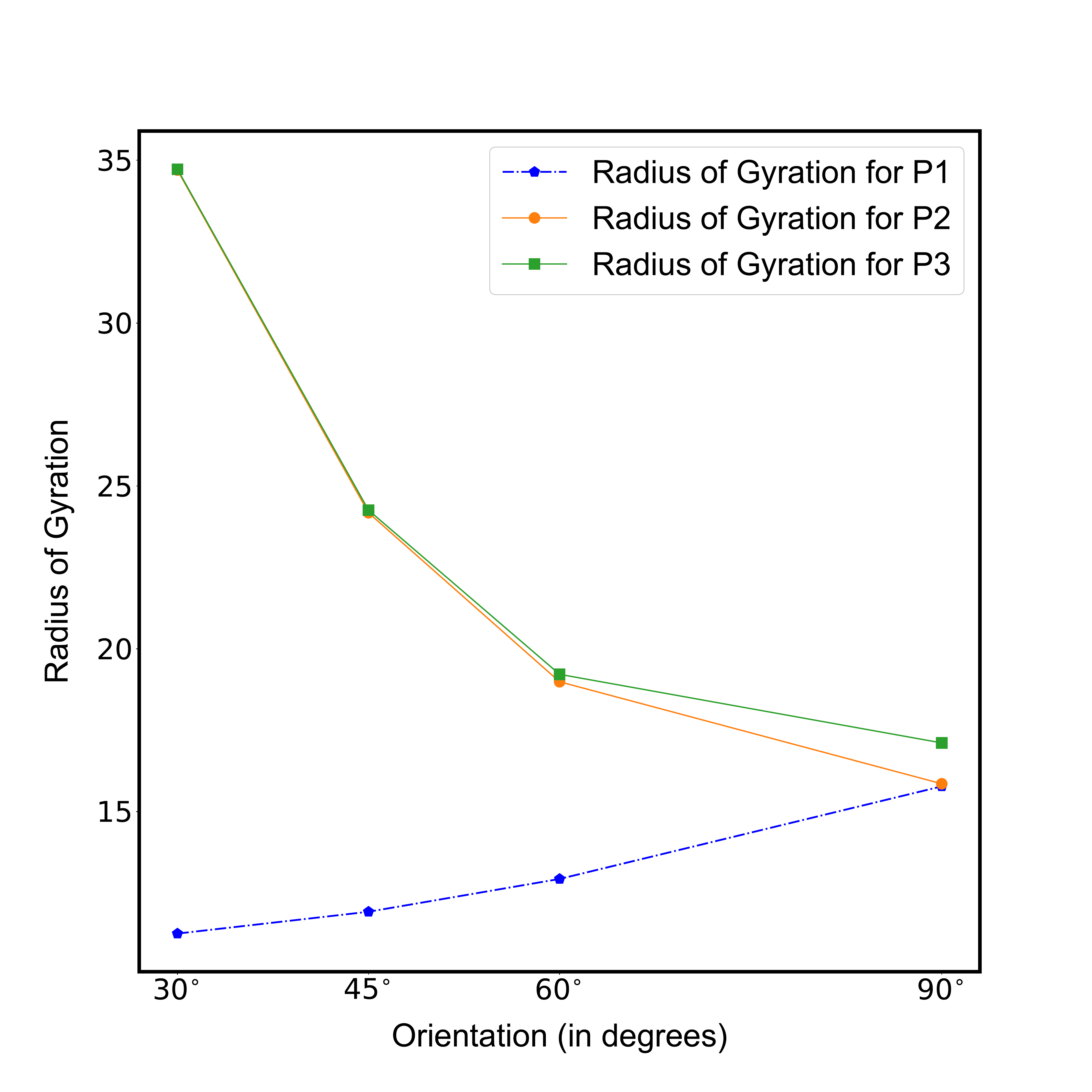}
	    \caption{The three radius of gyration corresponding to the three eigenvalues of inertia matrix along the three principal axis for all four configurations.}
	    \label{gyration}
	\end{figure}

	From the values of angles $\alpha$, $\beta$, and $\gamma$ presented in Table \ref{table_paxis}, it is observed that two principal axes P1, and P2 are aligned approximately along the acute and obtuse angle bisectors of the respective configuration (in all cases except when the wires
	are perpendicular to each other). The third principal axis (P3) is aligned along the z-axis for all four configurations (perpendicular to the plane of the nanowires). Therefore, it can be inferred that after the junction break-up, the nanoparticles possess rotation axes aligned along the angle bisectors of the initial angle between the nanowires.    
	Subsequently, we also calculate the three radii of gyration corresponding to the three eigenvalues of the inertia matrix corresponding to the three principal axes (details in the appendix). The radii of gyration values are presented in Fig.~\ref{gyration}. 
	We denote R1 as the radius of gyration corresponding to the eigenvalue of the inertia tensor for principal axis P1. Similarly, R2 corresponds to P2, and R3 corresponds to P3. From the plot, it is observed that the radii of gyrations R2 and R3 increase with decrease in the angle between nanowires. These radii of gyration to correspond to the principal axes perpendicular to the semi-major axis of the nanoparticles (i.e., along the z-axis and perpendicular to it along the y-axis). The other radius of gyration R1 corresponds to the principal axis along the x-axis (direction of elongation). This trend, along with the observation that the difference between the two radii of gyrations (P2 and P3) and the third
	radius (P1) increases with decrease in angle between the nanowires indicate that the nanoparticles diverge from sphericity.

	\section{Conclusions} \label{section5}	
	In this paper, we have shown, using a phase field model (implemented in both 2- and 3-D), that at the intersection of nanowires, sintering (curvature driven material flow) leads to the formation of junctions. These junctions act the initiators of nanowire break-up; the subsequent break-ups take place due to Rayleigh instability at the arms away from these junctions. The fragments coarsen due to the differences in sizes. Further, the effect of various parameters on the junction break-up can be summarised as follows:
	\begin{itemize}
		\item The radii of the nanowires play a key role in the kinetics; smaller the radii, faster is the kinetics. The kinetics is determined primarily by the radius of the smaller of the two nanowires that intersect. 
		\item The angle of intersection is the next key parameter; smaller the angle of intersection, slower the break-up at the junction. This is primarily due to the non-spherical morphology (and, relatively larger size) of the junction.
		\item The above two observations in terms of the size difference and the effect of intersection angle can be rationalised in terms of the mean curvatures (and, hence chemical potential) maps.
		\item The density of intersections in the simulation cell have very little or no effect on the kinetics.
	\end{itemize}
The mean curvatures and ISD maps based on the two principal curvatures can be used to rationalise the nanowire fragmentation process. 
Further, the shapes of the fragmented nanoparticles can be analysed quantitatively using the moment of inertia tensor and its eigenvalues 
and eigenvectors. Thus, phase-field models, and the simulation results based on them could be of use and importance in the study of stability 
of nanowire nets and their morphological evolution at long time scales. 
	
	\section*{Supplementary Material}
	See supplementary material for animations of nanowire junction break-up for different configurations. 
	\begin{acknowledgments}
		We thank the following high performance computing facilities for making the computational resources available to us: 
		(i) Dendrite and Space-Time, IIT Bombay, 
		(ii) Spinode – the DST-FIST HPC facility, Department of Metallurgical Engineering and Materials Science, IIT Bombay, and (iii) and C-DAC, Pune.
	\end{acknowledgments}
	
	\section*{Data availability statement}
	
	The codes used for the simulations are available at the following GitHub repository:\\
	https://github.com/abhinavroy1999/nanowire-fragmentation-code. 
	
	The data that support the 
	findings of this study are also available from the corresponding author upon reasonable request.

	\appendix
	
	\section{Effect of interfacial energy anisotropy}

	Experimentally, it is known that single-crystalline and poly-crystalline metallic nanowires possess faceted morphologies and this preference in surface orientation is attributed to surface energy anisotropy~\cite{Karim2007}. The diverse effects of surface energy anisotropy in FCC metallic nanowires have been investigated using Monte Carlo models~\cite{Gorshkov2020}. It has been reported that Au nanowires develop facets as a result surface diffusion in order to minimize the surface energy~\cite{Vigonski2017}.

	We note that for studying Rayleigh instability in these anisotropic systems, 3-D simulations are essential. Even though our formulation
	is valid for 3-D and we have a numerical implementation for 3-D,  given the highly non-linear nature of the evolution equations, and the variable mobility (to account for the faster surface diffusion), the time steps are extremely small and hence the system has to be evolved to much higher time steps to observe break-up. Hence, at the moment, we are not able to carry out 3D simulations in reasonable amounts of time. However, efforts of parallel implementation  of these models are in progress in our group; such implementations, when available, will help us study 3-D anisotropic systems. Hence, in this section we present some preliminary 2D simulation results on the effect of anisotropy in surface energy of the nanowires on morphological change and junction break-up.

	We study two different configurations with nanowires oriented at an angle of $90^{\circ}$ and $45^{\circ}$. In this case, we use the wire radius of value $\mathrm{R_1}, \mathrm{R_2} = 6$. The values of the isotropic and anisotropic contributions to the gradient energy coefficient, namely $\gamma_{I}$ and $\gamma_{A}$ are given in Table \ref{Table1}. The values are chosen such that the $<11>$ direction is energetically more favoured than the $<10>$ direction, i.e., the ratio of the interfacial energies (anisotropy ratio) $R_{\sigma} = \frac{\sigma_{<11>}}{\sigma_{<10>}}$ is chosen to be less than unity. Hence, it is expected that the morphological evolution of the nanowires will lead to surfaces on the nanowires whose surface normal vectors are along the energetically favoured $<11>$ direction.	As noted above, the $\Delta t$ values are to be chosen appropriately in this case in order to maintain the stability of the stiff numerical scheme.

\begin{figure}
		\centering
			\includegraphics[width=\textwidth]{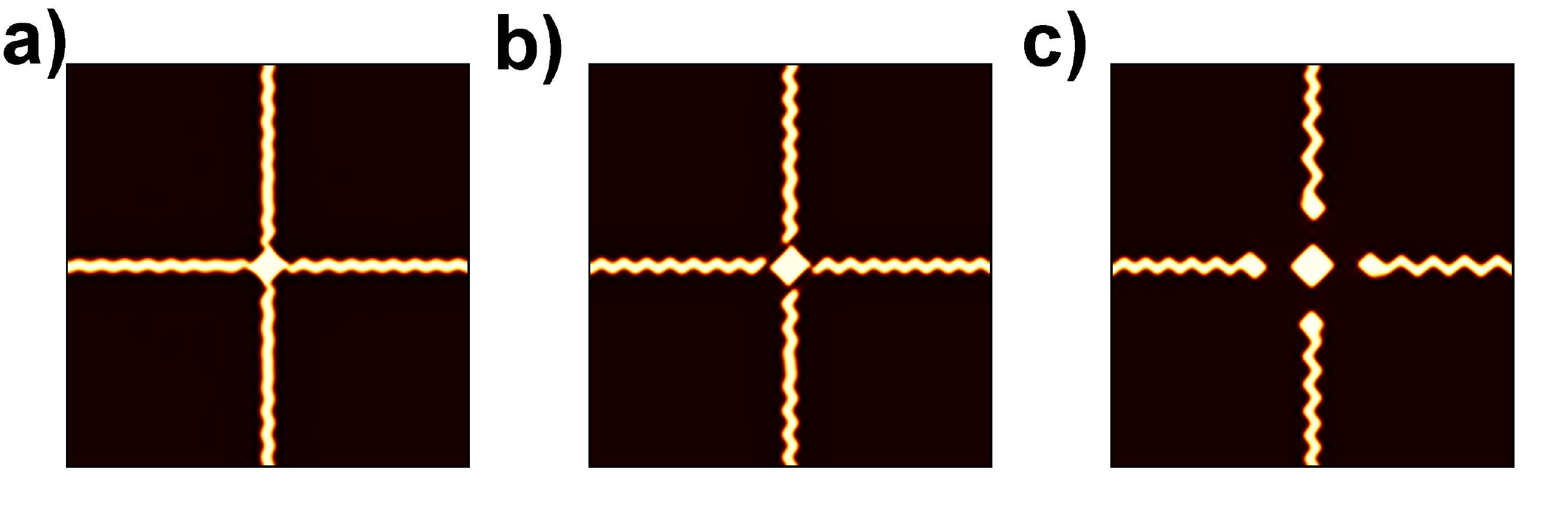}		
			\caption{Morphological evolution of two nanowires with cubic anisotropic interfacial free energy, and, intersecting at $90^{\circ}$ at different times: (a) 2000, (b) 3000 and (c) 20000 time units.
			 The $<11>$ interfaces are preferred over $<10>$ interfaces.}
		\label{Fig10}
	\end{figure}

\begin{figure}
		\centering
					\includegraphics[width=\textwidth]{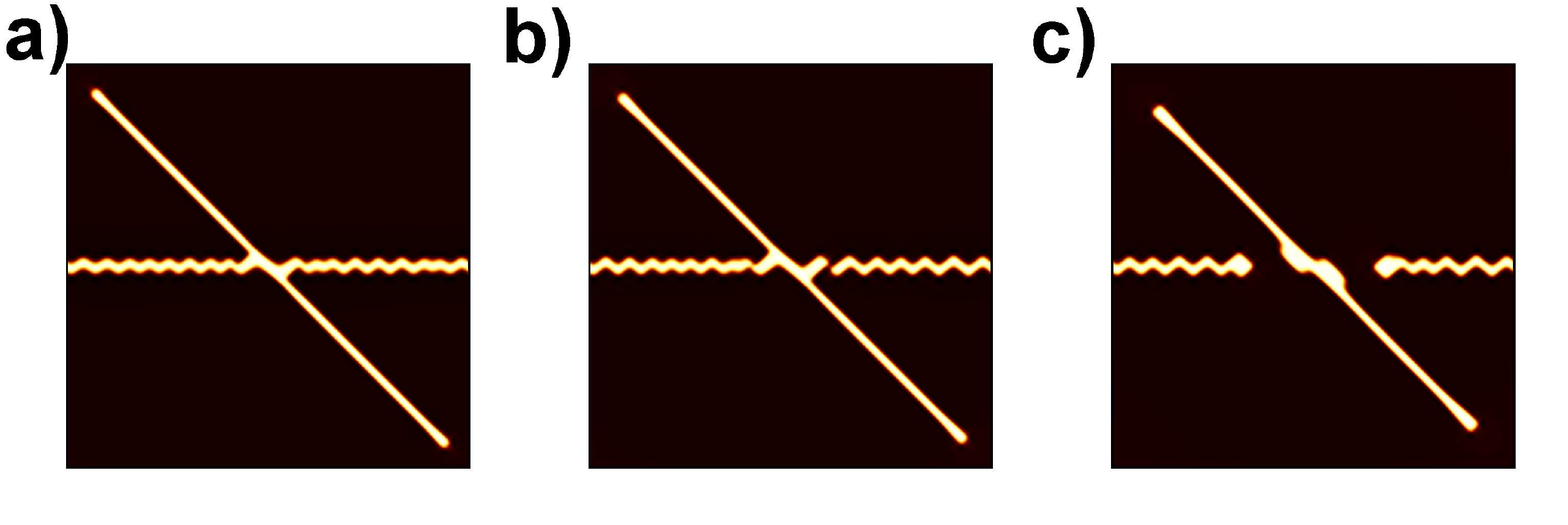}
		\caption{Morphological evolution of two nanowires with cubic anisotropic interfacial free energy, and, intersecting at $45^{\circ}$ at different times: (a) 5000, (b) 8000 and (c) 20000 time units. The $<11>$ interfaces are preferred over $<10>$ interfaces.}
		\label{Fig11}
	\end{figure}

	The 2D microstructure maps for the $90^{\circ}$ case is given in Fig.\ref{Fig10}, and the maps for the $45^{\circ}$ are given in Fig.\ref{Fig11}. From the microstructure maps, it is evident that for the  $45^{\circ}$ case, the wire aligned along the energetically favoured direction is stabilized, whereas the wire along the less energetically favoured direction undergoes change in morphology. The morphological change in the $90^{\circ}$ case is such that constrictions develop at regular intervals so as to give rise to surfaces with the surface normal vectors aligned parallel to the $<11>$ direction. In both the cases, we observe that the fragmentation of the nanowires occurs initially at the junction, followed by break-up along the nanowire arms. This observation leads to the conclusion that the order in which the nanowires fragment is independent of the anisotropy in interfacial energy, with the first break-up occurring at the junction. After the fragmentation the nanowires retract, and it is expected that in case of $90^{\circ}$ configuration, further fragmentation will occur along the nanowire arms. Further, in spite of the overall higher interfacial energy in the anisotropic cases, note that the kinetics of break-up is, relatively
	speaking, slower. This is primarily due to the stabilization of certain interfaces, we believe.
	

	In the past, theoretical treatment of surface energy anisotropy~\cite{Cahn1979, Wang2011} predicted that the weak axisymmetric perturbations will be stabilized depending on the nature of derivatives of interfacial energy (particularly the second derivative). Our phase field results above can be thought of as a validation of these predictions and an attempt to correlate the stability criteria with the underlying crystallographic symmetry. Here again, our simulation results are in good agreement with the kinetic Monte Carlo simulation results obtained by Vigonski et al.~\cite{Vigonski2017}, where they aligned the Au nanowires along $<110>$ direction and observed the fragmentation to occur initially at the junction. In their simulation, the nanowires initially having $\{110\}$ facets decomposed to more stable $\{111\}$ and $\{100\}$ facts. Our phase-field simulation results also reproduce same results; the nanowires develop energetically stable facets before undergoing
	fragmentation.

		To summarise, the preliminary 2-D simulations results indicate that in the case of systems with surface energy anisotropy, wires 
		with surfaces along the energetically favourable directions are unconditionally stable; and the anisotropy in surface energy has 
		a strong effect on kinetics. These results indicate that 3-D simulations of nanowires with interfacial anisotropy could be interesting.

    \section{Moment of inertia tensor}
   
    We define the moment of inertia matrix consisting of the nine components of the corresponding tensor as follows:
    \begin{equation}
    I = 
        \begin{pmatrix}
            I_{xx} & -I_{xy} & -I_{xz}\\
            -I_{xy} & I_{yy} & -I_{yz}\\
            -I_{xz} & -I_{yz} & I_{zz}
        \end{pmatrix}
    \end{equation}
    where, the components of the real symmetric inertia matrix are defined as:  
    \begin{equation}
    \begin{split}
         I_{xx} &= \sum (y^2 + z^2)dm\\
        I_{yy} &= \sum (x^2 + z^2)dm\\
        I_{zz} &= \sum (x^2 + y^2)dm\\
        I_{xy} &= I_{yx} = \sum (xy)dm\\
        I_{yz} &= I_{zy} = \sum (yz)dm\\
        I_{xz} &= I_{zx} = \sum (xz)dm  
    \end{split}
    \end{equation}
    In these terms, $dm$ is the mass of each individual particle constituting the body. We assign a value of unity to $dm$. 
    
    \section{Radius of gyration}

    The standard eigenvalue problem for real symmetric matrix is given as follows:
    \begin{equation}
      \begin{pmatrix}
            I_{xx} & -I_{xy} & -I_{xz}\\
            -I_{xy} & I_{yy} & -I_{yz}\\
            -I_{xz} & -I_{yz} & I_{zz}
        \end{pmatrix}
        \begin{pmatrix}
            Vx\\
            Vy\\
            Vz
        \end{pmatrix} = 
        \begin{pmatrix}
            I_{0} & 0 & 0\\
            0 & I_{0} & 0\\
            0 & 0 & I_{0}
        \end{pmatrix}
        \begin{pmatrix}
            Vx\\
            Vy\\
            Vz
        \end{pmatrix}
    \end{equation}
    where, $(V_x, V_y, V_z)$ are components of the eigenvectors. After solving the above eigenvalue problem, we obtain three eigenvalues $I_{0} = I_{1}, I_{2}, I_{3}$, and three eigenvectors corresponding to the three eigenvalues. We define the radius of gyration corresponding to these three eigenvalues of the moment of inertia matrix as:
    \begin{equation}
    \begin{split}
        R1 &= \sqrt{\frac{I_1}{M}}\\
        R2 &= \sqrt{\frac{I_2}{M}}\\
        R3 &= \sqrt{\frac{I_3}{M}}\\
    \end{split}    
    \end{equation}
    where, M is the mass of the body whose inertia tensor is calculated.
    
	\bibliography{NANOREFS}

\begin{thebibliography}{78}%
\makeatletter
\providecommand \@ifxundefined [1]{%
 \@ifx{#1\undefined}
}%
\providecommand \@ifnum [1]{%
 \ifnum #1\expandafter \@firstoftwo
 \else \expandafter \@secondoftwo
 \fi
}%
\providecommand \@ifx [1]{%
 \ifx #1\expandafter \@firstoftwo
 \else \expandafter \@secondoftwo
 \fi
}%
\providecommand \natexlab [1]{#1}%
\providecommand \enquote  [1]{``#1''}%
\providecommand \bibnamefont  [1]{#1}%
\providecommand \bibfnamefont [1]{#1}%
\providecommand \citenamefont [1]{#1}%
\providecommand \href@noop [0]{\@secondoftwo}%
\providecommand \href [0]{\begingroup \@sanitize@url \@href}%
\providecommand \@href[1]{\@@startlink{#1}\@@href}%
\providecommand \@@href[1]{\endgroup#1\@@endlink}%
\providecommand \@sanitize@url [0]{\catcode `\\12\catcode `\$12\catcode
  `\&12\catcode `\#12\catcode `\^12\catcode `\_12\catcode `\%12\relax}%
\providecommand \@@startlink[1]{}%
\providecommand \@@endlink[0]{}%
\providecommand \url  [0]{\begingroup\@sanitize@url \@url }%
\providecommand \@url [1]{\endgroup\@href {#1}{\urlprefix }}%
\providecommand \urlprefix  [0]{URL }%
\providecommand \Eprint [0]{\href }%
\providecommand \doibase [0]{http://dx.doi.org/}%
\providecommand \selectlanguage [0]{\@gobble}%
\providecommand \bibinfo  [0]{\@secondoftwo}%
\providecommand \bibfield  [0]{\@secondoftwo}%
\providecommand \translation [1]{[#1]}%
\providecommand \BibitemOpen [0]{}%
\providecommand \bibitemStop [0]{}%
\providecommand \bibitemNoStop [0]{.\EOS\space}%
\providecommand \EOS [0]{\spacefactor3000\relax}%
\providecommand \BibitemShut  [1]{\csname bibitem#1\endcsname}%
\let\auto@bib@innerbib\@empty
\bibitem [{\citenamefont {Lee}\ \emph {et~al.}(2012)\citenamefont {Lee},
  \citenamefont {Lee}, \citenamefont {Lee}, \citenamefont {Lee}, \citenamefont
  {Lee},\ and\ \citenamefont {Ko}}]{Lee2012}%
  \BibitemOpen
  \bibfield  {author} {\bibinfo {author} {\bibfnamefont {J.}~\bibnamefont
  {Lee}}, \bibinfo {author} {\bibfnamefont {P.}~\bibnamefont {Lee}}, \bibinfo
  {author} {\bibfnamefont {H.}~\bibnamefont {Lee}}, \bibinfo {author}
  {\bibfnamefont {D.}~\bibnamefont {Lee}}, \bibinfo {author} {\bibfnamefont
  {S.~S.}\ \bibnamefont {Lee}}, \ and\ \bibinfo {author} {\bibfnamefont
  {S.~H.}\ \bibnamefont {Ko}},\ }\href {\doibase 10.1039/c2nr31254a} {\bibfield
   {journal} {\bibinfo  {journal} {Nanoscale}\ }\textbf {\bibinfo {volume}
  {4}},\ \bibinfo {pages} {6408} (\bibinfo {year} {2012})}\BibitemShut
  {NoStop}%
\bibitem [{\citenamefont {Sannicolo}\ \emph {et~al.}(2016)\citenamefont
  {Sannicolo}, \citenamefont {Lagrange}, \citenamefont {Cabos}, \citenamefont
  {Celle}, \citenamefont {Simonato},\ and\ \citenamefont
  {Bellet}}]{Sannicolo2016}%
  \BibitemOpen
  \bibfield  {author} {\bibinfo {author} {\bibfnamefont {T.}~\bibnamefont
  {Sannicolo}}, \bibinfo {author} {\bibfnamefont {M.}~\bibnamefont {Lagrange}},
  \bibinfo {author} {\bibfnamefont {A.}~\bibnamefont {Cabos}}, \bibinfo
  {author} {\bibfnamefont {C.}~\bibnamefont {Celle}}, \bibinfo {author}
  {\bibfnamefont {J.~P.}\ \bibnamefont {Simonato}}, \ and\ \bibinfo {author}
  {\bibfnamefont {D.}~\bibnamefont {Bellet}},\ }\href {\doibase
  10.1002/smll.201602581} {\bibfield  {journal} {\bibinfo  {journal} {Small}\
  }\textbf {\bibinfo {volume} {12}},\ \bibinfo {pages} {6052} (\bibinfo {year}
  {2016})}\BibitemShut {NoStop}%
\bibitem [{\citenamefont {Lee}\ \emph {et~al.}(2020)\citenamefont {Lee},
  \citenamefont {Jang}, \citenamefont {Park}, \citenamefont {Park},
  \citenamefont {Park}, \citenamefont {Kim}, \citenamefont {Lee}, \citenamefont
  {Jeon}, \citenamefont {Lee}, \citenamefont {Ahn},\ and\ \citenamefont
  {Chung}}]{Lee2020}%
  \BibitemOpen
  \bibfield  {author} {\bibinfo {author} {\bibfnamefont {S.}~\bibnamefont
  {Lee}}, \bibinfo {author} {\bibfnamefont {J.}~\bibnamefont {Jang}}, \bibinfo
  {author} {\bibfnamefont {T.}~\bibnamefont {Park}}, \bibinfo {author}
  {\bibfnamefont {Y.~M.}\ \bibnamefont {Park}}, \bibinfo {author}
  {\bibfnamefont {J.~S.}\ \bibnamefont {Park}}, \bibinfo {author}
  {\bibfnamefont {Y.~K.}\ \bibnamefont {Kim}}, \bibinfo {author} {\bibfnamefont
  {H.~K.}\ \bibnamefont {Lee}}, \bibinfo {author} {\bibfnamefont {E.~C.}\
  \bibnamefont {Jeon}}, \bibinfo {author} {\bibfnamefont {D.~K.}\ \bibnamefont
  {Lee}}, \bibinfo {author} {\bibfnamefont {B.}~\bibnamefont {Ahn}}, \ and\
  \bibinfo {author} {\bibfnamefont {C.~H.}\ \bibnamefont {Chung}},\ }\href@noop
  {} {\bibfield  {journal} {\bibinfo  {journal} {ACS Applied Materials and
  Interfaces}\ }\textbf {\bibinfo {volume} {12}},\ \bibinfo {pages} {6169}
  (\bibinfo {year} {2020})}\BibitemShut {NoStop}%
\bibitem [{\citenamefont {Gonzalez-Garcia}\ \emph {et~al.}(2016)\citenamefont
  {Gonzalez-Garcia}, \citenamefont {Maurer}, \citenamefont {Reiser},
  \citenamefont {Kanelidis},\ and\ \citenamefont
  {Kraus}}]{Gonzalez-Garcia2016}%
  \BibitemOpen
  \bibfield  {author} {\bibinfo {author} {\bibfnamefont {L.}~\bibnamefont
  {Gonzalez-Garcia}}, \bibinfo {author} {\bibfnamefont {J.~H.}\ \bibnamefont
  {Maurer}}, \bibinfo {author} {\bibfnamefont {B.}~\bibnamefont {Reiser}},
  \bibinfo {author} {\bibfnamefont {I.}~\bibnamefont {Kanelidis}}, \ and\
  \bibinfo {author} {\bibfnamefont {T.}~\bibnamefont {Kraus}},\ }\href
  {\doibase 10.1016/j.proeng.2015.08.1120} {\bibfield  {journal} {\bibinfo
  {journal} {Procedia Engineering}\ }\textbf {\bibinfo {volume} {141}},\
  \bibinfo {pages} {152} (\bibinfo {year} {2016})}\BibitemShut {NoStop}%
\bibitem [{\citenamefont {Vazquez-Mena}\ \emph {et~al.}(2011)\citenamefont
  {Vazquez-Mena}, \citenamefont {Sannomiya}, \citenamefont {Villanueva},
  \citenamefont {Voros},\ and\ \citenamefont {Brugger}}]{Vazquez-Mena2011}%
  \BibitemOpen
  \bibfield  {author} {\bibinfo {author} {\bibfnamefont {O.}~\bibnamefont
  {Vazquez-Mena}}, \bibinfo {author} {\bibfnamefont {T.}~\bibnamefont
  {Sannomiya}}, \bibinfo {author} {\bibfnamefont {L.~G.}\ \bibnamefont
  {Villanueva}}, \bibinfo {author} {\bibfnamefont {J.}~\bibnamefont {Voros}}, \
  and\ \bibinfo {author} {\bibfnamefont {J.}~\bibnamefont {Brugger}},\ }\href
  {\doibase 10.1021/nn1019253} {\bibfield  {journal} {\bibinfo  {journal} {ACS
  Nano}\ }\textbf {\bibinfo {volume} {5}},\ \bibinfo {pages} {844} (\bibinfo
  {year} {2011})}\BibitemShut {NoStop}%
\bibitem [{\citenamefont {Song}\ \emph {et~al.}(2015)\citenamefont {Song},
  \citenamefont {Rim}, \citenamefont {Liu}, \citenamefont {Bob}, \citenamefont
  {Ye}, \citenamefont {Hsieh},\ and\ \citenamefont {Yang}}]{Song2015}%
  \BibitemOpen
  \bibfield  {author} {\bibinfo {author} {\bibfnamefont {T.~B.}\ \bibnamefont
  {Song}}, \bibinfo {author} {\bibfnamefont {Y.~S.}\ \bibnamefont {Rim}},
  \bibinfo {author} {\bibfnamefont {F.}~\bibnamefont {Liu}}, \bibinfo {author}
  {\bibfnamefont {B.}~\bibnamefont {Bob}}, \bibinfo {author} {\bibfnamefont
  {S.}~\bibnamefont {Ye}}, \bibinfo {author} {\bibfnamefont {Y.~T.}\
  \bibnamefont {Hsieh}}, \ and\ \bibinfo {author} {\bibfnamefont
  {Y.}~\bibnamefont {Yang}},\ }\href {\doibase 10.1021/acsami.5b06540}
  {\bibfield  {journal} {\bibinfo  {journal} {ACS Applied Materials and
  Interfaces}\ }\textbf {\bibinfo {volume} {7}},\ \bibinfo {pages} {24601}
  (\bibinfo {year} {2015})}\BibitemShut {NoStop}%
\bibitem [{\citenamefont {Langley}\ \emph {et~al.}(2014)\citenamefont
  {Langley}, \citenamefont {Lagrange}, \citenamefont {Giusti}, \citenamefont
  {Jim{\'{e}}nez}, \citenamefont {Br{\'{e}}chet}, \citenamefont {Nguyen},\ and\
  \citenamefont {Bellet}}]{Langley2014}%
  \BibitemOpen
  \bibfield  {author} {\bibinfo {author} {\bibfnamefont {D.~P.}\ \bibnamefont
  {Langley}}, \bibinfo {author} {\bibfnamefont {M.}~\bibnamefont {Lagrange}},
  \bibinfo {author} {\bibfnamefont {G.}~\bibnamefont {Giusti}}, \bibinfo
  {author} {\bibfnamefont {C.}~\bibnamefont {Jim{\'{e}}nez}}, \bibinfo {author}
  {\bibfnamefont {Y.}~\bibnamefont {Br{\'{e}}chet}}, \bibinfo {author}
  {\bibfnamefont {N.~D.}\ \bibnamefont {Nguyen}}, \ and\ \bibinfo {author}
  {\bibfnamefont {D.}~\bibnamefont {Bellet}},\ }\href@noop {} {\bibfield
  {journal} {\bibinfo  {journal} {Nanoscale}\ }\textbf {\bibinfo {volume}
  {6}},\ \bibinfo {pages} {13535} (\bibinfo {year} {2014})}\BibitemShut
  {NoStop}%
\bibitem [{\citenamefont {Shin}, \citenamefont {Yu},\ and\ \citenamefont
  {Song}(2007)}]{Shin2007}%
  \BibitemOpen
  \bibfield  {author} {\bibinfo {author} {\bibfnamefont {H.~S.}\ \bibnamefont
  {Shin}}, \bibinfo {author} {\bibfnamefont {J.}~\bibnamefont {Yu}}, \ and\
  \bibinfo {author} {\bibfnamefont {J.~Y.}\ \bibnamefont {Song}},\ }\href@noop
  {} {\bibfield  {journal} {\bibinfo  {journal} {Applied Physics Letters}\
  }\textbf {\bibinfo {volume} {91}},\ \bibinfo {pages} {173106} (\bibinfo
  {year} {2007})}\BibitemShut {NoStop}%
\bibitem [{\citenamefont {Rauber}\ \emph {et~al.}(2012)\citenamefont {Rauber},
  \citenamefont {Muench}, \citenamefont {Toimil-Molares},\ and\ \citenamefont
  {Ensinger}}]{Rauber2012}%
  \BibitemOpen
  \bibfield  {author} {\bibinfo {author} {\bibfnamefont {M.}~\bibnamefont
  {Rauber}}, \bibinfo {author} {\bibfnamefont {F.}~\bibnamefont {Muench}},
  \bibinfo {author} {\bibfnamefont {M.~E.}\ \bibnamefont {Toimil-Molares}}, \
  and\ \bibinfo {author} {\bibfnamefont {W.}~\bibnamefont {Ensinger}},\
  }\href@noop {} {\bibfield  {journal} {\bibinfo  {journal} {Nanotechnology}\
  }\textbf {\bibinfo {volume} {23}} (\bibinfo {year} {2012})}\BibitemShut
  {NoStop}%
\bibitem [{\citenamefont {Lagrange}\ \emph {et~al.}(2015)\citenamefont
  {Lagrange}, \citenamefont {Langley}, \citenamefont {Giusti}, \citenamefont
  {Jim{\'{e}}nez}, \citenamefont {Br{\'{e}}chet},\ and\ \citenamefont
  {Bellet}}]{Lagrange2015}%
  \BibitemOpen
  \bibfield  {author} {\bibinfo {author} {\bibfnamefont {M.}~\bibnamefont
  {Lagrange}}, \bibinfo {author} {\bibfnamefont {D.~P.}\ \bibnamefont
  {Langley}}, \bibinfo {author} {\bibfnamefont {G.}~\bibnamefont {Giusti}},
  \bibinfo {author} {\bibfnamefont {C.}~\bibnamefont {Jim{\'{e}}nez}}, \bibinfo
  {author} {\bibfnamefont {Y.}~\bibnamefont {Br{\'{e}}chet}}, \ and\ \bibinfo
  {author} {\bibfnamefont {D.}~\bibnamefont {Bellet}},\ }\href {\doibase
  10.1039/c5nr04084a} {\bibfield  {journal} {\bibinfo  {journal} {Nanoscale}\
  }\textbf {\bibinfo {volume} {7}},\ \bibinfo {pages} {17410} (\bibinfo {year}
  {2015})}\BibitemShut {NoStop}%
\bibitem [{\citenamefont {Rayleigh}(1878)}]{Rayleigh1878}%
  \BibitemOpen
  \bibfield  {author} {\bibinfo {author} {\bibfnamefont {L.}~\bibnamefont
  {Rayleigh}},\ }\href {\doibase 10.1112/plms/s1-10.1.4} {\bibfield  {journal}
  {\bibinfo  {journal} {Proceedings of the London Mathematical Society}\
  }\textbf {\bibinfo {volume} {s1-10}},\ \bibinfo {pages} {4} (\bibinfo {year}
  {1878})}\BibitemShut {NoStop}%
\bibitem [{\citenamefont {Nichols}\ and\ \citenamefont
  {Mullins}(1965)}]{Nichols1965}%
  \BibitemOpen
  \bibfield  {author} {\bibinfo {author} {\bibfnamefont {F.~A.}\ \bibnamefont
  {Nichols}}\ and\ \bibinfo {author} {\bibfnamefont {W.~W.}\ \bibnamefont
  {Mullins}},\ }\href {\doibase 10.1063/1.1714360} {\bibfield  {journal}
  {\bibinfo  {journal} {Journal of Applied Physics}\ }\textbf {\bibinfo
  {volume} {36}},\ \bibinfo {pages} {1826} (\bibinfo {year}
  {1965})}\BibitemShut {NoStop}%
\bibitem [{\citenamefont {Ma}(1998)}]{Ma1998}%
  \BibitemOpen
  \bibfield  {author} {\bibinfo {author} {\bibfnamefont {Q.}~\bibnamefont
  {Ma}},\ }\href {\doibase 10.1016/s1359-6454(97)00337-6} {\bibfield  {journal}
  {\bibinfo  {journal} {Acta Materialia}\ }\textbf {\bibinfo {volume} {46}},\
  \bibinfo {pages} {1669} (\bibinfo {year} {1998})}\BibitemShut {NoStop}%
\bibitem [{\citenamefont {Cahn}(1979)}]{Cahn1979}%
  \BibitemOpen
  \bibfield  {author} {\bibinfo {author} {\bibfnamefont {J.}~\bibnamefont
  {Cahn}},\ }\href {\doibase https://doi.org/10.1016/0036-9748(79)90205-9}
  {\bibfield  {journal} {\bibinfo  {journal} {Scripta Metallurgica}\ }\textbf
  {\bibinfo {volume} {13}},\ \bibinfo {pages} {1069} (\bibinfo {year}
  {1979})}\BibitemShut {NoStop}%
\bibitem [{\citenamefont {St{\"{o}}lken}\ and\ \citenamefont
  {Glaeser}(1992)}]{Stolken1992}%
  \BibitemOpen
  \bibfield  {author} {\bibinfo {author} {\bibfnamefont {J.~S.}\ \bibnamefont
  {St{\"{o}}lken}}\ and\ \bibinfo {author} {\bibfnamefont {A.~M.}\ \bibnamefont
  {Glaeser}},\ }\href {\doibase 10.1016/0956-716X(92)90209-W} {\bibfield
  {journal} {\bibinfo  {journal} {Scripta Metallurgica et Materialia}\ }\textbf
  {\bibinfo {volume} {27}},\ \bibinfo {pages} {449} (\bibinfo {year}
  {1992})}\BibitemShut {NoStop}%
\bibitem [{\citenamefont {Gurski}\ and\ \citenamefont
  {McFadden}(2003)}]{Gurski2003}%
  \BibitemOpen
  \bibfield  {author} {\bibinfo {author} {\bibfnamefont {K.~F.}\ \bibnamefont
  {Gurski}}\ and\ \bibinfo {author} {\bibfnamefont {G.~B.}\ \bibnamefont
  {McFadden}},\ }\href@noop {} {\bibfield  {journal} {\bibinfo  {journal}
  {Proceedings of the Royal Society A: Mathematical, Physical and Engineering
  Sciences}\ }\textbf {\bibinfo {volume} {459}},\ \bibinfo {pages} {2575}
  (\bibinfo {year} {2003})}\BibitemShut {NoStop}%
\bibitem [{\citenamefont {Kim}\ and\ \citenamefont {Thompson}(2015)}]{Kim2015}%
  \BibitemOpen
  \bibfield  {author} {\bibinfo {author} {\bibfnamefont {G.~H.}\ \bibnamefont
  {Kim}}\ and\ \bibinfo {author} {\bibfnamefont {C.~V.}\ \bibnamefont
  {Thompson}},\ }\href@noop {} {\bibfield  {journal} {\bibinfo  {journal} {Acta
  Materialia}\ }\textbf {\bibinfo {volume} {84}},\ \bibinfo {pages} {190}
  (\bibinfo {year} {2015})}\BibitemShut {NoStop}%
\bibitem [{\citenamefont {Gorshkov}, \citenamefont {Tereshchuk},\ and\
  \citenamefont {Sareh}(2020)}]{Gorshkov2020}%
  \BibitemOpen
  \bibfield  {author} {\bibinfo {author} {\bibfnamefont {V.~N.}\ \bibnamefont
  {Gorshkov}}, \bibinfo {author} {\bibfnamefont {V.~V.}\ \bibnamefont
  {Tereshchuk}}, \ and\ \bibinfo {author} {\bibfnamefont {P.}~\bibnamefont
  {Sareh}},\ }\href {\doibase 10.1039/c9ce01893j} {\bibfield  {journal}
  {\bibinfo  {journal} {CrystEngComm}\ }\textbf {\bibinfo {volume} {22}},\
  \bibinfo {pages} {2601} (\bibinfo {year} {2020})}\BibitemShut {NoStop}%
\bibitem [{\citenamefont {Molares}\ \emph {et~al.}(2004)\citenamefont
  {Molares}, \citenamefont {Balogh}, \citenamefont {Cornelius}, \citenamefont
  {Neumann},\ and\ \citenamefont {Trautmann}}]{Molares2004}%
  \BibitemOpen
  \bibfield  {author} {\bibinfo {author} {\bibfnamefont {M.~E.}\ \bibnamefont
  {Molares}}, \bibinfo {author} {\bibfnamefont {A.~G.}\ \bibnamefont {Balogh}},
  \bibinfo {author} {\bibfnamefont {T.~W.}\ \bibnamefont {Cornelius}}, \bibinfo
  {author} {\bibfnamefont {R.}~\bibnamefont {Neumann}}, \ and\ \bibinfo
  {author} {\bibfnamefont {C.}~\bibnamefont {Trautmann}},\ }\href {\doibase
  10.1063/1.1826237} {\bibfield  {journal} {\bibinfo  {journal} {Applied
  Physics Letters}\ }\textbf {\bibinfo {volume} {85}},\ \bibinfo {pages} {5337}
  (\bibinfo {year} {2004})}\BibitemShut {NoStop}%
\bibitem [{\citenamefont {Karim}\ \emph {et~al.}(2006)\citenamefont {Karim},
  \citenamefont {Toimil-Molares}, \citenamefont {Balogh}, \citenamefont
  {Ensinger}, \citenamefont {Cornelius}, \citenamefont {Khan},\ and\
  \citenamefont {Neumann}}]{Karim2006}%
  \BibitemOpen
  \bibfield  {author} {\bibinfo {author} {\bibfnamefont {S.}~\bibnamefont
  {Karim}}, \bibinfo {author} {\bibfnamefont {M.~E.}\ \bibnamefont
  {Toimil-Molares}}, \bibinfo {author} {\bibfnamefont {A.~G.}\ \bibnamefont
  {Balogh}}, \bibinfo {author} {\bibfnamefont {W.}~\bibnamefont {Ensinger}},
  \bibinfo {author} {\bibfnamefont {T.~W.}\ \bibnamefont {Cornelius}}, \bibinfo
  {author} {\bibfnamefont {E.~U.}\ \bibnamefont {Khan}}, \ and\ \bibinfo
  {author} {\bibfnamefont {R.}~\bibnamefont {Neumann}},\ }\href@noop {}
  {\bibfield  {journal} {\bibinfo  {journal} {Nanotechnology}\ }\textbf
  {\bibinfo {volume} {17}},\ \bibinfo {pages} {5954} (\bibinfo {year}
  {2006})}\BibitemShut {NoStop}%
\bibitem [{\citenamefont {Karim}\ \emph {et~al.}(2007)\citenamefont {Karim},
  \citenamefont {Toimil-Molares}, \citenamefont {Ensinger}, \citenamefont
  {Balogh}, \citenamefont {Cornelius}, \citenamefont {Khan},\ and\
  \citenamefont {Neumann}}]{Karim2007}%
  \BibitemOpen
  \bibfield  {author} {\bibinfo {author} {\bibfnamefont {S.}~\bibnamefont
  {Karim}}, \bibinfo {author} {\bibfnamefont {M.~E.}\ \bibnamefont
  {Toimil-Molares}}, \bibinfo {author} {\bibfnamefont {W.}~\bibnamefont
  {Ensinger}}, \bibinfo {author} {\bibfnamefont {A.~G.}\ \bibnamefont
  {Balogh}}, \bibinfo {author} {\bibfnamefont {T.~W.}\ \bibnamefont
  {Cornelius}}, \bibinfo {author} {\bibfnamefont {E.~U.}\ \bibnamefont {Khan}},
  \ and\ \bibinfo {author} {\bibfnamefont {R.}~\bibnamefont {Neumann}},\
  }\href@noop {} {\bibfield  {journal} {\bibinfo  {journal} {Journal of Physics
  D: Applied Physics}\ }\textbf {\bibinfo {volume} {40}},\ \bibinfo {pages}
  {3767} (\bibinfo {year} {2007})}\BibitemShut {NoStop}%
\bibitem [{\citenamefont {Huang}\ \emph {et~al.}(2010)\citenamefont {Huang},
  \citenamefont {Zhan}, \citenamefont {Wang}, \citenamefont {Zhang},
  \citenamefont {Xing}, \citenamefont {Guo}, \citenamefont {Leusink},
  \citenamefont {Zheng},\ and\ \citenamefont {Wu}}]{Huang2010}%
  \BibitemOpen
  \bibfield  {author} {\bibinfo {author} {\bibfnamefont {X.~H.}\ \bibnamefont
  {Huang}}, \bibinfo {author} {\bibfnamefont {Z.~Y.}\ \bibnamefont {Zhan}},
  \bibinfo {author} {\bibfnamefont {X.}~\bibnamefont {Wang}}, \bibinfo {author}
  {\bibfnamefont {Z.}~\bibnamefont {Zhang}}, \bibinfo {author} {\bibfnamefont
  {G.~Z.}\ \bibnamefont {Xing}}, \bibinfo {author} {\bibfnamefont {D.~L.}\
  \bibnamefont {Guo}}, \bibinfo {author} {\bibfnamefont {D.~P.}\ \bibnamefont
  {Leusink}}, \bibinfo {author} {\bibfnamefont {L.~X.}\ \bibnamefont {Zheng}},
  \ and\ \bibinfo {author} {\bibfnamefont {T.}~\bibnamefont {Wu}},\ }\href
  {\doibase 10.1063/1.3518470} {\bibfield  {journal} {\bibinfo  {journal}
  {Applied Physics Letters}\ }\textbf {\bibinfo {volume} {97}},\ \bibinfo
  {pages} {1} (\bibinfo {year} {2010})}\BibitemShut {NoStop}%
\bibitem [{\citenamefont {Beavers}, \citenamefont {Marotta},\ and\
  \citenamefont {Bottomley}(2010)}]{Beavers2010}%
  \BibitemOpen
  \bibfield  {author} {\bibinfo {author} {\bibfnamefont {K.~R.}\ \bibnamefont
  {Beavers}}, \bibinfo {author} {\bibfnamefont {N.~E.}\ \bibnamefont
  {Marotta}}, \ and\ \bibinfo {author} {\bibfnamefont {L.~A.}\ \bibnamefont
  {Bottomley}},\ }\href {\doibase 10.1021/cm901791u} {\bibfield  {journal}
  {\bibinfo  {journal} {Chemistry of Materials}\ }\textbf {\bibinfo {volume}
  {22}},\ \bibinfo {pages} {2184} (\bibinfo {year} {2010})}\BibitemShut
  {NoStop}%
\bibitem [{\citenamefont {Hsiung}\ \emph {et~al.}(2010)\citenamefont {Hsiung},
  \citenamefont {Liao}, \citenamefont {Gan}, \citenamefont {Wu}, \citenamefont
  {Hwang}, \citenamefont {Chen},\ and\ \citenamefont {Tsai}}]{Hsiung2010}%
  \BibitemOpen
  \bibfield  {author} {\bibinfo {author} {\bibfnamefont {C.~P.}\ \bibnamefont
  {Hsiung}}, \bibinfo {author} {\bibfnamefont {H.~W.}\ \bibnamefont {Liao}},
  \bibinfo {author} {\bibfnamefont {J.~Y.}\ \bibnamefont {Gan}}, \bibinfo
  {author} {\bibfnamefont {T.~B.}\ \bibnamefont {Wu}}, \bibinfo {author}
  {\bibfnamefont {J.~C.}\ \bibnamefont {Hwang}}, \bibinfo {author}
  {\bibfnamefont {F.}~\bibnamefont {Chen}}, \ and\ \bibinfo {author}
  {\bibfnamefont {M.~J.}\ \bibnamefont {Tsai}},\ }\href {\doibase
  10.1021/nn1010667} {\bibfield  {journal} {\bibinfo  {journal} {ACS Nano}\
  }\textbf {\bibinfo {volume} {4}},\ \bibinfo {pages} {5414} (\bibinfo {year}
  {2010})}\BibitemShut {NoStop}%
\bibitem [{\citenamefont {Oh}, \citenamefont {Lee},\ and\ \citenamefont
  {Lee}(2018)}]{Oh2018}%
  \BibitemOpen
  \bibfield  {author} {\bibinfo {author} {\bibfnamefont {H.}~\bibnamefont
  {Oh}}, \bibinfo {author} {\bibfnamefont {J.}~\bibnamefont {Lee}}, \ and\
  \bibinfo {author} {\bibfnamefont {M.}~\bibnamefont {Lee}},\ }\href {\doibase
  10.1016/j.apsusc.2017.08.102} {\bibfield  {journal} {\bibinfo  {journal}
  {Applied Surface Science}\ }\textbf {\bibinfo {volume} {427}},\ \bibinfo
  {pages} {65} (\bibinfo {year} {2018})}\BibitemShut {NoStop}%
\bibitem [{\citenamefont {Kolb}\ \emph {et~al.}(2005)\citenamefont {Kolb},
  \citenamefont {Hofmeister}, \citenamefont {Zacharias},\ and\ \citenamefont
  {G{\"{o}}sele}}]{Kolb2005}%
  \BibitemOpen
  \bibfield  {author} {\bibinfo {author} {\bibfnamefont {F.~M.}\ \bibnamefont
  {Kolb}}, \bibinfo {author} {\bibfnamefont {H.}~\bibnamefont {Hofmeister}},
  \bibinfo {author} {\bibfnamefont {M.}~\bibnamefont {Zacharias}}, \ and\
  \bibinfo {author} {\bibfnamefont {U.}~\bibnamefont {G{\"{o}}sele}},\ }\href
  {\doibase 10.1007/s00339-004-3188-7} {\bibfield  {journal} {\bibinfo
  {journal} {Applied Physics A}\ }\textbf {\bibinfo {volume} {80}},\ \bibinfo
  {pages} {1405} (\bibinfo {year} {2005})}\BibitemShut {NoStop}%
\bibitem [{\citenamefont {Bechelany}\ \emph {et~al.}(2012)\citenamefont
  {Bechelany}, \citenamefont {Riesterer}, \citenamefont {Brioude},
  \citenamefont {Cornu},\ and\ \citenamefont {Miele}}]{Bechelany2012}%
  \BibitemOpen
  \bibfield  {author} {\bibinfo {author} {\bibfnamefont {M.}~\bibnamefont
  {Bechelany}}, \bibinfo {author} {\bibfnamefont {J.~L.}\ \bibnamefont
  {Riesterer}}, \bibinfo {author} {\bibfnamefont {A.}~\bibnamefont {Brioude}},
  \bibinfo {author} {\bibfnamefont {D.}~\bibnamefont {Cornu}}, \ and\ \bibinfo
  {author} {\bibfnamefont {P.}~\bibnamefont {Miele}},\ }\href {\doibase
  10.1039/c2ce25636c} {\bibfield  {journal} {\bibinfo  {journal}
  {CrystEngComm}\ }\textbf {\bibinfo {volume} {14}},\ \bibinfo {pages} {7744}
  (\bibinfo {year} {2012})}\BibitemShut {NoStop}%
\bibitem [{\citenamefont {B{\"{u}}rki}\ and\ \citenamefont
  {Stafford}(2005)}]{Burki2005}%
  \BibitemOpen
  \bibfield  {author} {\bibinfo {author} {\bibfnamefont {J.}~\bibnamefont
  {B{\"{u}}rki}}\ and\ \bibinfo {author} {\bibfnamefont {C.~A.}\ \bibnamefont
  {Stafford}},\ }\href {\doibase 10.1007/s00339-005-3389-8} {\bibfield
  {journal} {\bibinfo  {journal} {Applied Physics A}\ }\textbf {\bibinfo
  {volume} {81}},\ \bibinfo {pages} {1519} (\bibinfo {year}
  {2005})}\BibitemShut {NoStop}%
\bibitem [{\citenamefont {Kassubek}\ \emph {et~al.}(2001)\citenamefont
  {Kassubek}, \citenamefont {Stafford}, \citenamefont {Grabert},\ and\
  \citenamefont {Goldstein}}]{Kassubek2001}%
  \BibitemOpen
  \bibfield  {author} {\bibinfo {author} {\bibfnamefont {F.}~\bibnamefont
  {Kassubek}}, \bibinfo {author} {\bibfnamefont {C.~A.}\ \bibnamefont
  {Stafford}}, \bibinfo {author} {\bibfnamefont {H.}~\bibnamefont {Grabert}}, \
  and\ \bibinfo {author} {\bibfnamefont {R.~E.}\ \bibnamefont {Goldstein}},\
  }\href {\doibase 10.1088/0951-7715/14/1/310} {\bibfield  {journal} {\bibinfo
  {journal} {Nonlinearity}\ }\textbf {\bibinfo {volume} {14}},\ \bibinfo
  {pages} {167} (\bibinfo {year} {2001})}\BibitemShut {NoStop}%
\bibitem [{\citenamefont {Vigonski}\ \emph {et~al.}(2017)\citenamefont
  {Vigonski}, \citenamefont {Jansson}, \citenamefont {Vlassov}, \citenamefont
  {Polyakov}, \citenamefont {Baibuz}, \citenamefont {Oras}, \citenamefont
  {Aabloo}, \citenamefont {Djurabekova},\ and\ \citenamefont
  {Zadin}}]{Vigonski2017}%
  \BibitemOpen
  \bibfield  {author} {\bibinfo {author} {\bibfnamefont {S.}~\bibnamefont
  {Vigonski}}, \bibinfo {author} {\bibfnamefont {V.}~\bibnamefont {Jansson}},
  \bibinfo {author} {\bibfnamefont {S.}~\bibnamefont {Vlassov}}, \bibinfo
  {author} {\bibfnamefont {B.}~\bibnamefont {Polyakov}}, \bibinfo {author}
  {\bibfnamefont {E.}~\bibnamefont {Baibuz}}, \bibinfo {author} {\bibfnamefont
  {S.}~\bibnamefont {Oras}}, \bibinfo {author} {\bibfnamefont {A.}~\bibnamefont
  {Aabloo}}, \bibinfo {author} {\bibfnamefont {F.}~\bibnamefont {Djurabekova}},
  \ and\ \bibinfo {author} {\bibfnamefont {V.}~\bibnamefont {Zadin}},\ }\href
  {\doibase 10.1088/1361-6528/aa9a1b} {\bibfield  {journal} {\bibinfo
  {journal} {Nanotechnology}\ }\textbf {\bibinfo {volume} {29}},\ \bibinfo
  {pages} {015704} (\bibinfo {year} {2017})}\BibitemShut {NoStop}%
\bibitem [{\citenamefont {Xue}\ \emph {et~al.}(2016)\citenamefont {Xue},
  \citenamefont {Xu}, \citenamefont {Zhao}, \citenamefont {Wang}, \citenamefont
  {Jiang}, \citenamefont {Yu}, \citenamefont {Wang}, \citenamefont {Xu},
  \citenamefont {Shi}, \citenamefont {Chen},\ and\ \citenamefont
  {i~Cabarrocas}}]{Xue2016}%
  \BibitemOpen
  \bibfield  {author} {\bibinfo {author} {\bibfnamefont {Z.}~\bibnamefont
  {Xue}}, \bibinfo {author} {\bibfnamefont {M.}~\bibnamefont {Xu}}, \bibinfo
  {author} {\bibfnamefont {Y.}~\bibnamefont {Zhao}}, \bibinfo {author}
  {\bibfnamefont {J.}~\bibnamefont {Wang}}, \bibinfo {author} {\bibfnamefont
  {X.}~\bibnamefont {Jiang}}, \bibinfo {author} {\bibfnamefont
  {L.}~\bibnamefont {Yu}}, \bibinfo {author} {\bibfnamefont {J.}~\bibnamefont
  {Wang}}, \bibinfo {author} {\bibfnamefont {J.}~\bibnamefont {Xu}}, \bibinfo
  {author} {\bibfnamefont {Y.}~\bibnamefont {Shi}}, \bibinfo {author}
  {\bibfnamefont {K.}~\bibnamefont {Chen}}, \ and\ \bibinfo {author}
  {\bibfnamefont {P.~R.}\ \bibnamefont {i~Cabarrocas}},\ }\href@noop {}
  {\bibfield  {journal} {\bibinfo  {journal} {Nature Communications}\ }\textbf
  {\bibinfo {volume} {7}},\ \bibinfo {pages} {1} (\bibinfo {year}
  {2016})}\BibitemShut {NoStop}%
\bibitem [{\citenamefont {Zhu}\ \emph {et~al.}(2019)\citenamefont {Zhu},
  \citenamefont {Chen}, \citenamefont {Wan}, \citenamefont {Peng},
  \citenamefont {Huang}, \citenamefont {Jiang}, \citenamefont {Li},\ and\
  \citenamefont {Chu}}]{Zhu2019}%
  \BibitemOpen
  \bibfield  {author} {\bibinfo {author} {\bibfnamefont {Y.}~\bibnamefont
  {Zhu}}, \bibinfo {author} {\bibfnamefont {J.}~\bibnamefont {Chen}}, \bibinfo
  {author} {\bibfnamefont {T.}~\bibnamefont {Wan}}, \bibinfo {author}
  {\bibfnamefont {S.}~\bibnamefont {Peng}}, \bibinfo {author} {\bibfnamefont
  {S.}~\bibnamefont {Huang}}, \bibinfo {author} {\bibfnamefont
  {Y.}~\bibnamefont {Jiang}}, \bibinfo {author} {\bibfnamefont
  {S.}~\bibnamefont {Li}}, \ and\ \bibinfo {author} {\bibfnamefont
  {D.}~\bibnamefont {Chu}},\ }\href {\doibase 10.1021/acsaelm.9b00218}
  {\bibfield  {journal} {\bibinfo  {journal} {ACS Applied Electronic
  Materials}\ }\textbf {\bibinfo {volume} {1}},\ \bibinfo {pages} {1275}
  (\bibinfo {year} {2019})}\BibitemShut {NoStop}%
\bibitem [{\citenamefont {Barnard}(2010)}]{Barnard2010}%
  \BibitemOpen
  \bibfield  {author} {\bibinfo {author} {\bibfnamefont {A.~S.}\ \bibnamefont
  {Barnard}},\ }\href@noop {} {\bibfield  {journal} {\bibinfo  {journal}
  {Reports on Progress in Physics}\ }\textbf {\bibinfo {volume} {73}} (\bibinfo
  {year} {2010})}\BibitemShut {NoStop}%
\bibitem [{\citenamefont {Chen}(2002)}]{Chen2002}%
  \BibitemOpen
  \bibfield  {author} {\bibinfo {author} {\bibfnamefont {L.-Q.}\ \bibnamefont
  {Chen}},\ }\href {\doibase 10.1146/annurev.matsci.32.112001.132041}
  {\bibfield  {journal} {\bibinfo  {journal} {Annual Review of Materials
  Research}\ }\textbf {\bibinfo {volume} {32}},\ \bibinfo {pages} {113}
  (\bibinfo {year} {2002})}\BibitemShut {NoStop}%
\bibitem [{\citenamefont {Steinbach}\ and\ \citenamefont
  {Shchyglo}(2011)}]{Steinbach2011}%
  \BibitemOpen
  \bibfield  {author} {\bibinfo {author} {\bibfnamefont {I.}~\bibnamefont
  {Steinbach}}\ and\ \bibinfo {author} {\bibfnamefont {O.}~\bibnamefont
  {Shchyglo}},\ }\href {\doibase 10.1016/j.cossms.2011.01.001} {\bibfield
  {journal} {\bibinfo  {journal} {Current Opinion in Solid State and Materials
  Science}\ }\textbf {\bibinfo {volume} {15}},\ \bibinfo {pages} {87} (\bibinfo
  {year} {2011})}\BibitemShut {NoStop}%
\bibitem [{\citenamefont {Lacasta}, \citenamefont {Hernandez-Machado},\ and\
  \citenamefont {Sancho}(1992)}]{Lacasta1992}%
  \BibitemOpen
  \bibfield  {author} {\bibinfo {author} {\bibfnamefont {A.~M.}\ \bibnamefont
  {Lacasta}}, \bibinfo {author} {\bibfnamefont {A.}~\bibnamefont
  {Hernandez-Machado}}, \ and\ \bibinfo {author} {\bibfnamefont {J.~M.}\
  \bibnamefont {Sancho}},\ }\href@noop {} {\bibfield  {journal} {\bibinfo
  {journal} {Physical Review B}\ }\textbf {\bibinfo {volume} {45}},\ \bibinfo
  {pages} {5276} (\bibinfo {year} {1992})}\BibitemShut {NoStop}%
\bibitem [{\citenamefont {Bray}\ and\ \citenamefont {Emmott}(1995)}]{Bray1995}%
  \BibitemOpen
  \bibfield  {author} {\bibinfo {author} {\bibfnamefont {A.~J.}\ \bibnamefont
  {Bray}}\ and\ \bibinfo {author} {\bibfnamefont {C.~L.}\ \bibnamefont
  {Emmott}},\ }\href {\doibase 10.1103/PhysRevB.52.R685} {\bibfield  {journal}
  {\bibinfo  {journal} {Physical Review B}\ }\textbf {\bibinfo {volume} {52}},\
  \bibinfo {pages} {685} (\bibinfo {year} {1995})}\BibitemShut {NoStop}%
\bibitem [{\citenamefont {Puri}, \citenamefont {Bray},\ and\ \citenamefont
  {Lebowitz}(1997)}]{Puri1997}%
  \BibitemOpen
  \bibfield  {author} {\bibinfo {author} {\bibfnamefont {S.}~\bibnamefont
  {Puri}}, \bibinfo {author} {\bibfnamefont {A.~J.}\ \bibnamefont {Bray}}, \
  and\ \bibinfo {author} {\bibfnamefont {J.~L.}\ \bibnamefont {Lebowitz}},\
  }\href {\doibase 10.1103/PhysRevE.56.758} {\bibfield  {journal} {\bibinfo
  {journal} {Physical Review E}\ }\textbf {\bibinfo {volume} {56}},\ \bibinfo
  {pages} {758} (\bibinfo {year} {1997})}\BibitemShut {NoStop}%
\bibitem [{\citenamefont {Zhu}\ \emph {et~al.}(1999)\citenamefont {Zhu},
  \citenamefont {Chen}, \citenamefont {Shen},\ and\ \citenamefont
  {Tikare}}]{Zhu1999}%
  \BibitemOpen
  \bibfield  {author} {\bibinfo {author} {\bibfnamefont {J.}~\bibnamefont
  {Zhu}}, \bibinfo {author} {\bibfnamefont {L.-Q.}\ \bibnamefont {Chen}},
  \bibinfo {author} {\bibfnamefont {J.}~\bibnamefont {Shen}}, \ and\ \bibinfo
  {author} {\bibfnamefont {V.}~\bibnamefont {Tikare}},\ }\href {\doibase
  10.1103/PhysRevE.60.3564} {\bibfield  {journal} {\bibinfo  {journal}
  {Physical Review E}\ }\textbf {\bibinfo {volume} {60}},\ \bibinfo {pages}
  {3564} (\bibinfo {year} {1999})}\BibitemShut {NoStop}%
\bibitem [{\citenamefont {Geslin}\ \emph {et~al.}(2019)\citenamefont {Geslin},
  \citenamefont {Buchet}, \citenamefont {Wada},\ and\ \citenamefont
  {Kato}}]{Geslin2019}%
  \BibitemOpen
  \bibfield  {author} {\bibinfo {author} {\bibfnamefont {P.~A.}\ \bibnamefont
  {Geslin}}, \bibinfo {author} {\bibfnamefont {M.}~\bibnamefont {Buchet}},
  \bibinfo {author} {\bibfnamefont {T.}~\bibnamefont {Wada}}, \ and\ \bibinfo
  {author} {\bibfnamefont {H.}~\bibnamefont {Kato}},\ }\href@noop {} {\bibfield
   {journal} {\bibinfo  {journal} {Physical Review Materials}\ }\textbf
  {\bibinfo {volume} {3}},\ \bibinfo {pages} {1} (\bibinfo {year}
  {2019})}\BibitemShut {NoStop}%
\bibitem [{\citenamefont {Salvalaglio}\ \emph {et~al.}(2020)\citenamefont
  {Salvalaglio}, \citenamefont {Selch}, \citenamefont {Voigt},\ and\
  \citenamefont {Wise}}]{Salvalaglio2020}%
  \BibitemOpen
  \bibfield  {author} {\bibinfo {author} {\bibfnamefont {M.}~\bibnamefont
  {Salvalaglio}}, \bibinfo {author} {\bibfnamefont {M.}~\bibnamefont {Selch}},
  \bibinfo {author} {\bibfnamefont {A.}~\bibnamefont {Voigt}}, \ and\ \bibinfo
  {author} {\bibfnamefont {S.~M.}\ \bibnamefont {Wise}},\ }\href@noop {}
  {\bibfield  {journal} {\bibinfo  {journal} {Mathematical Methods in the
  Applied Sciences}\ }\textbf {\bibinfo {volume} {44}},\ \bibinfo {pages}
  {5406} (\bibinfo {year} {2020})}\BibitemShut {NoStop}%
\bibitem [{\citenamefont {Andrews}\ \emph {et~al.}(2020)\citenamefont
  {Andrews}, \citenamefont {Elder}, \citenamefont {Voorhees},\ and\
  \citenamefont {Thornton}}]{Andrews2020}%
  \BibitemOpen
  \bibfield  {author} {\bibinfo {author} {\bibfnamefont {W.~B.}\ \bibnamefont
  {Andrews}}, \bibinfo {author} {\bibfnamefont {K.~L.}\ \bibnamefont {Elder}},
  \bibinfo {author} {\bibfnamefont {P.~W.}\ \bibnamefont {Voorhees}}, \ and\
  \bibinfo {author} {\bibfnamefont {K.}~\bibnamefont {Thornton}},\ }\href
  {\doibase 10.1103/PhysRevMaterials.4.103401} {\bibfield  {journal} {\bibinfo
  {journal} {Physical Review Materials}\ }\textbf {\bibinfo {volume} {4}},\
  \bibinfo {pages} {1} (\bibinfo {year} {2020})}\BibitemShut {NoStop}%
\bibitem [{\citenamefont {Joshi}, \citenamefont {Abinandanan},\ and\
  \citenamefont {Choudhury}(2016)}]{Joshi2016}%
  \BibitemOpen
  \bibfield  {author} {\bibinfo {author} {\bibfnamefont {C.}~\bibnamefont
  {Joshi}}, \bibinfo {author} {\bibfnamefont {T.~A.}\ \bibnamefont
  {Abinandanan}}, \ and\ \bibinfo {author} {\bibfnamefont {A.}~\bibnamefont
  {Choudhury}},\ }\href {\doibase 10.1016/j.actamat.2016.03.005} {\bibfield
  {journal} {\bibinfo  {journal} {Acta Materialia}\ }\textbf {\bibinfo {volume}
  {109}},\ \bibinfo {pages} {286} (\bibinfo {year} {2016})}\BibitemShut
  {NoStop}%
\bibitem [{\citenamefont {Joshi}\ \emph {et~al.}(2017)\citenamefont {Joshi},
  \citenamefont {Abinandanan}, \citenamefont {Mukherjee},\ and\ \citenamefont
  {Choudhury}}]{Joshi2017}%
  \BibitemOpen
  \bibfield  {author} {\bibinfo {author} {\bibfnamefont {C.}~\bibnamefont
  {Joshi}}, \bibinfo {author} {\bibfnamefont {T.~A.}\ \bibnamefont
  {Abinandanan}}, \bibinfo {author} {\bibfnamefont {R.}~\bibnamefont
  {Mukherjee}}, \ and\ \bibinfo {author} {\bibfnamefont {A.}~\bibnamefont
  {Choudhury}},\ }\href {\doibase 10.1016/j.commatsci.2017.07.026} {\bibfield
  {journal} {\bibinfo  {journal} {Computational Materials Science}\ }\textbf
  {\bibinfo {volume} {139}},\ \bibinfo {pages} {75} (\bibinfo {year}
  {2017})}\BibitemShut {NoStop}%
\bibitem [{\citenamefont {Mukherjee}\ \emph {et~al.}(2011)\citenamefont
  {Mukherjee}, \citenamefont {Chakrabarti}, \citenamefont {Anumol},
  \citenamefont {Abinandanan},\ and\ \citenamefont
  {Ravishankar}}]{Mukherjee2011}%
  \BibitemOpen
  \bibfield  {author} {\bibinfo {author} {\bibfnamefont {R.}~\bibnamefont
  {Mukherjee}}, \bibinfo {author} {\bibfnamefont {T.}~\bibnamefont
  {Chakrabarti}}, \bibinfo {author} {\bibfnamefont {E.~A.}\ \bibnamefont
  {Anumol}}, \bibinfo {author} {\bibfnamefont {T.~A.}\ \bibnamefont
  {Abinandanan}}, \ and\ \bibinfo {author} {\bibfnamefont {N.}~\bibnamefont
  {Ravishankar}},\ }\href {\doibase 10.1021/nn103036q} {\bibfield  {journal}
  {\bibinfo  {journal} {ACS Nano}\ }\textbf {\bibinfo {volume} {5}},\ \bibinfo
  {pages} {2700} (\bibinfo {year} {2011})}\BibitemShut {NoStop}%
\bibitem [{\citenamefont {Chakrabarti}, \citenamefont {Verma},\ and\
  \citenamefont {Manna}(2017)}]{Chakrabarti2017}%
  \BibitemOpen
  \bibfield  {author} {\bibinfo {author} {\bibfnamefont {T.}~\bibnamefont
  {Chakrabarti}}, \bibinfo {author} {\bibfnamefont {N.}~\bibnamefont {Verma}},
  \ and\ \bibinfo {author} {\bibfnamefont {S.}~\bibnamefont {Manna}},\
  }\href@noop {} {\bibfield  {journal} {\bibinfo  {journal} {Materials and
  Design}\ }\textbf {\bibinfo {volume} {119}},\ \bibinfo {pages} {425}
  (\bibinfo {year} {2017})}\BibitemShut {NoStop}%
\bibitem [{\citenamefont {Debierre}\ \emph {et~al.}(2003)\citenamefont
  {Debierre}, \citenamefont {Karma}, \citenamefont {Celestini},\ and\
  \citenamefont {Gu{\'{e}}rin}}]{Debierre2003}%
  \BibitemOpen
  \bibfield  {author} {\bibinfo {author} {\bibfnamefont {J.~M.}\ \bibnamefont
  {Debierre}}, \bibinfo {author} {\bibfnamefont {A.}~\bibnamefont {Karma}},
  \bibinfo {author} {\bibfnamefont {F.}~\bibnamefont {Celestini}}, \ and\
  \bibinfo {author} {\bibfnamefont {R.}~\bibnamefont {Gu{\'{e}}rin}},\
  }\href@noop {} {\bibfield  {journal} {\bibinfo  {journal} {Physical Review
  E}\ }\textbf {\bibinfo {volume} {68}},\ \bibinfo {pages} {1} (\bibinfo {year}
  {2003})}\BibitemShut {NoStop}%
\bibitem [{\citenamefont {Ji}\ \emph {et~al.}(2018)\citenamefont {Ji},
  \citenamefont {Ghaffari}, \citenamefont {Li},\ and\ \citenamefont
  {Chen}}]{Ji2018}%
  \BibitemOpen
  \bibfield  {author} {\bibinfo {author} {\bibfnamefont {Y.}~\bibnamefont
  {Ji}}, \bibinfo {author} {\bibfnamefont {B.}~\bibnamefont {Ghaffari}},
  \bibinfo {author} {\bibfnamefont {M.}~\bibnamefont {Li}}, \ and\ \bibinfo
  {author} {\bibfnamefont {L.~Q.}\ \bibnamefont {Chen}},\ }\href {\doibase
  10.1016/j.commatsci.2018.04.051} {\bibfield  {journal} {\bibinfo  {journal}
  {Computational Materials Science}\ }\textbf {\bibinfo {volume} {151}},\
  \bibinfo {pages} {84} (\bibinfo {year} {2018})}\BibitemShut {NoStop}%
\bibitem [{\citenamefont {Abinandanan}\ and\ \citenamefont
  {Haider}(2001)}]{Abinandanan2002}%
  \BibitemOpen
  \bibfield  {author} {\bibinfo {author} {\bibfnamefont {T.~A.}\ \bibnamefont
  {Abinandanan}}\ and\ \bibinfo {author} {\bibfnamefont {F.}~\bibnamefont
  {Haider}},\ }\href {\doibase 10.1080/01418610110038420} {\bibfield  {journal}
  {\bibinfo  {journal} {Philosophical Magazine A}\ }\textbf {\bibinfo {volume}
  {81}},\ \bibinfo {pages} {2457} (\bibinfo {year} {2001})}\BibitemShut
  {NoStop}%
\bibitem [{\citenamefont {Wang}(2011)}]{Wang2011}%
  \BibitemOpen
  \bibfield  {author} {\bibinfo {author} {\bibfnamefont {N.}~\bibnamefont
  {Wang}},\ }\emph {\bibinfo {title} {Phase-field studies of materials
  interfaces}},\ \href@noop {} {\bibinfo {type} {{Ph.D Thesis}}},\ \bibinfo
  {school} {Northeastern University} (\bibinfo {year} {2011})\BibitemShut
  {NoStop}%
\bibitem [{\citenamefont {Roy}\ \emph {et~al.}(2017)\citenamefont {Roy},
  \citenamefont {Nani}, \citenamefont {Lahiri},\ and\ \citenamefont
  {Gururajan}}]{Roy2017}%
  \BibitemOpen
  \bibfield  {author} {\bibinfo {author} {\bibfnamefont {A.}~\bibnamefont
  {Roy}}, \bibinfo {author} {\bibfnamefont {E.~S.}\ \bibnamefont {Nani}},
  \bibinfo {author} {\bibfnamefont {A.}~\bibnamefont {Lahiri}}, \ and\ \bibinfo
  {author} {\bibfnamefont {M.~P.}\ \bibnamefont {Gururajan}},\ }\href {\doibase
  10.1080/14786435.2017.1348633} {\bibfield  {journal} {\bibinfo  {journal}
  {Philosophical Magazine}\ }\textbf {\bibinfo {volume} {97}},\ \bibinfo
  {pages} {2705} (\bibinfo {year} {2017})}\BibitemShut {NoStop}%
\bibitem [{\citenamefont {Cahn}\ and\ \citenamefont
  {Hilliard}(1958)}]{Cahn1958}%
  \BibitemOpen
  \bibfield  {author} {\bibinfo {author} {\bibfnamefont {J.~W.}\ \bibnamefont
  {Cahn}}\ and\ \bibinfo {author} {\bibfnamefont {J.~E.}\ \bibnamefont
  {Hilliard}},\ }\href {\doibase 10.1063/1.1744102} {\bibfield  {journal}
  {\bibinfo  {journal} {The Journal of Chemical Physics}\ }\textbf {\bibinfo
  {volume} {28}},\ \bibinfo {pages} {258} (\bibinfo {year} {1958})}\BibitemShut
  {NoStop}%
\bibitem [{\citenamefont {Dziwnik}, \citenamefont {M{\"{u}}nch},\ and\
  \citenamefont {Wagner}(2017)}]{Dziwnik2017}%
  \BibitemOpen
  \bibfield  {author} {\bibinfo {author} {\bibfnamefont {M.}~\bibnamefont
  {Dziwnik}}, \bibinfo {author} {\bibfnamefont {A.}~\bibnamefont
  {M{\"{u}}nch}}, \ and\ \bibinfo {author} {\bibfnamefont {B.}~\bibnamefont
  {Wagner}},\ }\href {\doibase 10.1088/1361-6544/aa5e5d} {\bibfield  {journal}
  {\bibinfo  {journal} {Nonlinearity}\ }\textbf {\bibinfo {volume} {30}},\
  \bibinfo {pages} {1465} (\bibinfo {year} {2017})}\BibitemShut {NoStop}%
\bibitem [{\citenamefont {Jiang}\ \emph {et~al.}(2012)\citenamefont {Jiang},
  \citenamefont {Bao}, \citenamefont {Thompson},\ and\ \citenamefont
  {Srolovitz}}]{Jiang2012}%
  \BibitemOpen
  \bibfield  {author} {\bibinfo {author} {\bibfnamefont {W.}~\bibnamefont
  {Jiang}}, \bibinfo {author} {\bibfnamefont {W.}~\bibnamefont {Bao}}, \bibinfo
  {author} {\bibfnamefont {C.~V.}\ \bibnamefont {Thompson}}, \ and\ \bibinfo
  {author} {\bibfnamefont {D.~J.}\ \bibnamefont {Srolovitz}},\ }\href {\doibase
  10.1016/j.actamat.2012.07.002} {\bibfield  {journal} {\bibinfo  {journal}
  {Acta Materialia}\ }\textbf {\bibinfo {volume} {60}},\ \bibinfo {pages}
  {5578} (\bibinfo {year} {2012})}\BibitemShut {NoStop}%
\bibitem [{\citenamefont {Verma}\ and\ \citenamefont
  {Mukherjee}(2020)}]{Verma2020}%
  \BibitemOpen
  \bibfield  {author} {\bibinfo {author} {\bibfnamefont {M.}~\bibnamefont
  {Verma}}\ and\ \bibinfo {author} {\bibfnamefont {R.}~\bibnamefont
  {Mukherjee}},\ }\href {\doibase 10.1016/j.jallcom.2020.155163} {\bibfield
  {journal} {\bibinfo  {journal} {Journal of Alloys and Compounds}\ }\textbf
  {\bibinfo {volume} {835}},\ \bibinfo {pages} {155163} (\bibinfo {year}
  {2020})}\BibitemShut {NoStop}%
\bibitem [{\citenamefont {Cheynis}\ \emph {et~al.}(2012)\citenamefont
  {Cheynis}, \citenamefont {Bussmann}, \citenamefont {Leroy}, \citenamefont
  {Passanante},\ and\ \citenamefont {M{\"{u}}ller}}]{Cheynis2012}%
  \BibitemOpen
  \bibfield  {author} {\bibinfo {author} {\bibfnamefont {F.}~\bibnamefont
  {Cheynis}}, \bibinfo {author} {\bibfnamefont {E.}~\bibnamefont {Bussmann}},
  \bibinfo {author} {\bibfnamefont {F.}~\bibnamefont {Leroy}}, \bibinfo
  {author} {\bibfnamefont {T.}~\bibnamefont {Passanante}}, \ and\ \bibinfo
  {author} {\bibfnamefont {P.}~\bibnamefont {M{\"{u}}ller}},\ }\href {\doibase
  10.1504/IJNT.2012.045344} {\bibfield  {journal} {\bibinfo  {journal}
  {International Journal of Nanotechnology}\ }\textbf {\bibinfo {volume} {9}},\
  \bibinfo {pages} {396} (\bibinfo {year} {2012})}\BibitemShut {NoStop}%
\bibitem [{\citenamefont {Liu}\ and\ \citenamefont {Plawsky}(2017)}]{Liu2017}%
  \BibitemOpen
  \bibfield  {author} {\bibinfo {author} {\bibfnamefont {S.}~\bibnamefont
  {Liu}}\ and\ \bibinfo {author} {\bibfnamefont {J.~L.}\ \bibnamefont
  {Plawsky}},\ }\href {\doibase 10.1021/acs.langmuir.7b03259} {\bibfield
  {journal} {\bibinfo  {journal} {Langmuir}\ }\textbf {\bibinfo {volume}
  {33}},\ \bibinfo {pages} {14066} (\bibinfo {year} {2017})}\BibitemShut
  {NoStop}%
\bibitem [{\citenamefont {Liu}\ \emph {et~al.}(2019)\citenamefont {Liu},
  \citenamefont {Berbezier}, \citenamefont {Favre}, \citenamefont {Ronda},
  \citenamefont {David}, \citenamefont {Abbarchi}, \citenamefont {Gaillard},
  \citenamefont {Frisch}, \citenamefont {Croset},\ and\ \citenamefont
  {Aqua}}]{LiuEtAl2019}%
  \BibitemOpen
  \bibfield  {author} {\bibinfo {author} {\bibfnamefont {K.}~\bibnamefont
  {Liu}}, \bibinfo {author} {\bibfnamefont {I.}~\bibnamefont {Berbezier}},
  \bibinfo {author} {\bibfnamefont {L.}~\bibnamefont {Favre}}, \bibinfo
  {author} {\bibfnamefont {A.}~\bibnamefont {Ronda}}, \bibinfo {author}
  {\bibfnamefont {T.}~\bibnamefont {David}}, \bibinfo {author} {\bibfnamefont
  {M.}~\bibnamefont {Abbarchi}}, \bibinfo {author} {\bibfnamefont
  {P.}~\bibnamefont {Gaillard}}, \bibinfo {author} {\bibfnamefont
  {T.}~\bibnamefont {Frisch}}, \bibinfo {author} {\bibfnamefont
  {B.}~\bibnamefont {Croset}}, \ and\ \bibinfo {author} {\bibfnamefont {J.-N.}\
  \bibnamefont {Aqua}},\ }\href {\doibase 10.1103/PhysRevMaterials.3.023403}
  {\bibfield  {journal} {\bibinfo  {journal} {Physical Review Materials}\
  }\textbf {\bibinfo {volume} {3}},\ \bibinfo {pages} {023403} (\bibinfo {year}
  {2019})}\BibitemShut {NoStop}%
\bibitem [{\citenamefont {Cahn}, \citenamefont {Elliott},\ and\ \citenamefont
  {Novick-Cohen}(1996)}]{Cahn1996}%
  \BibitemOpen
  \bibfield  {author} {\bibinfo {author} {\bibfnamefont {J.~W.}\ \bibnamefont
  {Cahn}}, \bibinfo {author} {\bibfnamefont {C.~M.}\ \bibnamefont {Elliott}}, \
  and\ \bibinfo {author} {\bibfnamefont {A.}~\bibnamefont {Novick-Cohen}},\
  }\href {\doibase 10.1017/s0956792500002369} {\bibfield  {journal} {\bibinfo
  {journal} {European Journal of Applied Mathematics}\ }\textbf {\bibinfo
  {volume} {7}},\ \bibinfo {pages} {287} (\bibinfo {year} {1996})}\BibitemShut
  {NoStop}%
\bibitem [{\citenamefont {Gugenberger}, \citenamefont {Spatschek},\ and\
  \citenamefont {Kassner}(2008)}]{Gugenberger2008}%
  \BibitemOpen
  \bibfield  {author} {\bibinfo {author} {\bibfnamefont {C.}~\bibnamefont
  {Gugenberger}}, \bibinfo {author} {\bibfnamefont {R.}~\bibnamefont
  {Spatschek}}, \ and\ \bibinfo {author} {\bibfnamefont {K.}~\bibnamefont
  {Kassner}},\ }\href {\doibase 10.1103/PhysRevE.78.016703} {\bibfield
  {journal} {\bibinfo  {journal} {Physical Review E}\ }\textbf {\bibinfo
  {volume} {78}},\ \bibinfo {pages} {1} (\bibinfo {year} {2008})},\ \Eprint
  {http://arxiv.org/abs/0711.1809} {0711.1809} \BibitemShut {NoStop}%
\bibitem [{\citenamefont {Dziwnik}(2019)}]{Dziwnik2019}%
  \BibitemOpen
  \bibfield  {author} {\bibinfo {author} {\bibfnamefont {M.}~\bibnamefont
  {Dziwnik}},\ }\href {\doibase 10.1002/pamm.201900396} {\bibfield  {journal}
  {\bibinfo  {journal} {Proceedings in Applied Mathematics and Mechanics}\
  }\textbf {\bibinfo {volume} {19}},\ \bibinfo {pages} {29} (\bibinfo {year}
  {2019})}\BibitemShut {NoStop}%
\bibitem [{\citenamefont {Shin}, \citenamefont {Choi},\ and\ \citenamefont
  {Kim}(2019)}]{Shin2019}%
  \BibitemOpen
  \bibfield  {author} {\bibinfo {author} {\bibfnamefont {J.}~\bibnamefont
  {Shin}}, \bibinfo {author} {\bibfnamefont {Y.}~\bibnamefont {Choi}}, \ and\
  \bibinfo {author} {\bibfnamefont {J.}~\bibnamefont {Kim}},\ }\href@noop {}
  {\bibfield  {journal} {\bibinfo  {journal} {Mathematical Problems in
  Engineering}\ }\textbf {\bibinfo {volume} {2019}} (\bibinfo {year}
  {2019})}\BibitemShut {NoStop}%
\bibitem [{\citenamefont {Hoffrogge}\ \emph {et~al.}(2021)\citenamefont
  {Hoffrogge}, \citenamefont {Mukherjee}, \citenamefont {Nani}, \citenamefont
  {Amos}, \citenamefont {Wang}, \citenamefont {Schneider},\ and\ \citenamefont
  {Nestler}}]{Hoffrogge2021}%
  \BibitemOpen
  \bibfield  {author} {\bibinfo {author} {\bibfnamefont {P.~W.}\ \bibnamefont
  {Hoffrogge}}, \bibinfo {author} {\bibfnamefont {A.}~\bibnamefont
  {Mukherjee}}, \bibinfo {author} {\bibfnamefont {E.~S.}\ \bibnamefont {Nani}},
  \bibinfo {author} {\bibfnamefont {P.~G.~K.}\ \bibnamefont {Amos}}, \bibinfo
  {author} {\bibfnamefont {F.}~\bibnamefont {Wang}}, \bibinfo {author}
  {\bibfnamefont {D.}~\bibnamefont {Schneider}}, \ and\ \bibinfo {author}
  {\bibfnamefont {B.}~\bibnamefont {Nestler}},\ }\href@noop {} {\bibfield
  {journal} {\bibinfo  {journal} {Physical Review E}\ }\textbf {\bibinfo
  {volume} {103}},\ \bibinfo {pages} {1} (\bibinfo {year} {2021})}\BibitemShut
  {NoStop}%
\bibitem [{\citenamefont {Chen}\ and\ \citenamefont {Shen}(1998)}]{Chen1998}%
  \BibitemOpen
  \bibfield  {author} {\bibinfo {author} {\bibfnamefont {L.-Q.}\ \bibnamefont
  {Chen}}\ and\ \bibinfo {author} {\bibfnamefont {J.}~\bibnamefont {Shen}},\
  }\href {\doibase 10.1016/s0010-4655(97)00115-x} {\bibfield  {journal}
  {\bibinfo  {journal} {Computer Physics Communications}\ }\textbf {\bibinfo
  {volume} {108}},\ \bibinfo {pages} {147} (\bibinfo {year}
  {1998})}\BibitemShut {NoStop}%
\bibitem [{\citenamefont {Frigo}\ and\ \citenamefont
  {Johnson}(2005)}]{Frigo2005}%
  \BibitemOpen
  \bibfield  {author} {\bibinfo {author} {\bibfnamefont {M.}~\bibnamefont
  {Frigo}}\ and\ \bibinfo {author} {\bibfnamefont {S.~G.}\ \bibnamefont
  {Johnson}},\ }\href {\doibase 10.1109/JPROC.2004.840301} {\bibfield
  {journal} {\bibinfo  {journal} {Proceedings of the IEEE}\ }\textbf {\bibinfo
  {volume} {93}},\ \bibinfo {pages} {216} (\bibinfo {year} {2005})}\BibitemShut
  {NoStop}%
\bibitem [{\citenamefont {Roy}\ and\ \citenamefont
  {Gururajan}(2015)}]{Roy2015}%
  \BibitemOpen
  \bibfield  {author} {\bibinfo {author} {\bibfnamefont {A.}~\bibnamefont
  {Roy}}\ and\ \bibinfo {author} {\bibfnamefont {M.~P.}\ \bibnamefont
  {Gururajan}},\ }\href@noop {} {\bibfield  {journal} {\bibinfo  {journal} {PTM
  2015 - Proceedings of the International Conference on Solid-Solid Phase
  Transformations in Inorganic Materials 2015}\ ,\ \bibinfo {pages} {325}}
  (\bibinfo {year} {2015})}\BibitemShut {NoStop}%
\bibitem [{\citenamefont {Mullins}(1957)}]{Mullins1957}%
  \BibitemOpen
  \bibfield  {author} {\bibinfo {author} {\bibfnamefont {W.~W.}\ \bibnamefont
  {Mullins}},\ }\href {\doibase 10.1063/1.1722742} {\bibfield  {journal}
  {\bibinfo  {journal} {Journal of Applied Physics}\ }\textbf {\bibinfo
  {volume} {28}},\ \bibinfo {pages} {333} (\bibinfo {year} {1957})}\BibitemShut
  {NoStop}%
\bibitem [{\citenamefont {Gururajan}(2006)}]{gururajan2006}%
  \BibitemOpen
  \bibfield  {author} {\bibinfo {author} {\bibfnamefont {M.}~\bibnamefont
  {Gururajan}},\ }\emph {\bibinfo {title} {Elastic Inhomogeneity Effects on
  Microstructures}},\ \href@noop {} {Ph.D. thesis},\ \bibinfo  {school} {Indian
  Institute of Science Bangalore} (\bibinfo {year} {2006})\BibitemShut
  {NoStop}%
\bibitem [{\citenamefont {Bellew}\ \emph {et~al.}(2015)\citenamefont {Bellew},
  \citenamefont {Manning}, \citenamefont {{Gomes da Rocha}}, \citenamefont
  {Ferreira},\ and\ \citenamefont {Boland}}]{Bellew2015}%
  \BibitemOpen
  \bibfield  {author} {\bibinfo {author} {\bibfnamefont {A.~T.}\ \bibnamefont
  {Bellew}}, \bibinfo {author} {\bibfnamefont {H.~G.}\ \bibnamefont {Manning}},
  \bibinfo {author} {\bibfnamefont {C.}~\bibnamefont {{Gomes da Rocha}}},
  \bibinfo {author} {\bibfnamefont {M.~S.}\ \bibnamefont {Ferreira}}, \ and\
  \bibinfo {author} {\bibfnamefont {J.~J.}\ \bibnamefont {Boland}},\ }\href
  {\doibase 10.1021/acsnano.5b05469} {\bibfield  {journal} {\bibinfo  {journal}
  {ACS Nano}\ }\textbf {\bibinfo {volume} {9}},\ \bibinfo {pages} {11422}
  (\bibinfo {year} {2015})}\BibitemShut {NoStop}%
\bibitem [{\citenamefont {Chockalingam}\ \emph {et~al.}(2016)\citenamefont
  {Chockalingam}, \citenamefont {Kouznetsova}, \citenamefont {van~der Sluis},\
  and\ \citenamefont {Geers}}]{Chockalingam2016}%
  \BibitemOpen
  \bibfield  {author} {\bibinfo {author} {\bibfnamefont {K.}~\bibnamefont
  {Chockalingam}}, \bibinfo {author} {\bibfnamefont {V.~G.}\ \bibnamefont
  {Kouznetsova}}, \bibinfo {author} {\bibfnamefont {O.}~\bibnamefont {van~der
  Sluis}}, \ and\ \bibinfo {author} {\bibfnamefont {M.~G.}\ \bibnamefont
  {Geers}},\ }\href {\doibase 10.1016/j.cma.2016.07.002} {\bibfield  {journal}
  {\bibinfo  {journal} {Computer Methods in Applied Mechanics and Engineering}\
  }\textbf {\bibinfo {volume} {312}},\ \bibinfo {pages} {492} (\bibinfo {year}
  {2016})}\BibitemShut {NoStop}%
\bibitem [{\citenamefont {Jahangir}, \citenamefont {Devaraj},\ and\
  \citenamefont {Malhotra}(2020)}]{Jahangir2020}%
  \BibitemOpen
  \bibfield  {author} {\bibinfo {author} {\bibfnamefont {M.~N.}\ \bibnamefont
  {Jahangir}}, \bibinfo {author} {\bibfnamefont {H.}~\bibnamefont {Devaraj}}, \
  and\ \bibinfo {author} {\bibfnamefont {R.}~\bibnamefont {Malhotra}},\ }\href
  {\doibase 10.1021/acs.jpcc.0c05716} {\bibfield  {journal} {\bibinfo
  {journal} {Journal of Physical Chemistry C}\ }\textbf {\bibinfo {volume}
  {124}},\ \bibinfo {pages} {19849} (\bibinfo {year} {2020})}\BibitemShut
  {NoStop}%
\bibitem [{\citenamefont {Mendoza}, \citenamefont {Alkemper},\ and\
  \citenamefont {Voorhees}(2003)}]{Mendoza2003}%
  \BibitemOpen
  \bibfield  {author} {\bibinfo {author} {\bibfnamefont {R.}~\bibnamefont
  {Mendoza}}, \bibinfo {author} {\bibfnamefont {J.}~\bibnamefont {Alkemper}}, \
  and\ \bibinfo {author} {\bibfnamefont {P.}~\bibnamefont {Voorhees}},\
  }\href@noop {} {\bibfield  {journal} {\bibinfo  {journal} {Metallurgical and
  Materials Transactions A}\ }\textbf {\bibinfo {volume} {34}},\ \bibinfo
  {pages} {481} (\bibinfo {year} {2003})}\BibitemShut {NoStop}%
\bibitem [{\citenamefont {Kwon}, \citenamefont {Thornton},\ and\ \citenamefont
  {Voorhees}(2010)}]{Kwon2010}%
  \BibitemOpen
  \bibfield  {author} {\bibinfo {author} {\bibfnamefont {Y.}~\bibnamefont
  {Kwon}}, \bibinfo {author} {\bibfnamefont {K.}~\bibnamefont {Thornton}}, \
  and\ \bibinfo {author} {\bibfnamefont {P.}~\bibnamefont {Voorhees}},\
  }\href@noop {} {\bibfield  {journal} {\bibinfo  {journal} {Philosophical
  Magazine}\ }\textbf {\bibinfo {volume} {90}},\ \bibinfo {pages} {317}
  (\bibinfo {year} {2010})}\BibitemShut {NoStop}%
\bibitem [{\citenamefont {Park}, \citenamefont {Voorhees},\ and\ \citenamefont
  {Thornton}(2014)}]{Park2014}%
  \BibitemOpen
  \bibfield  {author} {\bibinfo {author} {\bibfnamefont {C.-L.}\ \bibnamefont
  {Park}}, \bibinfo {author} {\bibfnamefont {P.~W.}\ \bibnamefont {Voorhees}},
  \ and\ \bibinfo {author} {\bibfnamefont {K.}~\bibnamefont {Thornton}},\
  }\href@noop {} {\bibfield  {journal} {\bibinfo  {journal} {Computational
  Materials Science}\ }\textbf {\bibinfo {volume} {85}},\ \bibinfo {pages} {46}
  (\bibinfo {year} {2014})}\BibitemShut {NoStop}%
\bibitem [{\citenamefont {Park}\ \emph {et~al.}(2017)\citenamefont {Park},
  \citenamefont {Gibbs}, \citenamefont {Voorhees},\ and\ \citenamefont
  {Thornton}}]{Park2017}%
  \BibitemOpen
  \bibfield  {author} {\bibinfo {author} {\bibfnamefont {C.-L.}\ \bibnamefont
  {Park}}, \bibinfo {author} {\bibfnamefont {J.}~\bibnamefont {Gibbs}},
  \bibinfo {author} {\bibfnamefont {P.}~\bibnamefont {Voorhees}}, \ and\
  \bibinfo {author} {\bibfnamefont {K.}~\bibnamefont {Thornton}},\ }\href@noop
  {} {\bibfield  {journal} {\bibinfo  {journal} {Acta Materialia}\ }\textbf
  {\bibinfo {volume} {132}},\ \bibinfo {pages} {13} (\bibinfo {year}
  {2017})}\BibitemShut {NoStop}%
\bibitem [{\citenamefont {Li}, \citenamefont {Han},\ and\ \citenamefont
  {Ruan}(2018)}]{Li2018}%
  \BibitemOpen
  \bibfield  {author} {\bibinfo {author} {\bibfnamefont {D.}~\bibnamefont
  {Li}}, \bibinfo {author} {\bibfnamefont {T.}~\bibnamefont {Han}}, \ and\
  \bibinfo {author} {\bibfnamefont {H.}~\bibnamefont {Ruan}},\ }\href {\doibase
  10.1021/acsomega.8b01320} {\bibfield  {journal} {\bibinfo  {journal} {ACS
  Omega}\ }\textbf {\bibinfo {volume} {3}},\ \bibinfo {pages} {7191} (\bibinfo
  {year} {2018})}\BibitemShut {NoStop}%
\bibitem [{\citenamefont {Hu}\ \emph {et~al.}(2019)\citenamefont {Hu},
  \citenamefont {Liang}, \citenamefont {Sun}, \citenamefont {Zheng},
  \citenamefont {Duan},\ and\ \citenamefont {Zhuang}}]{Hu2019}%
  \BibitemOpen
  \bibfield  {author} {\bibinfo {author} {\bibfnamefont {Y.}~\bibnamefont
  {Hu}}, \bibinfo {author} {\bibfnamefont {C.}~\bibnamefont {Liang}}, \bibinfo
  {author} {\bibfnamefont {X.}~\bibnamefont {Sun}}, \bibinfo {author}
  {\bibfnamefont {J.}~\bibnamefont {Zheng}}, \bibinfo {author} {\bibfnamefont
  {J.}~\bibnamefont {Duan}}, \ and\ \bibinfo {author} {\bibfnamefont
  {X.}~\bibnamefont {Zhuang}},\ }\href@noop {} {\bibfield  {journal} {\bibinfo
  {journal} {Nanomaterials}\ }\textbf {\bibinfo {volume} {9}} (\bibinfo {year}
  {2019})}\BibitemShut {NoStop}%
\bibitem [{\citenamefont {Hoshen}\ and\ \citenamefont
  {Kopelman}(1976)}]{hoshen1976}%
  \BibitemOpen
  \bibfield  {author} {\bibinfo {author} {\bibfnamefont {J.}~\bibnamefont
  {Hoshen}}\ and\ \bibinfo {author} {\bibfnamefont {R.}~\bibnamefont
  {Kopelman}},\ }\href@noop {} {\bibfield  {journal} {\bibinfo  {journal}
  {Physical Review B}\ }\textbf {\bibinfo {volume} {14}},\ \bibinfo {pages}
  {3438} (\bibinfo {year} {1976})}\BibitemShut {NoStop}%
\end{thebibliography}%
\end{document}